\pdfoutput=1

\documentclass[a4paper,11pt]{article}

\usepackage[refcommands]{heppub} 
\usepackage{bbm}
\usepackage{subfigure}
\graphicspath{ {./images/} }
\usepackage{mathtools}

\usepackage{tikz}
\usetikzlibrary{arrows,decorations.markings}
\tikzset{myptr/.style={decoration={markings,mark=at position 1 with 
			{\arrow[scale=2]{>}}},postaction={decorate}}}
\definecolor{ublue}{HTML}{0063A6}
\definecolor{ured}{HTML}{DD4814}
\definecolor{uviolet}{HTML}{A71C49}

\usepackage{bm}
\usepackage{graphicx}
\usepackage{amsmath,amssymb,mathtools} 
\usepackage{xspace}
\usepackage{wrapfig}
\usepackage{slashed}
\usepackage[normalem]{ulem}
\usepackage[utf8]{inputenc}
\usepackage{mathtools}
\usepackage{stmaryrd}
\usepackage{appendix}
\usepackage{upgreek}
\usepackage{multirow}

\usepackage{scrextend}

\usepackage{tabularx}
\usepackage{tensor}

\newcommand{\mb}[1]{\boldsymbol{#1}}

\definecolor{darkgreen}{rgb}{0.0, 0.4, 0.0}

\newcommand{\df}{\mathrm{d}}

\newcommand{\bn}{{\bar{n}}}

\newcommand{\ra}{\rightarrow}
\newcommand{\lra}{\leftrightarrow}

\newcommand{\eps}{\epsilon}

\newcommand{\bslash}{{b\!\!\!\slash}}

\newcommand{\cP}{{\mathcal P}}

\newcommand{\tnslash}{\tilde{n}\!\!\!\slash}
\newcommand{\bnslash}{\bar{n}\!\!\!\slash}

\def\dfbar{{\:\mathchar'26\mkern-12mu \df}}

\newcommand{\nn}{\nonumber}

\newcommand{\lqcd}{\Lambda_\mathrm{QCD}}

\newcommand{\tn}{ {\tilde{n}} }

\newcommand{\SCETb}{\ensuremath{{\rm SCET}_{\rm II}}\xspace}

\newcommand{\Pythiaxx}{\texttt{Pythia\xspace8.3}\xspace}
\newcommand{\Pythia}{\texttt{Pythia}\xspace}

\newcommand{\Herwig}{\texttt{Herwig}\xspace}
\newcommand{\Herwigxx}{\texttt{Herwig\xspace7.3}\xspace}

\def\df{\textrm{d}}
\def\nn{\nonumber}

\newcommand{\ee}{{e^+e^-}}

\allowdisplaybreaks[2]

\DeclareMathOperator{\Tr}{Tr}

\newcommand{\im}{\mathrm{i}}
\newcommand{\cG}{\mathcal{G}}
\newcommand{\cF}{\mathcal{F}}

\newcommand{\cC}{\mathcal{C}}
\newcommand{\cO}{\mathcal{O}}

\newcommand{\cH}{\mathcal{H}}

\usepackage[mathscr]{eucal}

\newcommand{\cD}{\mathcal{D}}

\newcommand{\cE}{\mathcal{E}}

\DeclareMathOperator*{\sumint}{%
    \mathchoice%
    {\ooalign{$\displaystyle\sum$\cr\hidewidth$\displaystyle\int$\hidewidth\cr}}
    {\ooalign{\raisebox{.14\height}{\scalebox{.7}{$\textstyle\sum$}}\cr\hidewidth$\textstyle\int$\hidewidth\cr}}
    {\ooalign{\raisebox{.2\height}{\scalebox{.6}{$\scriptstyle\sum$}}\cr$\scriptstyle\int$\cr}}
    {\ooalign{\raisebox{.2\height}{\scalebox{.6}{$\scriptstyle\sum$}}\cr$\scriptstyle\int$\cr}}
}

\newcommand{\comment}[1]{}

\newcommand{\td}{\mathrm{d}}

\renewcommand{\v}[1]{\boldsymbol{#1}}

\renewcommand{\>}{\right\rangle}
\newcommand{\<}{\left\langle}
\newcommand{\lkl}{\left|}

\usepackage{breakurl}

\preprint{
	\begin{flushright}
		DESY-26-012\\
		UWThPh 2026-2
\end{flushright}}

\title{High precision heavy-boson-jet substructure with energy correlators}

\author[a]{Jack Holguin,}
\affiliation[a]{Department of Physics and Astronomy,
	University of Manchester, \\ Manchester M13 9PL, United Kingdom}
\emailAdd{jack.holguin@manchester.ac.uk}

\author[b]{Ian Moult,}
\affiliation[b]{Department of Physics, Yale University, New Haven, CT 06511}
\emailAdd{ian.moult@yale.edu}

\author[c]{Aditya Pathak,}
\affiliation[c]{Deutsches Elektronen-Synchrotron DESY, Notkestr. 85, 22607 Hamburg, Germany}
\emailAdd{aditya.pathak@desy.de}

\author[d]{Massimiliano Procura,}
\affiliation[d]{University of Vienna, Faculty of Physics, Boltzmanngasse 5, A-1090 Vienna, Austria}
\emailAdd{mprocura@univie.ac.at}

\author[a]{Siddharth Sule}
\emailAdd{siddharth.sule@manchester.ac.uk}

\abstract{Energy-correlator-based jet substructure has gained significant attention in recent years. One of the notable applications has been the study of multi-scale jets, where distinct physical scales manifest as features localised in different angular regions of the correlator. In this article, we present the first high-precision study of energy correlators on the simplest multi-scale jets: heavy boson jets. In such systems, the boson mass $M$ introduces an additional scale, generating a sharp peak at angles $\sim M/p_T^{\rm jet}$. We show that this feature can be computed directly by boosting the EEC spectrum measured in $e^+e^- \rightarrow {\rm hadrons}$ at the $Z$ pole. We identify that the peak arises from boosting the well-studied Sudakov factorisation governing the back-to-back limit of the two-point correlator. As a result, the feature is controlled by Sudakov resummation, not a Breit-Wigner-like structure in the $Z$ decay, and is therefore calculable with exceptional precision. We provide predictions at N$^3$LL$'$ accuracy for both $pp$ $Z$-tagged jets and $e^+e^-$ di-$Z$ production, and compare them to \Herwig and \Pythia simulations, finding close agreement. We also demonstrate that the boosted-$Z$ spectrum can be constructed directly by boosting OPAL measurements at the $Z$ pole. In this light, energy-correlator jet substructure on the hadronic decays of heavy bosons at the LHC provide access to clean, lepton-collider-like measurements across a wide range of effective centre-of-mass energies set by the boson jet transverse momentum.
}

\begin{document}
	
\maketitle

\section{Introduction}

Recent years have seen substantial growth in the application of energy-energy correlators (EECs) to the study of collider jet substructure. The renewed interest began with the analysis of EECs in massless quark and gluon jets~\cite{Dixon:2019uzg,Komiske:2022enw}, where the distributions exhibit a striking power-law scaling. This behaviour reflects the near-critical dynamics of an asymptotically free gauge theory and provides a direct window into the fractal structure of successive QCD branchings. The precise perturbative calculability of this regime has enabled EECs to deliver the most precise extraction of the QCD coupling from a jet substructure observable~\cite{CMS:2024mlf}. 

Following these developments, significant attention has turned toward situations in which the asymptotically-free scaling is broken by the presence of additional physical scales~\cite{Craft:2022kdo,Holguin:2022epo,Holguin:2023bjf,Andres:2022ovj,Andres:2023ymw,Barata:2023bhh,Devereaux:2023vjz}. The first such studies focused on heavy-flavour~\cite{Craft:2022kdo} and top jets~\cite{Holguin:2022epo}, where the quark mass leaves a visible imprint on the EEC spectrum. In top jets, the decay generates a threshold-like peak correlated with the kinematics of the top decay products. For charm and bottom jets, by contrast, the heavy-quark mass suppresses collinear radiation through the dead-cone effect, producing a characteristic angular feature where radiation is suppressed below angles $\sim m_Q/p_T$. More recently, EEC measurements have been extended to heavy-ion collisions, where the quark-gluon plasma also introduces additional scales --- most simply the medium’s temperature and finite extent --- leading to enhanced subleading-power corrections with clear departures from the power-law scaling~\cite{Andres:2024xvk}.

There are several key motivations for studying EECs in multi-scale jet environments. Many of the most phenomenologically interesting multi-scale processes suffer from large backgrounds, posing significant challenges for precision studies. EECs naturally suppress soft contamination from hadronisation and the underlying event, while backgrounds from mis-tagged jets follow a simple power-law behaviour. As a result, such backgrounds remain both calculable and subtractable, even in complex collider environments~\cite{CMS:2025ydi,CMS:2025jam,Liang-Gilman:2025gjl}. These features make EECs particularly well suited to multi-scale settings, where clean access to intrinsic jet substructure is essential for meaningful theory-experiment comparisons.

Furthermore, and perhaps more significantly, EEC observables possess a remarkably simple theoretical structure. They are highly inclusive, being effectively determined by inclusive matrix elements for generalised fragmentation~\cite{Lee:2025okn,Chang:2025kgq,Guo:2025zwb,Kang:2025zto}. This inclusivity makes the corresponding theoretical predictions both conceptually simple and analytically tractable. As a consequence, EECs can be computed directly on charged-particle final states, requiring only minimal input from low moments of track functions~\cite{Jaarsma:2023ell}. It is precisely this calculational simplicity that underlies recent proposals to use EECs as precision probes of the top-quark mass~\cite{Holguin:2022epo,Holguin:2023bjf,Holguin:2024tkz,Xiao:2024rol}. When an EEC is measured on a jet originating from a heavy decaying particle, a pronounced threshold-peak structure is observed, breaking the simple scaling behaviour characteristic of light jets. This peak is associated with a time-like resonance localised at an angular scale $\sim m_X/p_T$, where $m_X$ is the mass of the heavy particle. In these studies, event generators were used to demonstrate that the top-quark mass can be extracted from this peak and that such extractions are remarkably robust against hadronisation, underlying-event contamination, and soft radiation. However, the underlying physics responsible for this robustness has so far only been discussed qualitatively. One of the aims of this work is therefore to provide a more explicit and quantitative understanding of these features in a controlled setting, and to provide a precise theoretical description of the threshold peak.

In this paper, we investigate EEC measurements on hadronically decaying boosted $Z$ bosons. $Z$-tagged jets provide an ideal testbed for multi-scale jet physics: they are experimentally accessible at the LHC, theoretically clean owing to the colour-singlet nature of the $Z$, and kinematically simple because the decay is two-body at leading order. We consider two complementary collider environments. The first is the measurement of the EEC on $Z$ jets produced in $pp$ collisions, a programme directly applicable to the LHC and HL-LHC. The second is the analogous measurement in $e^+ e^- \rightarrow ZZ \rightarrow {\rm hadrons} + \ell^+ \ell^-$, relevant for future lepton colliders such as FCC-ee. The $e^+ e^-$ setup allows the EEC to be measured inclusively on the full hadronic $Z$ decay, without the need for jet reconstruction, thereby granting access to both the low- and high-boost regimes. Finally, the approach presented here can be equally well applied to other jets of other  heavy  colour singlet resonances such as the Higgs and the $W$.

Because the $Z$ is colour neutral, the factorisation of its hadronic decay is particularly simple: the boosted-frame EEC can be computed entirely from the distribution measured in the $Z$ rest frame. This enables a high-precision prediction using results already available in the literature. In this work we present calculations at N$^3$LL$'$ accuracy, including leading non-perturbative effects, which could even be extended to N$^4$LL accuracy using recent advances~\cite{Duhr:2022yyp,Moult:2022xzt,Electron-PositronAlliance:2025fhk,Jaarsma:2025tck}. Through this analysis, we identify the apparent threshold peak found in massive-jet EECs not as a typical resonance structure --- \emph{e.g.}\ a Breit–Wigner --- but as a consequence of Sudakov logarithm resummation in the decay rest frame, governed by a single soft function. As a result, these features are calculable to exceptionally high precision, with the leading non-perturbative effects also computable via their encoding in the Collins-Soper kernel, which can be extracted from lattice QCD~\cite{Avkhadiev:2024mgd,Avkhadiev:2023poz,Shanahan:2021tst,Shanahan:2020zxr,Shanahan:2019zcq}, together with the leading non-perturbative power correction~\cite{Jaarsma:2023ell,Electron-PositronAlliance:2025fhk}.

The ethos of this work closely resembles that of Feige \emph{et al.}~\cite{Feige:2012vc}, who proposed studying the substructure of boosted $Z$ boson jets to obtain an environment accessible within any collider system (with sufficiently high centre-of-mass energy) which retains the clean theoretical calculability of the $e^+ e^- \rightarrow {\rm hadrons}$ process at the $Z$ pole. To this end, Feige \emph{et al.} studied the 2-subjettiness distribution and provided predictions at NNLL accuracy by boosting the rest-frame $Z$-pole result. The energy-correlator observables we study elegantly realise this ambition. To demonstrate this, we also provide predictions for both the $pp$ and $e^+ e^-$ observables we propose, computed by directly boosting OPAL measurements of the EEC performed at the $Z$ pole~\cite{OPAL:1990reb}. We find these OPAL-based predictions to be in exceptionally close agreement with event-generator predictions for the same observables. This also serves to provide further validation of our approach. Our results therefore offer new insight into the origin and predictability of heavy-jet EEC features, laying essential groundwork for future precision studies of EECs in top-quark jets, while also putting forward a consistent programme of experimental collider physics for the study of jets originating from heavy neutral decays across a broad range of collision systems.

The paper is organized as follows. In the next section we introduce the two measurement scenarios in detail: the EEC of a boosted, hadronically decaying colour-singlet jet, and the EEC in semi-leptonic $e^+e^- \rightarrow ZZ$ events. For each case we outline the relevant factorisation. \Sec{HEEC} demonstrates how high-precision predictions are obtained from the factorisation by exploiting the symmetries of light-ray operators and existing precision results for the $e^+e^- \rightarrow {\rm hadrons}$ EEC at the $Z$ pole. Using this factorisation and these symmetries, we provide a complete numerical evaluation of predictions for the EEC observables we introduce and compare them against results from the \Herwig and \Pythia event generators. We find robust agreement with the event generator results across both the $e^+e^-$ and $pp$ scenarios. 
In \secn{refact}, we discuss extending our approach to higher point energy correlators and present a complementary effective-field-theory construction, which generalises naturally to the case of massive coloured-particle decays, such as for applications to the top quark mass measurement. 
Finally, we conclude in \secn{conclusions} and provide a brief outlook for further developments.

\section{$Z$ production factorisation and the hadronic EEC tensor}
\label{sec:Obs}

\subsection{$pp \rightarrow Z + X \rightarrow {\rm jet} + X$}

\label{sec:ppObs}
Of primary interest in this work is the measurement of the energy-energy correlator (EEC) on the substructure of a sample of $Z$-tagged jets at the LHC. This observable is both experimentally accessible and provides a theoretically clean testing ground for the study of EECs on jets formed from massive, narrow-width decays. Throughout this work, we assume that the tagging of a $Z$-boson jet is completely efficient and that the jet radius ($R$) is large compared to the typical angular separation between the boosted $Z$ decay products.

The observable is computable as
\begin{align}
    \frac{\td \Sigma(R)}{\td \chi \, \td p^{Z}_{T} \, \td \eta_Z} = \sum_{h_i, h_j \in \text{$Z$ jet}} \int \df\Phi_{h_i,h_j} ~ \frac{\td \sigma_{pp \rightarrow h_i h_j}  \left(R\right)}{\td p^{Z}_{T} \, \td \eta_Z \,  \df \Phi_{h_i,h_j}} ~ \frac{p_{T, i} \, p_{T, j}}{(p^{Z}_{T})^2} ~ \delta\left(\Delta R_{ij} - \chi\right),
\end{align}
where $p_T^Z$ is the measured transverse momentum of the $Z$-jet, $R$ is the jet radius, and $\eta_Z$ is the pseudorapidity of the jet axis. The quantity $\td \sigma_{pp \rightarrow h_i,h_j}$ denotes the differential inclusive hadronic cross section to produce hadrons $h_i$ and $h_j$ inside the tagged $Z$-jet, inclusive over all other final-state radiation. The measure $\td \Phi_{h_i,h_j}$ is the Lorentz-invariant phase-space element for hadrons $h_i$ and $h_j$, and is more generally defined for $N$ hadrons $h_1 \ldots h_N$ by
\begin{align}\label{eq:Phi_N}
    \df \Phi_{h_1,\ldots ,h_N} \equiv \Bigg(\prod_{j=1}^N  \dfbar^4 p_{h_j}\: \delta_+(p_{h_j}^2-m_{h_j}^2) \Bigg) \, ,
\end{align}
where we employ the shorthand $\dfbar^d p = \df^d p/(2\pi)^d$ and $\delta_+(p^2) = (2\pi)\delta(p^2)\Theta(p^0)$. Here $p_{T,i}$ denotes the transverse momentum of $h_i$, and $\Delta R_{ij}$ is the longitudinally boost-invariant angular separation between hadrons $h_i$ and $h_j$, defined as $$\Delta R_{ij} = \sqrt{(\eta_i - \eta_j)^2 + (\phi_i - \phi_j)^2}.$$

The differential cross section for the production and subsequent decay of a $Z$ boson within a small-$R$ jet can be factorised, at leading power in $M_Z/p_T^Z \ll 1$ and $R \ll 1$, into two hadronic tensors (see \Refcite{Feige:2012vc} for hadronically decaying $Z$ bosons and \Refcite{Ebert:2020dfc} for details in the case of leptonic $Z$ decays):
\begin{align}
    \frac{\td \sigma_{pp \rightarrow h_i h_j}\left(R\right)}{\td p^{Z}_{T} \, \td \eta_Z \,  \df \Phi_{h_i,h_j}} = \int \td^4 q ~ \frac{1}{2 E_{\rm cm}^2} H^{ij}_{ZZ \, \mu\nu}(q, \v{p}_i , \v{p}_j, p_T^Z, \eta_Z, R) ~ W_{ZZ}^{\mu\nu}  (q, P_a , P_b) 	\,, \label{eq:start}
\end{align}
where the hadronic tensor $W_{ZZ}^{\mu\nu}$ describes the production of an off-shell $Z$ boson with momentum $q$ from incoming hadrons with momenta $P_a$ and $P_b$, and $E_{\rm cm}$ denotes the hadronic centre-of-mass energy. The tensor $H_{ZZ \, \mu\nu}^{ij}$ encodes the production of two hadrons ($h_i,h_j$) from a hadronic $Z$ decay within a jet of radius $R$. The subscript $ZZ$ indicates that interference contributions from $\gamma$ and $W^\pm$ exchange are neglected for simplicity. Power corrections to this factorisation formula arise dominantly from interference between the $Z$ decay products and the rest of the event, as well as from contamination by hadrons inside the jet that do not originate from the $Z$ decay; these effects are suppressed by $M_Z/p_T^Z$ and $R^2$, respectively.

The hadronic tensor describing the production is standard and is given by
\begin{align}\label{eq:Wmunu_prod}
    W^{\mu\nu} ( q, P_a, P_b) &= \sum_X  \langle pp | J_Z^{\dagger \mu }(0) | X \rangle \langle X | J_{Z}^\nu (0)| pp \rangle \, \delta^{(4)} \big(P_a + P_b - q - p_X\big) \\
    &= \sum_X \int \td^4 z \: e^{ \im q\cdot z} \langle pp | J^{\dagger\mu}_Z (z) | X \rangle \langle X | J_Z^\nu (0) | pp \rangle \nn\, ,
\end{align}
where $J_{Z}^\mu$ is the current describing the coupling of the $Z$ boson to quarks,
\begin{align}\label{eq:JZdef}
    J_Z^\mu = -|e| \sum_f \bar q_f \gamma^\mu \big(v_f - a_f \gamma_5\big) q_f \, .
\end{align}

The basic matrix-element structure of the hadronic tensor describing the decay is also standard. However, it can be simplified for the specific EEC measurement considered here. Explicitly,
\begin{align}
    & H^{\rm EEC}_{ZZ \, \mu\nu}(q,\chi,p_T^Z, \eta_Z,R)  =  \sum_{i,j \in Z\, {\rm jet}} \int \df\Phi_{h_i,h_j} ~ H^{ij}_{ZZ \, \mu\nu}(q, \v{p}_i , \v{p}_j , p_T^Z, \eta_Z, R) ~ p_{T, i} \, p_{T, j} ~ \delta\left(\Delta R_{ij} - \chi\right) \nonumber \\
    & = \int \td m_T \td \phi_Z \sumint_{X_n} 
    \mathcal{P}_{Z\ra X_n}^{\mu\nu} ~ m_T \, p_T^Z \,  \delta^{(4)}\Bigg(p_Z - \!\! \sum_{k \in X_n \, {\rm in \, jet } \, R} p_{k} \Bigg) \sum_{ij\in X_n \, {\rm in \, jet } \, R} p_{T , i} p_{T , j} ~ \delta(\chi - \Delta R_{ij})  \, , \label{eq:HEEC}
\end{align}
where $X_n$ denotes a generic $Z$ decay final state, which need not be fully contained within a jet of radius $R$ and therefore need not satisfy the measurement cuts imposed by the EEC measurement and jet clustering. For compactness, we have suppressed the explicit dependence of the squared matrix element for the $Z$ decay on the $Z$ momentum, $q$, and on the phase space of its decay products, $\Phi_{X_n}$. Explicitly, the squared matrix element for a $Z$ decay into a final state $X_n$ is given by
\begin{align}\label{eq:P_Z_to_Xn_def}
    \mathcal{P}_{Z\ra X_n}^{\mu\nu} (q , \Phi_{X_n})= \int \df^4 z \: e^{+\im q \cdot z} \: \langle 0 | Z^{\mu} (z) | X_n \rangle \langle X_n | Z^{\nu} (0) | 0 \rangle \, ,
\end{align}
summed over the $Z$ polarisations. We fix
\begin{align}
    p_Z = \big(m_T \cosh \eta_Z , \; \cos(\phi_Z) \, p^Z_T , \; \sin(\phi_Z) \, p^Z_T , \; m_T \sinh \eta_Z\big)  \, .
\end{align}
Notice that \eq{HEEC} integrates over $m_T$, and we have not enforced that the $Z$ jet originates from an on-shell $Z$ boson. Note also that the EEC measurement is not normalised within the definition of $H^{\rm EEC}_{ZZ \, \mu\nu}$ and therefore increases the mass dimension of the EEC hadronic tensor by $[{\rm GeV}^2]$ relative to $H^{ij}_{ZZ}$. This choice will be convenient in later discussions, where dimensional analysis will be useful.

We can expand in the limit $\chi \sim M_Z/p_T^Z \ll R \ll 1$, assuming that the decaying $Z$ boson is close to the mass shell, to recover the small-angle factorisation limit of the EEC within a narrow jet~\cite{Lee:2024icn,Lee:2024tzc,Generet:2025vth} formed from a massive decay~\cite{Craft:2022kdo,Holguin:2022epo}, where the entire decay is contained within the jet. In this regime, since the $Z$ is a colour singlet, it does not radiate coloured particles prior to its decay, and consequently there is no jet function and no associated evolution in $R$. As a result, we can fix $p_T^Z = q_T$ and $\eta_Z = y_q$, and the hadronic tensor simplifies to
\begin{align}
    H^{\rm EEC}_{ZZ \, \mu\nu}(q,\chi,p_T^Z, \eta_Z,R) &= \Bigg(\sumint_{X_n}   \mathcal{P}_{Z\ra X_n}^{\mu\nu} \big(q, \Phi_{X_n} \big) ~ \delta(p^Z_T - q_T ) \delta(\eta_Z - y_q )  \\
    &\quad \times   \sum_{ij\in X_n} p_{T , i} p_{T , j} ~ \delta(\chi - \Delta R_{ij})\Bigg) \bigg(1  + \mathcal{O}\left(\frac{\chi^2}{R^2}, \frac{M_Z}{R \, p^Z_T}\right)  \bigg)\, . \nn 
\end{align}

We can now strip the hadronic tensor of the fixed jet parameters $p_T^Z$, $\eta_Z$, and $R$, and define
\begin{align}
    & H^{\rm EEC}_{ZZ \, \mu\nu}(q,\chi) = 
    \sumint_{X_n}  \mathcal{P}_{Z\ra X_n}^{\mu\nu} \big(q,\Phi_{X_n} \big)  \sum_{ij\in X_n} p_{T , i} p_{T , j} ~ \delta(\chi - \Delta R_{ij})  \, .
\end{align}
With this definition, the observable can be written to leading power in $\chi/R$, $R$, $M_Z/(R p_T^Z)$, and $M_Z/p_T^Z$ in the form
\begin{align}\label{eq:HEEC_fact_0}
    &\frac{\td \Sigma}{\td \chi \, \td p^{Z}_{T} \, \td \eta_Z} = \int \td^4 q ~ \frac{H^{\rm EEC}_{ZZ \, \mu\nu}(q, \chi) ~ W_{ZZ}^{\mu\nu}  (q, P_a , P_b)}{2 E_{\rm cm}^2 ~ (p^{Z}_{T})^2}  ~ \delta(q_T - p_T^Z) ~ \delta(y_q - \eta_Z).
\end{align}

The tensor $H^{\rm EEC}_{ZZ \, \mu\nu}(q,\chi)$ can be further simplified near the $Z$ resonance due to the inclusivity of the EEC measurement. This allows both $H^{\rm EEC}_{ZZ \, \mu\nu}$ and $W_{ZZ}^{\mu\nu}$ to be projected onto the physical polarisation states, and permits the extraction of a Breit-Wigner distribution associated with the $Z$ decay. We begin from \eq{P_Z_to_Xn_def}, where the $Z$ decay operator was left implicit in the Hamiltonian in the Heisenberg picture. We now switch to the interaction picture and explicitly include a single insertion of the decay operator on both sides:
\begin{align}
    \cP_{Z\ra X_n\,{\mu\nu}} (q,\Phi_{X_n}) &= \int \df^d z \int \df^d x \int \df^d y \: e^{+\im q\cdot z} \: \\
    &\quad  \times\big \langle 0\big | \overline{\rm T} \Big\{Z_{\mu}(z) Z_\alpha (x)J^{\dagger\alpha}_{Z}(x)\Big\}\big | X_n\big \rangle \big \langle X_n \big | {\rm T} \Big\{Z_\nu(0) Z_\beta(y) J_{Z}^\beta (y) \Big\} \big | 0 \big \rangle 	\nn\\
    &= \nn
    \int \df^d z \int \df^d x \int \df^d y \int \dfbar^d p \int \dfbar^d k \: e^{+\im q\cdot z} e^{+ \im p\cdot (x + y-z)} e^{-\im k \cdot y} \\
    &\nn \quad \times \frac{\Big(g_{\mu\alpha} - \frac{p_\mu p_\alpha}{M_Z^2}\Big)}{p^2 - M_Z^2 - \im \Gamma_Z M_Z } \frac{\Big(g_{\nu\beta} -\frac{k_\nu k_\beta}{M_Z^2}\Big)}{k^2 - M_Z^2 + \im \Gamma_ZM_Z }  \big \langle 0\big | J^{\dagger\alpha}_{Z}(x)\big | X_n\big \rangle \big \langle X_n \big | J^\beta_{Z} (0) \big | 0 \big \rangle 	\,. 
\end{align}
In obtaining the second equality, we have written the time-ordered products of the $Z$ fields in terms of propagators, shifted $x \ra x + y$, and used translation invariance of the hadronic matrix elements. This allows the $y$ and $z$ integrals to be performed, setting $k = p = q$, which yields
\begin{align}
    \cP_{Z\ra X_n\,\mu\nu}\big(q,\Phi_{X_n}\big) = \frac{\Big(g_{\mu\alpha} - \frac{q_\mu q_\alpha}{M_Z^2}\Big)\Big(g_{\nu\beta} - \frac{q_\nu q_\beta}{M_Z^2}\Big)}{\big[q^2 - M_Z^2\big]^2 +  \big[\Gamma_Z M_Z\big]^2 }   \int \df^d x \: e^{\im q\cdot x} \, \langle 0\big | J^{\dagger\alpha }_{Z}(x)\big | X_n\big \rangle \big \langle X_n \big | J^\beta_{Z} (0) \big | 0 \big \rangle 	\,. 
\end{align}
This leads to the final leading-power expression for the observable,
\begin{align}
    \frac{\td \Sigma}{\td \chi \, \td p^{Z}_{T} \, \td \eta_Z} = \int \td^4 q ~ \frac{\df \sigma_{\rm EEC}}{\td^4 q \, \td\chi } ~ \delta(q_T - p_T^Z) ~ \delta(y_q - \eta_Z), \label{eq:Sigmapp}
\end{align}
where
\begin{align}	
    \frac{\df \sigma_{\rm EEC}}{\td^4 q \, \td\chi } &= \frac{ W_{\rm incl} (q^2, P_a \cdot q, P_b \cdot q)}{2E_{\rm cm}^2}  \frac{1}{\big[q^2 - M_Z^2\big]^2 +  \big[\Gamma_Z M_Z\big]^2 } \frac{H_{\rm EEC} (q,\chi)}{q_T^2} . \label{eq:sigmaEEC}
\end{align}
Here $W_{\rm incl}$ is the production tensor inclusively summed over polarisations,
\begin{align}
    W_{\rm incl} (q^2, P_a \cdot q, P_b \cdot q)  =  \bigg(\frac{q^\mu q^\nu}{M_Z^2} - g^{\mu\nu}\bigg) W_{\mu\nu} (q, P_a, P_b) \, ,
\end{align}
and
\begin{align}\label{eq:HEECdef}
    & H_{\rm EEC} (q,\chi) = \bigg(\frac{q_\mu q_\nu}{M_Z^2} - g_{\mu\nu}\bigg) \times \\
    &  \sumint_{X_n} \sum_{ij\in X_n} p_{T \, i} p_{T , j} ~ \delta(\chi - \Delta R_{ij}) \int \td^4 x \: e^{\im q \cdot x} \, \langle 0\big | J^{\dagger\mu}_{Z}(x)\big | X_n\big \rangle \big \langle X_n \big | J^\nu_{Z} (0) \big | 0 \big \rangle . \nonumber 
\end{align}

In summary, the computation of this observable reduces to a standard initial-state hadronic tensor, which is well documented in the literature~\cite{Ebert:2020dfc}, together with the computation of $H_{\rm EEC}$ for an off-shell $Z$ decay that is independent of the hadronic initial state. This result is convoluted with a Breit-Wigner distribution, allowing the narrow-width approximation to be employed to restrict the $Z$ momentum to the vicinity of the mass shell, thereby further simplifying the calculation. The remainder of this paper will focus on $H_{\rm EEC}$ and its properties.

\subsection{$e^+ e^- \rightarrow Z Z \rightarrow {\rm hadrons} + l^+ l^-$}

\label{sec:eeObs}

For us, it is informative to also consider the measurement of the energy-energy correlator (EEC) on the substructure of a sample of di-$Z$ events produced at a lepton collider. In the limit of large centre-of-mass energy, where both $Z$ bosons receive a large boost, the collinear limit of this measurement matches exactly the substructure measurement  performed on a $pp \rightarrow Z + X$ sample and considered above. However, by using symmetries, this process is also readily computable outside the collinear limit and at lower centre-of-mass energy values. It therefore provides a useful benchmark. Experimentally, this measurement is not expected to be accessible in the near future, although it could be studied at FCC-ee when operating at the $Z H$ and $t \bar{t}$ thresholds.

This observable is computable as:
\begin{align}
    \frac{\td \Sigma_{ee}(\Gamma)}{\td \theta} = \sum_{i,j,X} & \int^{\Gamma}_{-\Gamma} \td \epsilon_1 \int^{\Gamma}_{-\Gamma} \td \epsilon_2 \int \td \sigma_{e^{+}e^{-}\rightarrow ij+X+l^+ l^-} ~ \frac{E_{i} E_{j}}{Q^2} ~ \delta\left(\theta_{ij} - \theta\right) \nonumber \\
    & \times \delta\left((p_{l^+} + p_{l^-})^2 - M_Z^2 - \epsilon_1 \right) \delta\left((p_{i} + p_{j} + p_X )^2 - M_Z^2 - \epsilon_2 \right),
\end{align}
where $\td \sigma_{e^{+}e^{-}\rightarrow ij+X+l^+ l^-}$ is the differential partonic cross section to produce hadrons $i,j$ alongside two leptons $l^+, l^-$ and anything else ($X$). $E_{i}$ is the energy of hadron $i$ and $\theta_{ij} $ is the angle between hadrons $i,j$. The integration over $\epsilon_{1,2}$ tags the leptonic system and hadronic system as each having originated from a $Z$ decay, allowing for a width $\Gamma$. Finally, we choose to fix $Q = \sqrt{s}/2$ so that we have the sum rule
\begin{align}
    \int \frac{\td \Sigma_{ee}(\Gamma)}{\td \theta} ~ \td \theta = \sigma_{ee}^{\rm tot} \, ,
\end{align}
up to electro-weak corrections to the $Z$ production.

This observable can be decomposed in a fashion identical to the previous $pp\rightarrow Z + X$ example. Under the assumption that $\Gamma > \Gamma_Z$, but remaining small enough so as to isolate the $ZZ$ process from the backgrounds, the integrals over the taggers $\epsilon_{2}$ drop out. The final result is 
\begin{align}
    \frac{\td \Sigma_{ee}(\Gamma)}{\td \theta} = \int \td^4 q ~ \frac{\df \sigma^{ee}_{\rm EEC}(\Gamma)}{\td^4 q \, \td\theta } ~ \delta\Big(q_0 - Q\Big) \, .
\end{align}
where $q$ is the momentum of the hadronically decaying $Z$ and where
\begin{align}	
    \frac{\df \sigma^{ee}_{\rm EEC}(\Gamma)}{\td^4 q \, \td\theta } &= \frac{ L_{\rm incl} (q^2, P_{e^+} \cdot q, P_{e^-} \cdot q, \Gamma)}{2E_{\rm cm}^2}  \frac{1}{\big[q^2 - M_Z^2\big]^2 +  \big[\Gamma_Z M_Z\big]^2 } \frac{H^{ee}_{\rm EEC} (q,\theta)}{q_0^{\, 2}} \, \label{eq:eefactorised}.
\end{align}
We have defined the hadronic tensor for the EEC measurement on the $Z$ decay as, 
\begin{align}\label{eq:HeeDef}	
    H^{ee}_{\rm EEC} (q,\theta) = & \bigg(\frac{q_\mu q_\nu}{M_Z^2} - g_{\mu\nu}\bigg) \times \\
    & \sumint_{X_n}  \sum_{ij\in X_n} E_{i} E_{j} ~ \delta(\theta - \theta_{ij}) \int \td^4 x \: e^{\im  q \cdot x} \, \langle 0\big | J^{\dagger\mu}_{Z}(x)\big | X_n\big \rangle \big \langle X_n \big | J^\nu_{Z} (0)  \big | 0 \big \rangle , \nonumber 
\end{align}
and we have introduced $L_{\rm incl}$ as the leptonic initial state tensor, by analogy to $W_{\rm incl}$. It is defined through
\begin{align}
    &L^{\mu\nu} ( q, P_{e^+}, P_{e^-},\Gamma) = \int \df \Phi_{l^+,l^-} \int^{\Gamma}_{-\Gamma} \td \epsilon_1 ~ \delta\left((p_{l^+} + p_{l^-})^2 - M_Z^2 - \epsilon_1 \right) \times  \, \nonumber \\
    &\sum_X \langle e^+e^- | J_Z^{\dagger \mu }(0) | X + l^+ l^- \rangle \langle X + l^+ l^-| J_{Z}^\nu (0)| e^+e^- \rangle \; \delta^{(4)}(q + p_X + p_{l^+} + p_{l^-}-P_{e^+} + P_{e^-} ) .
\end{align}
In this case, no expansions in $\theta$ or the $Z$ energy are required to derive the result in \eq{eefactorised}, and so there is no constraint on either the measured correlator angle or the size of the $Z$ boost.

In summary, the computation of this $e^+e^-$ observable also reduces to a tensor for the production, a Breit-Wigner distribution for the momentum of the decaying $Z$ boson, and the computation of $H^{ee}_{\rm EEC}$. $H^{ee}_{\rm EEC}$ appears to be very similar to $H_{\rm EEC}$ from the $pp$ case. Indeed they are approximately equal if one considers sufficiently-fat boosted jets in the $pp$ case, where $p_{T , i}/p_T^Z \approx E_i/Q$ for massless hadrons.

\section{Constraining $H_{\rm EEC}$ and $H^{ee}_{\rm EEC}$ from symmetries}

\label{sec:HEEC}

We will now discuss the computation of $H_{\rm EEC}(q,\chi)$ and $H^{ee}_{\rm EEC}(q,\chi)$ from the symmetries of the energy-flow operator within massless QCD and making use of the narrow width approximation to place $q$ on-shell at the $Z$ mass. This will enable us to relate $H_{\rm EEC}(q,\chi)$ and $H^{ee}_{\rm EEC}(q,\chi)$ to both measurements and computations of the EEC in $e^+ e^- \rightarrow {\rm hadrons}$ at the $Z$-pole, extensively studied in the literature.

To begin, we re-arrange $H^{ee}_{\rm EEC}(q,\chi)$ into an equal-time correlator (Wightman function):
\begin{align}	\label{eq:Hee}
    & H^{ee}_{\rm EEC} (q,\theta) =  \\
    & \bigg(\frac{q_\mu q_\nu}{M_Z^2} - g_{\mu\nu}\bigg) \int \td^4 x \: e^{\im q\cdot x} \, \langle 0\big | J^{\dagger\mu}_{Z}(x) \sum_{X_n} 
    \big | X_n\big \rangle \big \langle X_n \big | \sum_{ij\in X_n} E_{i} E_{j} ~  J^\nu_{Z} (0) \big | 0 \big \rangle ~ \delta(\theta - \theta_{ij}) , \nonumber  \\
    & = \bigg(\frac{q_\mu q_\nu}{M_Z^2} - g_{\mu\nu}\bigg) \int \td^2 \mb n_1 \td^2 \mb n_2 \int \td^4 x \: e^{\im q\cdot x} \, \langle 0\big | J^{\dagger\mu}_{Z}(x)  \, \cE(n_1) \cE(n_2) \,  J^\nu_{Z} (0) \big | 0 \big \rangle ~ \delta\left(\theta - \theta_{12}\right)\nn
\end{align}
where $n_1$ and $n_2$ are normalised light-like vectors along the unit vectors $\mb n_1$ and $\mb n_2$ such that $n_i^\mu = (1, \mb n_i)$, 
$\theta_{12} = \cos^{-1}(1-n_1 \cdot n_2)$, $| 0 \rangle$ is the vacuum at past infinity, and $| X_n \rangle$ the outgoing state. Here we have used that the sum/integral over the final state $X_n$ reduces to the identity, and have introduced the energy-flow operator which is defined so that
\begin{align}
    \mathcal{E}(n) \lkl X :{\rm out} \> = \sum_{i \in X} E_i \, \delta^2(\Omega_n - \Omega_{p_i}) \, \lkl X :{\rm out} \> \,, \label{eq:EFOheuristic}
\end{align} 
where $i$ is an outgoing particle with momentum $p_i$ and $\Omega_{n_i}$ is the solid angle in the direction of $\mb n_i$. Formally, $\mathcal{E}(n)$ is a light-ray operator and has a definition in terms of the energy-momentum tensor integrated over the trajectory of a null vector,
\begin{align}
    \cE(n) = 
    \lim_{x^+\rightarrow \infty} \left(\frac{x \cdot \bar{n}} {n\cdot \bar{n}}\right)^2 \int^{\infty}_{-\infty} \mathrm{d} (x \cdot n) ~ \frac{\bar{n}_{\mu}\bar{n}_{\nu}}{(n \cdot \bar{n})^2} ~ T^{\mu\nu} (x), \label{eq:EFOcovariant}
\end{align}
where $\bar{n} =  (1, - \mb n_i)$.

From this point on, we will use an abridged notation:
\begin{align}	\label{eq:n1n2_corr}
    \langle \cE(n_1) \cE(n_2) \rangle_q \equiv \frac{1}{\sigma_0}\bigg(\frac{q_\mu q_\nu}{M_Z^2} - g_{\mu\nu}\bigg) \int \td^4 x \: e^{\im q\cdot x} \, \langle 0\big | J^{\dagger\mu}_{Z}(x)  \, \cE(n_1) \cE(n_2) \,  J^\nu_{Z} (0) \big | 0 \big \rangle \, ,
\end{align}
where
\begin{align}
    \sigma_0 \equiv \bigg(\frac{q_\mu q_\nu}{M_Z^2} - g_{\mu\nu}\bigg) \int \td^4 x \: e^{\im q\cdot x} \, \langle 0\big | J^{\dagger\mu}_{Z}(x)  \,  J^\nu_{Z} (0) \big | 0 \big \rangle \, ,
\end{align}

Crucial to our discussion is the symmetry of the energy-flow operator under reparametrisation,
\begin{align}
    \cE(\rho n) = \rho^{-3} \cE(n)\, , \label{eq:symms} 
\end{align}
which fixes its transformation under longitudinal boosts acting as $n \mapsto \rho n$ and is determined by the operator’s collinear spin~\cite{Chen:2022jhb}. More generally, under longitudinal boosts a light-ray operator transforms with an eigenvalue $\rho^{-\delta}$, where $\delta$ is the operator’s \textit{celestial dimension}. For the energy-flow operator $\cE$, $\delta = 3$. This can be read off from \eq{EFOcovariant} by noting that the energy-momentum tensor is physical and is therefore invariant under reparametrisation. The collinear spin of a massless-QCD operator is a renormalisation-group invariant, and consequently so is the celestial dimension of a light-ray operator. For $\cE$, this again follows from \eq{EFOcovariant}: the reparametrisation symmetry is exact and is preserved by renormalisation, since the energy-momentum tensor itself is a renormalisation-group invariant.

The EEC is a scalar --- not a Lorentz scalar --- only depending on the four dynamical non-vanishing scalar quantities: $n_{1} \cdot n_{2}$, $q\cdot n_{1}$, $q\cdot n_{2}$, and $q^{2}$; and additionally $\Lambda_{\rm QCD}$ and $M_Z$. From these quantities, there is a single dynamical cross-ratio which is invariant under $n \rightarrow \rho n$:
\begin{align}\label{eq:bar_z_def}
    \bar z = \frac{q^2 \, n_{1} \cdot n_{2}}{2 \, q\cdot n_{1} ~ q\cdot n_{2}},
\end{align}
leading to three dimensionless quantities upon which the EEC can depend, $\bar z $, $M_Z^2 / q^2$ and $\Lambda_{\rm QCD}^2 / q^2$. Note additionally that $\bar z$ is a Lorentz scalar, despite that $n_{1} \cdot n_{2}$, $q\cdot n_{1}$, $q\cdot n_{2}$ are individually frame dependent. 

Consistency with the reparametrisation symmetry of $\cE$ requires that the EEC has the functional form,
\begin{align}\label{eq:shape}
    \< \cE( n_{1}) \cE(n_{2}) \>_{q} = \frac{q^{2}\mathcal{F}_{\cE\cE}\left(\bar z , \; M_Z^2/q^{2} , \; \Lambda_{\rm QCD}^2/q^{2}\right)}{4 \pi^{2} ~ (n_{1} \cdot n_{2})^{3} } \, ,
\end{align}
where $\mathcal{F}_{\cE\cE}$ is the \textit{EEC shape function} and is a reparametrisation invariant~\cite{Belitsky:2013xxa,Belitsky:2013bja}. The EEC shape function is manifestly Lorentz invariant and so can be easily used to determine the EEC in different frames of reference.\footnote{Note that the terminology EEC shape function is not to be confused with the non-perturbative shape function that appears in di-jet event shapes.}

For the purpose of this work, $\mathcal{F}_{\cE\cE}$ for the EEC measured on an off-shell $Z$ decay can be uniquely determined by considering the EEC measured on colour-singlet collisions in the centre-of-mass frame with energies near the $Z$ pole such that an intermediate $Z$ dominates over all other channels. In this frame, $\bar{z} = z = (1-\cos \theta)/2$ where $\theta$ is the typically measured EEC angle. This enables us to exploit the well known calculations and measurements of the EEC on a $e^+ e^- \rightarrow Z \rightarrow {\rm hadrons}$ at $\sqrt{s} = 91.2 ~ {\rm GeV}$. In particular, we determine the EEC shape function from the relation
\begin{align}
    \mathcal{F}_{\cE\cE}\left( z , \frac{\Lambda_{\rm QCD}^2}{M_Z^2} \right) = \frac{8 z^3}{M_Z^2} \int\td^2 \mb n_1 \td^2 \mb n_2 ~ \< \cE(n_{1}) \cE(n_{2}) \>_{Z{\rm\hbox{-}pole} } ~ \delta(z - n_1 \cdot n_2 /2) \, . \label{eq:FEEdef}
\end{align}
We will drop the $M_Z$ and $\Lambda_{\rm QCD}^2$ arguments from now on for compactness.

With the EEC shape function, the full EEC measured on a $e^+ e^- \rightarrow Z Z \rightarrow {\rm hadrons} + l^+ l^-$ process is given by 
\begin{align}
    \frac{\td \Sigma_{ee}(\Gamma)}{\td \theta} =  \int &\td^4 q ~  \frac{ L_{\rm incl} (q^2, P_{e^+} \cdot q, P_{e^-} \cdot q, \Gamma)}{2E_{\rm cm}^2}  \frac{1}{\big[q^2 - M_Z^2\big]^2 +  \big[\Gamma_Z M_Z\big]^2 } ~ \delta(q_0 - Q) \nonumber \\
    & \times \int \df^2 \mb n_1 \df^2 \mb n_2   ~ \frac{M_Z^{2} ~ \sigma_0}{4 \pi^{2} q_0^{\, 2} ~ (n_{1} \cdot n_{2})^{3} }  ~ \mathcal{F}_{\cE\cE}\left(\frac{q^2 ~ n_{1} \cdot n_{2}}{2 \, q \cdot n_{1} ~ q \cdot n_{2}} \right) ~ \delta\left(\theta - \theta_{12}\right) \, , 
\end{align}
again where $\theta_{12} = \cos^{-1}(1-n_1 \cdot n_2)$. Using the narrow width approximation
\begin{align}
    \lim_{\Gamma_Z/M_Z \rightarrow 0} \frac{1}{\big[q^2 - M_Z^2\big]^2 +  \big[\Gamma_Z M_Z\big]^2 } = \frac{\pi}{\Gamma_Z M_Z} \, \delta\left(q^2 - M_Z^2 \right) \, ,
\end{align}
the complete result can be written as 
\begin{align}
    \frac{\td \Sigma_{ee}(\Gamma)}{\td \theta} = N \int \td^2 \mb n  \int \td^2 \mb n_1 \td^2 \mb n_2   ~ \frac{L_{\rm incl}(\mb n) ~ M_Z^{2}}{4 \pi^{2} q_0^{\, 2} ~ (n_{1} \cdot n_{2})^{3} }  ~ \mathcal{F}_{\cE\cE}\left(\frac{M^{2}_Z ~ n_{1} \cdot n_{2}}{2 \, q \cdot n_{1} ~ q \cdot n_{2}} \right) ~ \delta\left(\theta - \theta_{12}\right) \, , \label{eq:boostonee}
\end{align}
where $N$ is a normalisation factor and 
$$q=\left(Q,\sqrt{Q^2 - M_Z^2} ~ \mb n\right)^{\rm T},$$
where $\mb n$ is a unit vector along the $Z$ direction in the laboratory frame and $L_{\rm incl}(\mb n) = L_{\rm incl}(M_Z^2, P_{e^+} \cdot q, P_{e^-} \cdot q, \Gamma)$. We have suppressed the $\Gamma$ dependence once again under the assumption that $\Gamma > \Gamma_Z$, but remaining small enough so as to isolate the $ZZ$ process.

We can also use the same EEC shape function in the boosted-jet limit of EEC of a $pp \rightarrow Z + X \rightarrow {\rm jet} + X$ process. With the narrow width approximation, we find that
\begin{align} \label{eq:boostonpp}
    &\frac{\td \Sigma}{\td \chi \, \td p^{Z}_{T} \, \td \eta_Z} =  \\
    &\mathcal{N}(p^{Z}_{T}, \eta_Z) \int \frac{\td \phi_{Z}}{2\pi} \int \df^2 \mb n_1 \df^2 \mb n_2 ~ \frac{ M_Z^{2}}{4 \pi^{2} (p^{Z}_{T})^2 ~ (n_{1} \cdot n_{2})^{3} }  ~ \mathcal{F}_{\cE\cE}\left(\frac{M^{2}_Z ~ n_{1} \cdot n_{2}}{2 \, p_Z \cdot n_{1} ~ p_Z \cdot n_{2}} \right) ~ \delta\left(\chi - \chi_{12}\right) \, ,\nonumber
\end{align}
where 
$$p_Z=\left(\sqrt{M_Z^2 + (p^{Z}_{T})^2})\cosh \eta_Z , p^Z_T \cos \phi_{Z} , p^Z_T \sin \phi_{Z}   , \sqrt{M_Z^2 + (p^{Z}_{T})^2} \sinh \eta_Z\right)^{\rm T}.$$
In this instance, since the measurement is differential in the $Z$ jet rapidity and transverse momentum, all dependence on the production can be factored into $\mathcal{N}$. This formula will receive additional power corrections which scale as $R^2$ due to the approximation $p_{T,i}/p_T^Z \approx E/Q$ in the small angle limit where $\theta \approx \chi$. \Eq{boostonpp} can be readily evaluated using numerical techniques and with the EEC shape function determinable from results in the literature. 

\subsection{The rest frame shape function}

\begin{figure}
    \centering
    \includegraphics[width=0.49\linewidth]{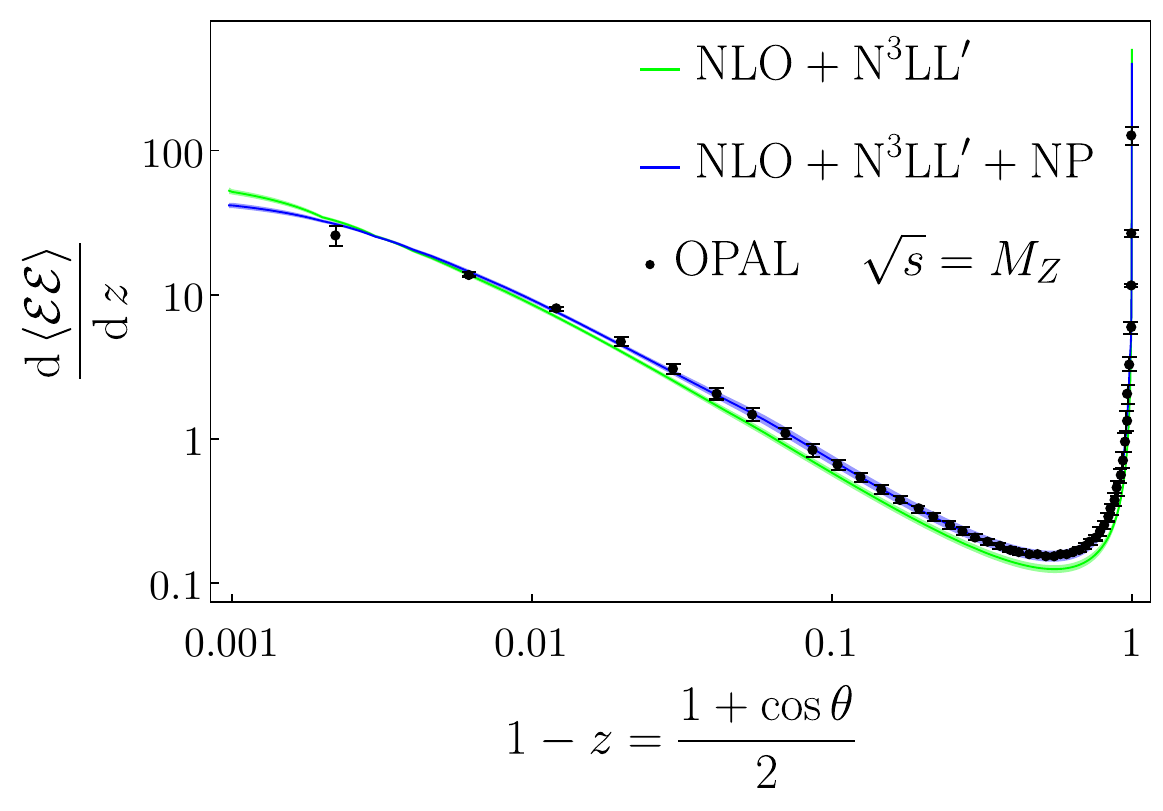}
    \includegraphics[width=0.49\linewidth]{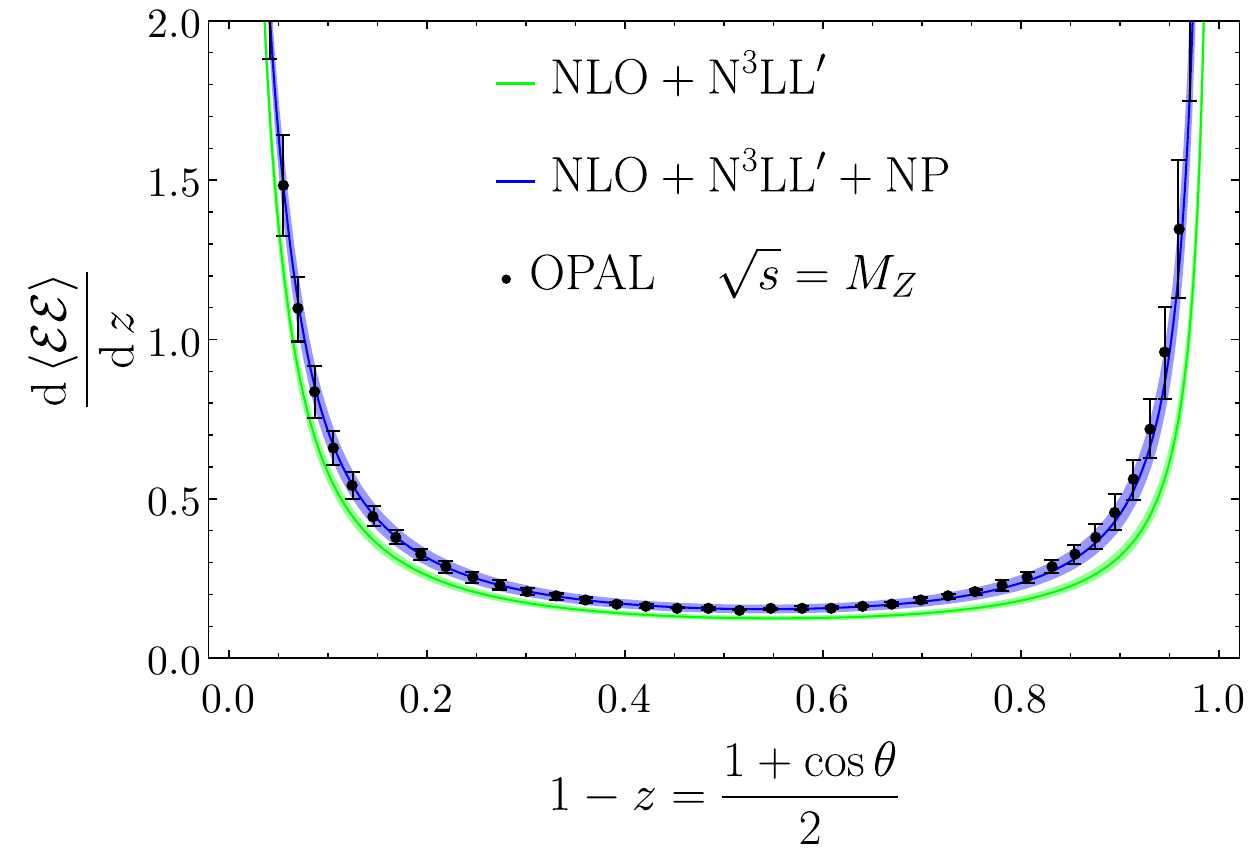}
    \caption{The EEC spectrum computed in $e^+e^-$ at the $Z$ pole as a function of $z$, compared with OPAL data~\cite{OPAL:1990reb}. The left panel uses logarithmic axes to emphasise the behaviour near the back-to-back region, while the right panel shows the same distributions on linear axes. Results are shown both with (blue) and without (green) non-perturbative contributions as described in the text. For the perturbative curves, the bands correspond to scale variations by a factor of 2. The bands on the non-perturbative curves are found by adding in quadrature the bands from scale variation with the bands from the uncertainties on the non-perturbative ingredients.}
    \label{fig:rest_frame}
\end{figure}

We now recap the computation of the EEC in $e^+e^- \rightarrow Z \rightarrow {\rm hadrons}$ at the $Z$ pole, expressed as a function of $z = n_1 \cdot n_2 /2$. The EEC is given by
\begin{align}
    \frac{\td \Sigma_{\ee \rightarrow {\rm hadrons}}}{\td z} =\frac{1}{Q^2} \int \df^2 \mb n_1 \df^2 \mb n_2 ~ \langle \cE(n_{1}) \cE(n_{2}) \rangle_{q_{\rm rest}} ~ \delta(z - n_1 \cdot n_2 /2)\,.
\end{align}
The correlator exhibits three distinct regimes: the collinear limit $z\rightarrow 0$, the fixed–order region $z \sim 0.5$, and the back-to-back limit $z \rightarrow 1$. The collinear limit is not relevant for our purposes, as it probes the region $\theta, \chi \ll M_Z/p_T^Z$. In this regime only the highly collimated structure around the individual products of the $Z$ decay is resolved, rather than the broader decay pattern.

In the fixed–order region one finds
\begin{align}
    \frac{\td \Sigma^{z\sim 0.5}_{\ee \rightarrow {\rm hadrons}}}{\td z} = \frac{\td \Sigma_{\ee \rightarrow {\rm hadrons}}^{\rm pert.}}{\td z} + \frac{1}{\left[z^{3/2}(1-z)^{3/2}\right]_{+}}\frac{\Omega}{Q} +\mathcal{O}\!\left(\frac{\Lambda_{\rm QCD}^2}{Q^2}\right)\,,
\end{align}
where $\Sigma_{\ee \rightarrow {\rm hadrons}}^{\rm pert.}$ denotes the perturbative prediction, computed numerically to NNLO~\cite{Tulipant:2017ybb} and analytically at NLO~\cite{Dixon:2018qgp}. The parameter $\Omega \sim \Lambda_{\rm QCD}$ is a semi-universal non-perturbative quantity which may be extracted independently from event-shape data~\cite{Korchemsky:1999kt,Schindler:2023cww,Lee:2024esz} such as thrust~\cite{Abbate:2010xh}.\footnote{We have included the plus-prescription for the power correction’s divergent endpoints $z = 0, 1$ for consistency with prior results [44]. However, these endpoints never enter our analysis, and so the plus-prescription can be dropped.}

In the back-to-back limit, the EEC diverges at fixed order, and large logarithms of $(1-z)$ must be resummed. The resummation causes $\td \Sigma_{\ee \rightarrow {\rm hadrons}}/\td z$ to approach a constant as $z \rightarrow 1$. Historically this has been achieved via several methods~\cite{Dokshitzer:1999sh,Ebert:2020sfi,Aglietti:2024xwv,Electron-PositronAlliance:2025fhk}, but the most precise modern calculations are based on a factorisation theorem derived from the limit where a colour-singlet produces two fragmenting partons~\cite{Moult:2018jzp}. Most recently, this has been computed at N$^4$LL matched to N$^2$LO~\cite{Electron-PositronAlliance:2025fhk}.

This factorisation receives additional non-perturbative contributions from the soft function $S_{\rm EEC}$. Early studies by Dokshitzer \emph{et al.}~\cite{Dokshitzer:1999sh} used a dispersive model in which non-perturbative effects were introduced through a gluer model for non-perturbative power corrections and an analytically extended running coupling that remains finite at all scales~\cite{Dokshitzer:1995qm}. More recent approaches incorporate the leading non-perturbative corrections through the non-perturbative component of the Collins-Soper kernel governing the rapidity evolution of $S_{\rm EEC}$, which may be constrained using lattice QCD~\cite{Avkhadiev:2024mgd,Avkhadiev:2023poz,Shanahan:2021tst,Shanahan:2020zxr,Shanahan:2019zcq}.

\begin{figure}
    \centering
    \includegraphics[width=0.49\linewidth]{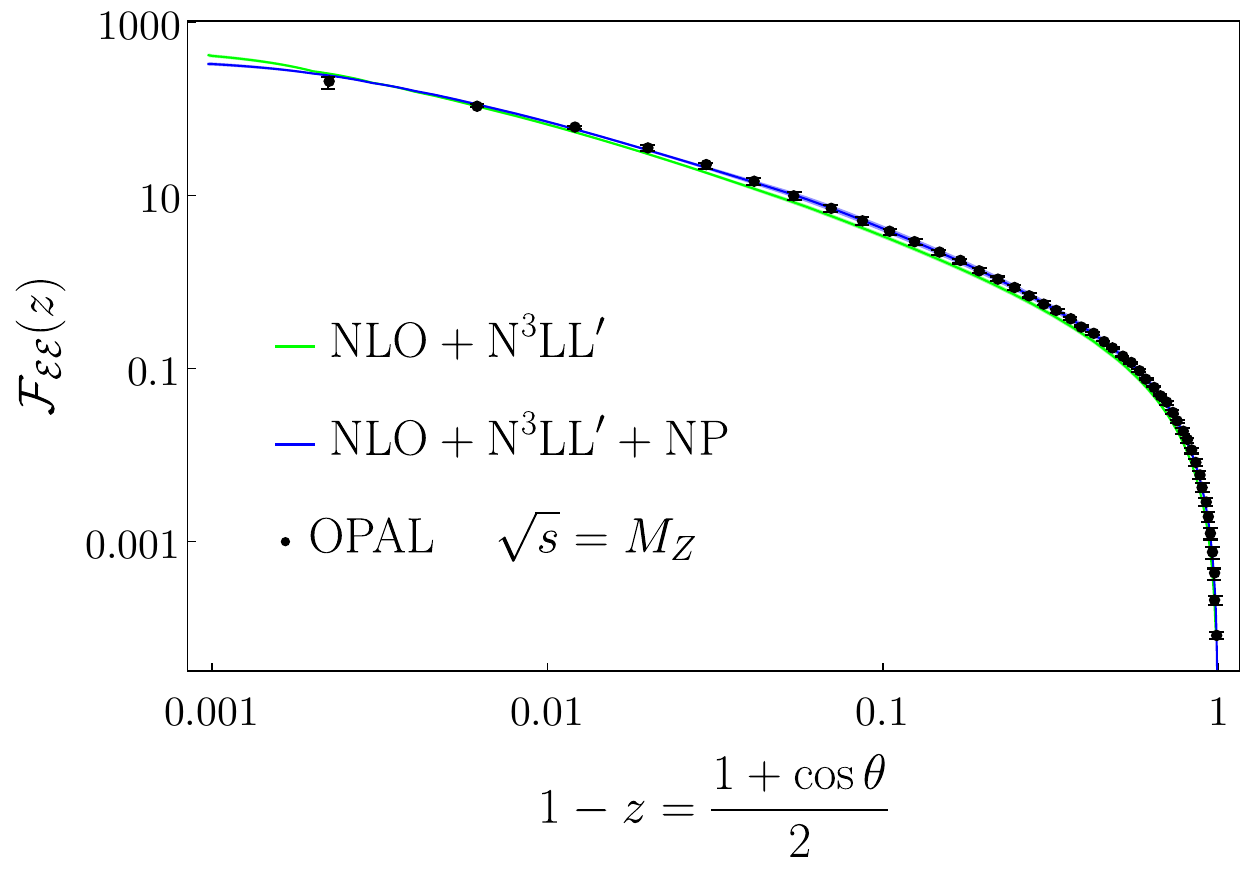}
    \includegraphics[width=0.49\linewidth]{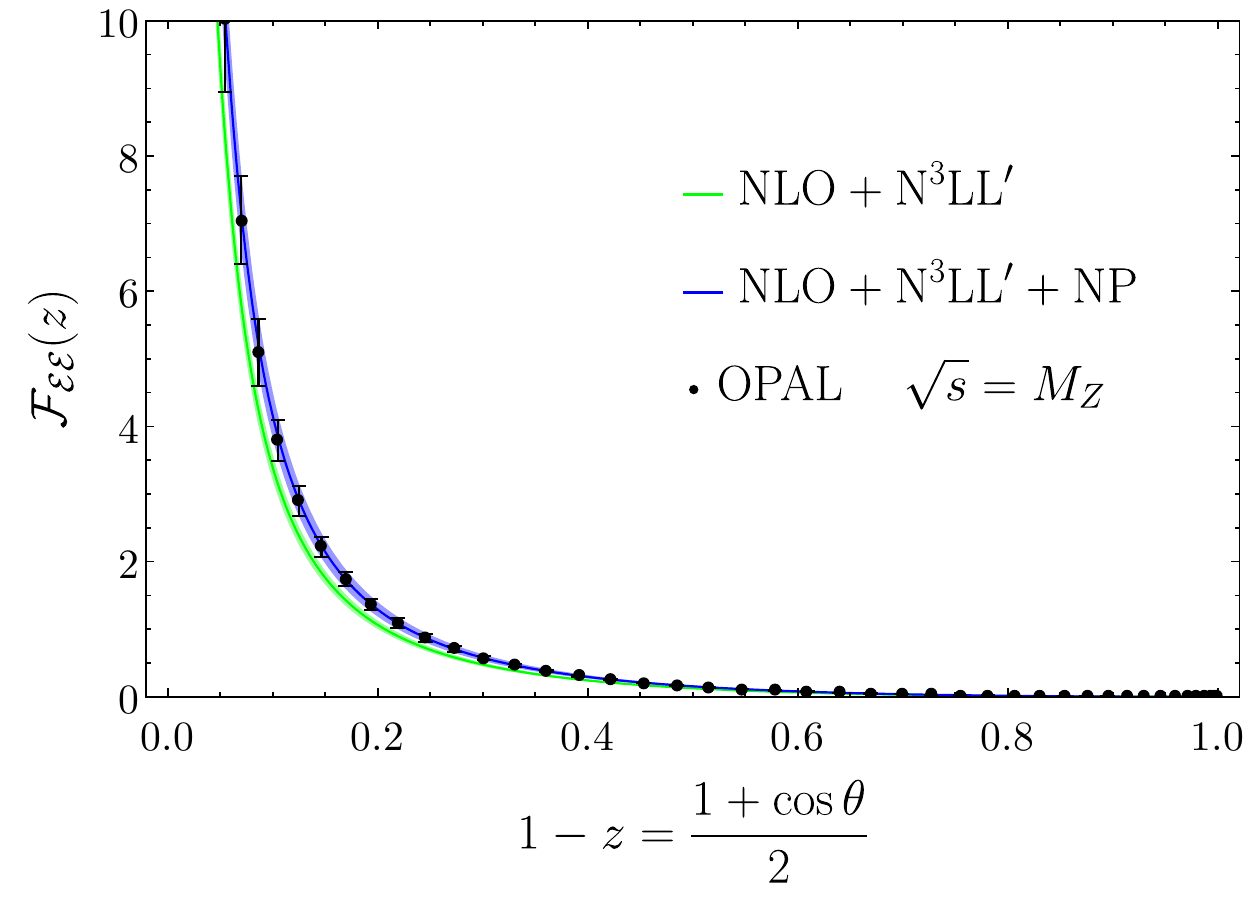}
    \caption{The EEC shape function extracted from the $e^+ e^-$ EEC spectrum at the $Z$ pole. As per \fig{rest_frame}, left panel displays the distribution on logarithmic axes to highlight the back-to-back behaviour, while the right panel shows the same result using linear axes. Curves for both the purely perturbative (green) and perturbative plus non-perturbative corrections (blue) are presented. For the perturbative curves, the bands correspond to scale variations by a factor of 2. The bands on the non-perturbative curves quadrature the bands from scale variation with the bands from the uncertainties on the non-perturbative ingredients.}
    \label{fig:CSF}
\end{figure}

Our aim is not to provide the most precise available computation of the EEC in the $Z$ rest frame, but rather to construct a calculation that is sufficiently accurate to identify the essential features of the EEC within a boosted $Z$ jet. To this end, we take the analytic NLO expression for $\td \Sigma^{z \sim 0.5}_{\ee \rightarrow {\rm hadrons}}/\td z$~\cite{Dixon:2018qgp} and perform subtractive matching to the numerical N$^3$LL$'$ result available in the literature~\cite{Ebert:2020sfi,Aglietti:2024xwv}. We carry out this procedure both with and without the inclusion of non-perturbative effects in the fixed-order region and in the back-to-back (b2b) limit, in order to assess their impact on the EEC when embedded within a boosted $Z$ jet.

Specifically, we consider two predictions for the $\Sigma_{\ee \rightarrow {\rm hadrons}}$ spectrum:
\begin{align}
    \frac{\td \Sigma^{\rm NLO + N^3LL'}_{\ee \rightarrow {\rm hadrons}}}{\td z}
    = \frac{\td \Sigma^{\rm NLO}_{\ee \rightarrow {\rm hadrons}}}{\td z}
    + \frac{\td \Sigma^{\rm N^3LL' \, b2b}_{\ee \rightarrow {\rm hadrons}}}{\td z}
    - \frac{\td \Sigma^{\rm NLO}_{\ee \rightarrow {\rm hadrons}}}{\td z} \bigg|_{z \rightarrow 1} \, ,
\end{align}
and
\begin{align}
    \frac{\td \Sigma^{\rm NLO + N^3LL' + NP}_{\ee \rightarrow {\rm hadrons}}}{\td z}
    = f_{\rm NP}(z)\,
    \frac{\td \Sigma^{\rm NLO + N^3LL'}_{\ee \rightarrow {\rm hadrons}}}{\td z} \, .
\end{align}
Here $f_{\rm NP}(z)$ is a non-perturbative envelope, given by
\begin{align}
    f_{\rm NP}(z) &= \left(1 - g(z,a,b)\right)\,
    \frac{\td \Sigma^{z \sim 0.5}_{\ee \rightarrow {\rm hadrons}}}{\td z}
    \bigg/
    \frac{\td \Sigma^{\rm NLO}_{\ee \rightarrow {\rm hadrons}}}{\td z}\\
    &\quad
    + g(z,a,b) \,
    \frac{\td \Sigma^{\rm Dokshitzer \, \emph{et al.}}_{\ee \rightarrow {\rm hadrons}}}{\td z}
    \bigg/
    \frac{\td \Sigma^{\rm NLL \, b2b}_{\ee \rightarrow {\rm hadrons}}}{\td z} \, . \nonumber
\end{align}
The function $g(z,a,b)$ is a standard unit smooth-transition function which provides a smooth interpolation between the two regimes over the interval $a < z < b$. It is given by
\begin{align}
    g(z,a,b) = \frac{f(\tilde{z})}{f(\tilde{z})+f(1-\tilde{z})} \Bigg|_{\tilde{z} = \frac{z-a}{b-a}} \, , ~~~ {\rm where} ~~~ f(x) = e^{-1/x} \, \Theta(x) \, .
\end{align}
The quantity $\Sigma^{\rm Dokshitzer \, \emph{et al.}}_{\ee \rightarrow {\rm hadrons}}$ denotes the NLL prediction supplemented with the dispersive model for non-perturbative power corrections in the back-to-back limit~\cite{Dokshitzer:1999sh}. The parameters $a = 0.93$ and $b = 0.998$ are chosen to ensure a smooth, kink-free interpolation while preserving the sum rule
\begin{align}
    \int \td z \,
    \frac{\td \Sigma^{\rm NLO + N^3LL' + NP}_{\ee \rightarrow {\rm hadrons}}}{\td z}
    = 1 \, .
\end{align}
Reasonable variations in $a$ and $b$ lie within the quoted theoretical uncertainties.

Finally, we use the $R$-scheme parameter value $\Omega = 0.32 \pm 0.05~\mathrm{GeV}$ from Ref.~\cite{Abbate:2010xh}, together with the standard value $\alpha_{\rm s}(M_Z) = 0.118$. Taken together, these values are slightly larger than the single best-fit results reported in the literature~\cite{Schindler:2023cww} in the $\overline{\rm MS}$ scheme. Since we treat power corrections at leading (classical) accuracy and the fixed-order spectrum at NLO, we do not attempt to distinguish between different schemes for $\Omega$. The larger value of $\Omega$ effectively absorbs part of the missing NNLO contributions, thereby improving the phenomenological accuracy of our proof-of-principle predictions, which are limited to NLO fixed-order accuracy. However, rigorous predictions for practical phenomenology should employ the most precise available perturbative ingredients (N$^4$LL matched to N$^2$LO~\cite{Electron-PositronAlliance:2025fhk}) together with parameters extracted in a renormalisation-scheme-consistent framework.

\Fig{rest_frame} shows the resulting spectra, both with and without non-perturbative contributions, together with OPAL data at the $Z$ pole~\cite{OPAL:1990reb}.

\subsection{Numerical evaluation of $\Sigma_{ee}$ and $\Sigma$}

In this section we numerically evaluate the $\Sigma$ and $\Sigma_{ee}$ spectra within the narrow width approximation using the EEC shape function which we have constructed from the literature (discussed in the previous section). We also extract the EEC shape function directly from OPAL data and use it to evaluate $\Sigma$ and $\Sigma_{ee}$.

\begin{figure}
    \centering
    \includegraphics[width=0.6\linewidth]{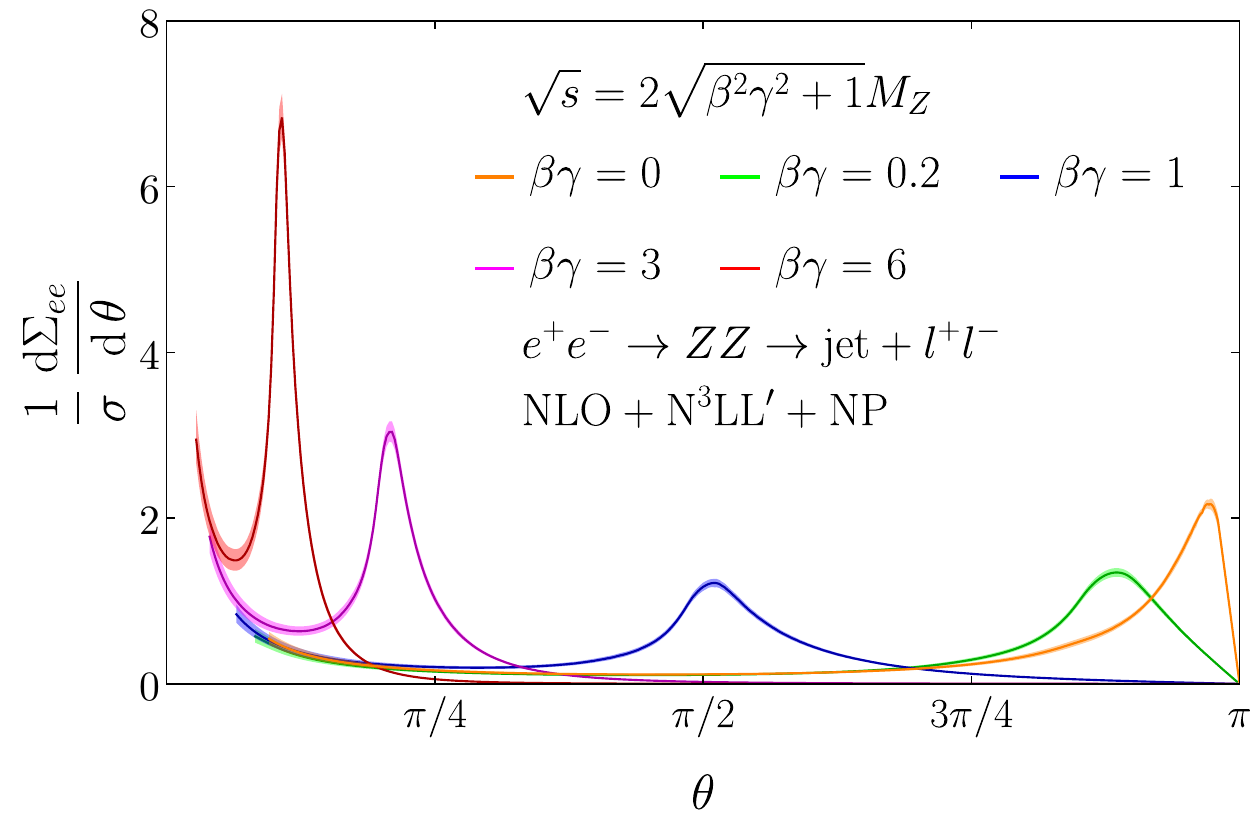}
    \caption{The spectrum $\td \Sigma_{ee} / \td \theta$ for $e^+e^- \rightarrow ZZ \rightarrow {\rm hadrons} + \ell^+\ell^-$ at several centre-of-mass energies. The curves shown correspond to the evaluation of \eq{boostonee} with a shape function computed with NLO+N$^3$LL+NP accuracy (the blue curve in \fig{CSF}). The boost of the $Z$ bosons shifts the Sudakov peak to smaller angles, following the relation $z_{\rm peak} \approx 1/(\beta\gamma)^2$ for large boosts.}
    \label{fig:eeZZboosts}
\end{figure}

We extract the EEC shape function from \fig{rest_frame} using \eq{FEEdef}, yielding \fig{CSF}. This is then used as input to evaluate \eq{boostonee} and \eq{boostonpp}. We begin by considering the process $e^+ e^- \rightarrow Z Z \rightarrow {\rm hadrons} + \ell^+ \ell^-$ and compute $\Sigma_{ee}$ from \eq{boostonee}. The results for several centre-of-mass energies are presented in \fig{eeZZboosts}. The Jacobian between the variables $z$ and $\theta$ causes the Sudakov shoulder near the back-to-back configuration ($\theta \rightarrow \pi$) to appear as a narrow peak when the $Z$ bosons are produced at rest ($\beta \gamma = 0$). As the centre-of-mass energy increases, and the $Z$ bosons become increasingly boosted, this peak moves to smaller angles. The Born peak position is given by $z_{\rm peak} \approx 1/\gamma^2$, where $z=(1-\cos\theta)/2$ as explained in \secn{discussion}. For large boosts, this corresponds to $z_{\rm peak} \approx 1/(\beta\gamma)^2 = (M_Z / P_T^Z)^2$ in terms of the natural hadron collider variables. This motivates our choice of presentation: we plot the mixed variables $\td \Sigma_{ee}/\td \theta$ as a function of~$z$ and express the boost in terms of $\beta\gamma$, allowing direct comparison between $\Sigma_{ee}$ and its hadron collider counterpart, $\Sigma$.

\begin{figure}
    \centering
    \includegraphics[width=0.49\linewidth]{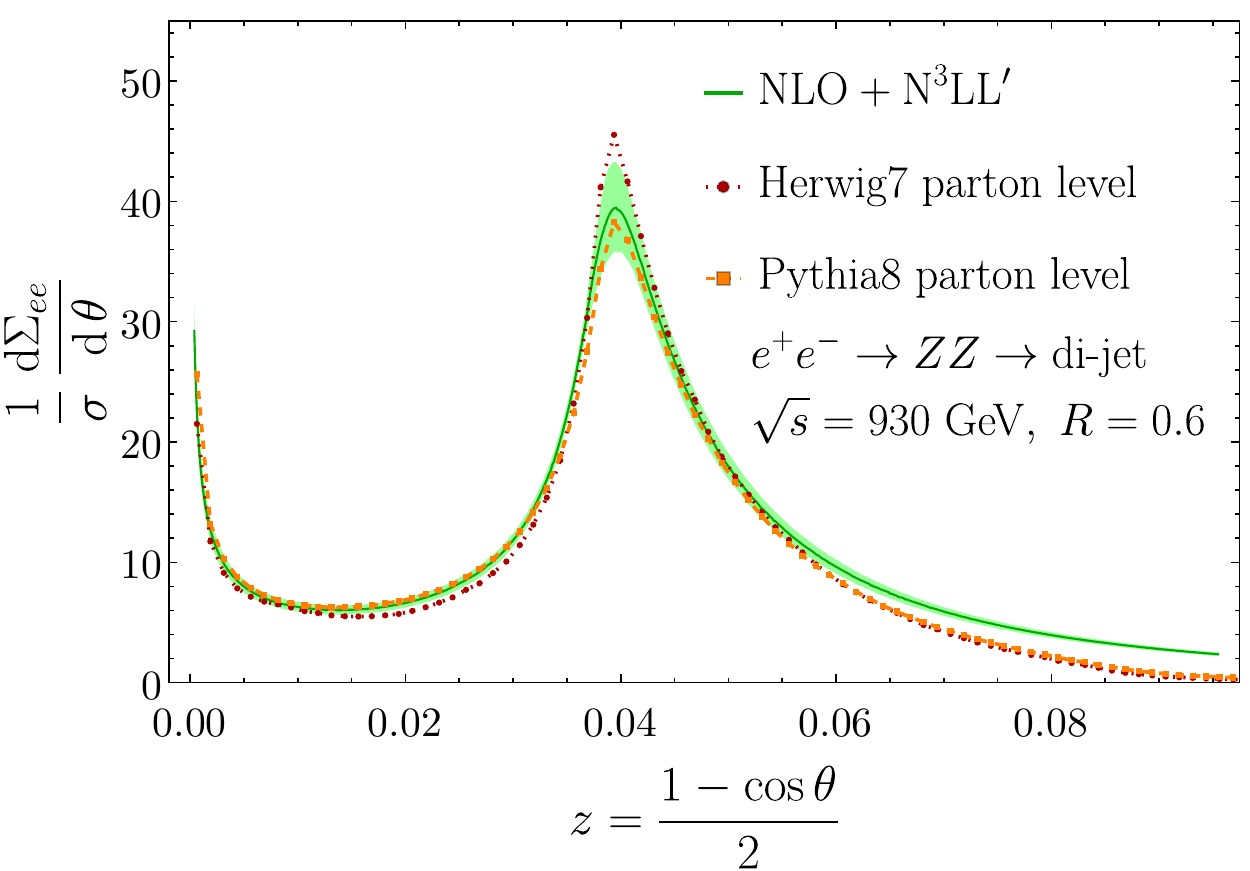}
    \includegraphics[width=0.49\linewidth]{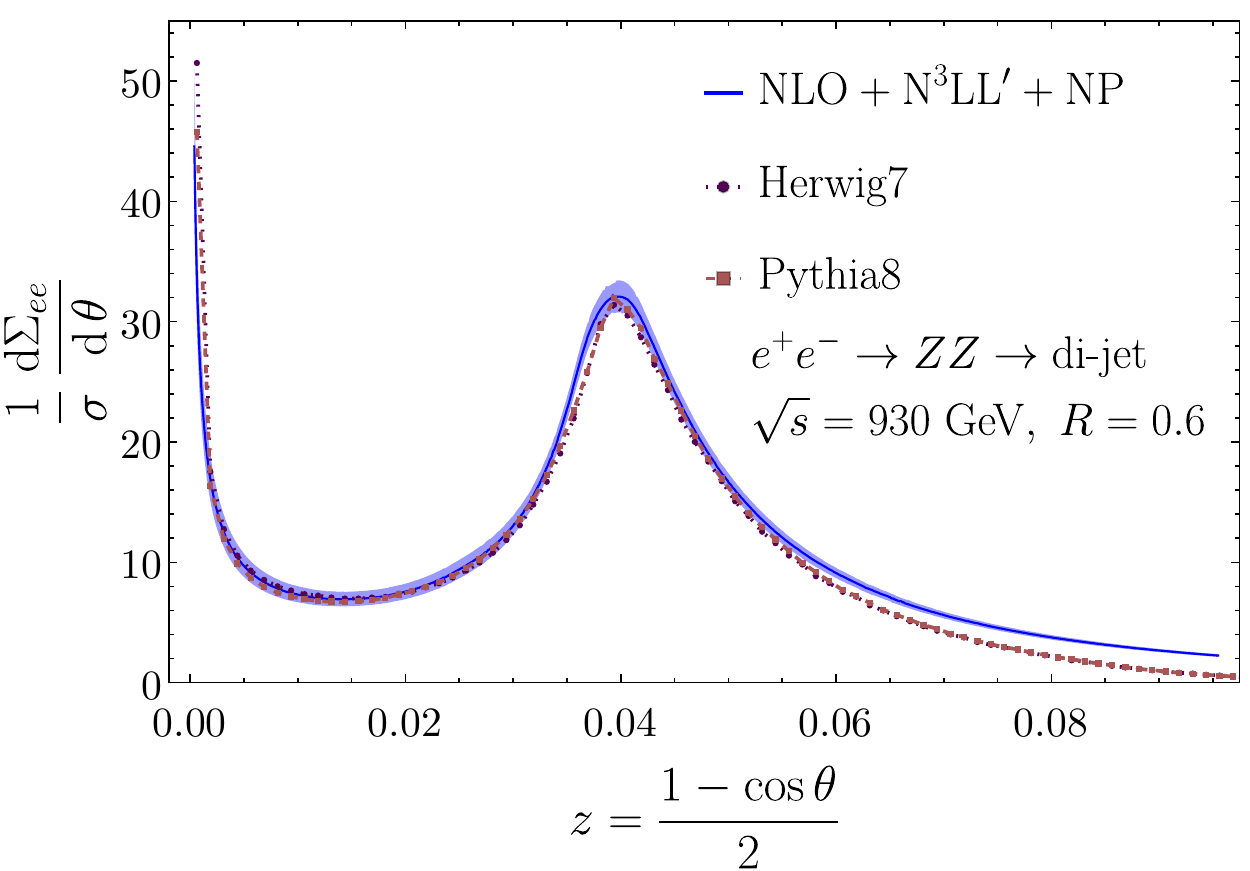}
    \caption{Comparison of the prediction for $\td \Sigma_{ee} / \td \theta$ using the factorisation in \eq{boostonee} at $\sqrt{s}=930~\text{GeV}$ against \Herwig and \Pythia simulations. 
        Left: perturbative comparison with hadronisation turned off. 
        Right: full comparison including non-perturbative effects. 
        The \Herwig and \Pythia curves are produced from events with a $e^+ e^- \rightarrow ZZ \rightarrow q \bar q q'\bar q'$ hard process. To increase computational speed, jets are reconstructed using anti-$k_T$ clustering with $R=0.6$, and the EEC is computed on the constituents of the two leading jets within the fiducial selection. The jet selection has no effect on the measurement in the Sudakov region.}
    \label{fig:eeZZ}
\end{figure}

In \fig{eeZZ} we show the result at $\sqrt{s}=930~\text{GeV}$, corresponding to $\beta\gamma = 5$ for the Born $Z$ bosons. We compare with the \Herwig and \Pythia event generators. Events were generated using \Herwigxx~\cite{Bewick:2023tfi} and \Pythiaxx~\cite{Bierlich:2022pfr}, and the EEC was computed with \texttt{Rivet~3}~\cite{Bierlich:2019rhm}. The hard process was generated at leading order for the topology $e^+ e^- \rightarrow ZZ \rightarrow q \bar q q'\bar q'$. Jets were clustered using anti-$k_T$ in \texttt{fastjet}~\cite{Cacciari:2011ma} and the EEC computed on their constituents, emulating an inclusive $Z$-jet type measurement. Computing the EEC only on the jet constituents provides a significant computational speed-up without distorting our results since jet radius corrections maximally scale as $R^2$, only affecting $z \gtrsim R^2/4$, which lies outside the region of interest. Jet clustering is also required for the $pp$ observable $\Sigma$, so it is instructive to first assess its impact in the clean $e^+e^-$ environment. We use $R=0.6$ and require the jet momentum to lie in the interval $[4.85 M_Z ,\, 5.1 M_Z]$. In each event, the EEC is computed on the constituents of the two leading jets which pass the jet momentum cut.

\Fig{eeZZ} shows remarkably good agreement between the curves computed from \eq{boostonee} and the event-generator results. The left panel compares perturbative predictions: the green curve from \fig{CSF}, boosted to $\beta\gamma=5$ and neglecting both non-perturbative power corrections and non-perturbative effects in the back-to-back resummation, is compared with parton-level event-generator results obtained with hadronisation switched off. The right panel includes non-perturbative effects, comparing the boosted blue curve from \fig{CSF} with full hadron-level simulations. Notably, \Herwig and \Pythia exhibit very good agreement once hadronisation is included, despite differing by as much as $20\%$ at parton level. This likely reflects the fact that both hadronisation models are tuned to $e^+e^-$ data at the $Z$ pole and therefore yield consistent predictions for this observable. Jet-radius effects become relevant only for $z > 0.06$, leading to a suppression of the event-generator curves relative to the predictions from \eq{boostonee}.

The origin of the parton-level discrepancy is difficult to interpret quantitatively given the substantial differences between the shower formalisms employed by the two generators. In particular, the showers use different evolution schemes (dipole versus angular-ordered), each of which is formally accurate only at leading-logarithmic and leading-colour level; and they also rely on different tuned parameter choices, including the shower strong coupling and infrared cut-off. Consequently, neither shower should be regarded as a faithful representation of a systematically accurate perturbative calculation. Rather, the spread between their predictions provides a reasonable estimate of the range within which the full perturbative result is expected to lie. 

\begin{figure}
    \centering
    \includegraphics[width=0.6\linewidth]{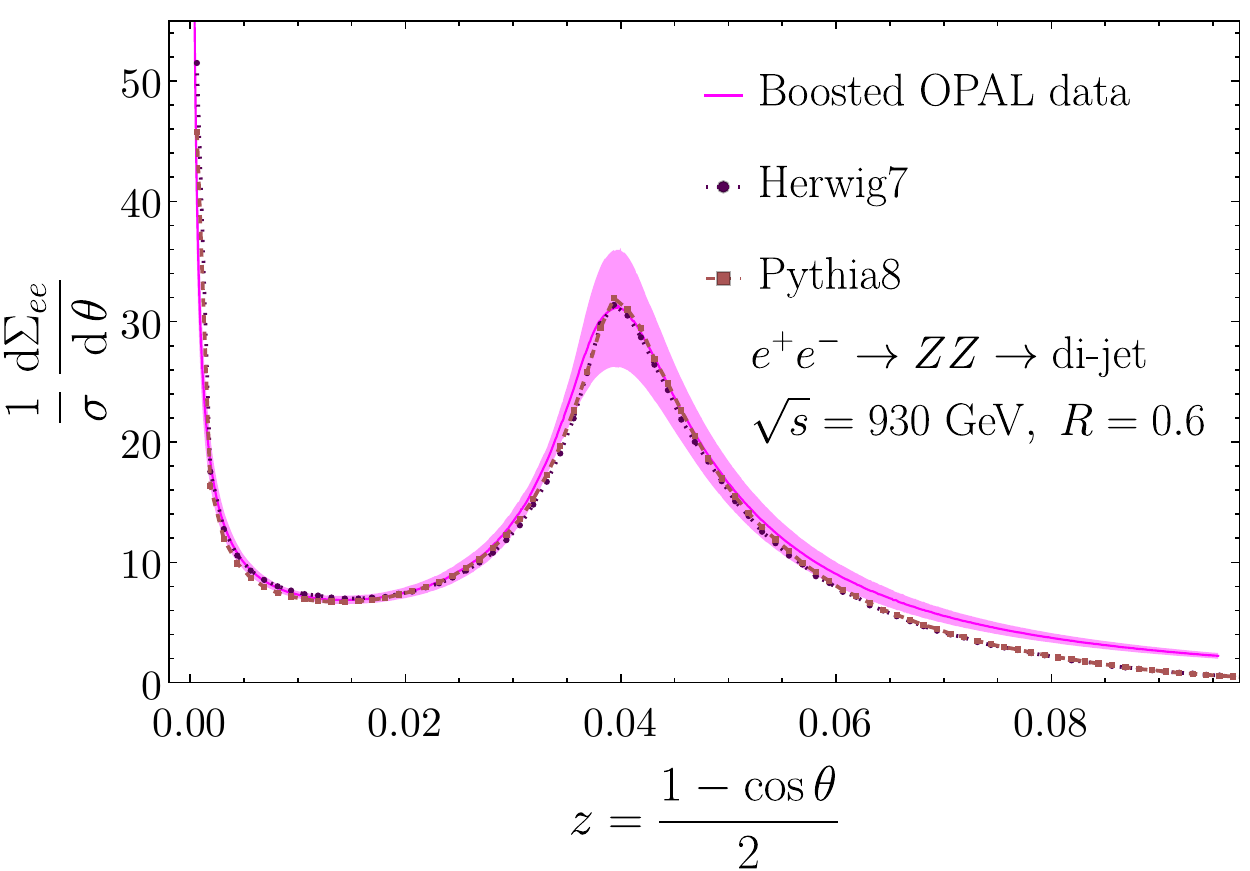}
    \caption{As in the right panel of \fig{eeZZ}, but using the EEC shape function extracted directly from OPAL data taken at the $Z$ pole. The magenta band represents the propagated experimental uncertainties from the extraction.}
    \label{fig:eeZZOPAL}
\end{figure}

Finally, in \fig{eeZZOPAL} we show the results using a shape function extracted directly from OPAL data. Interpolation was required in order to perform the boost. The magenta band represents the propagated interpolated experimental uncertainties. We find that event-generator predictions lie within this band across the full range $z < 0.06$, beyond which jet radius effects become significant.

\begin{figure}
    \centering
    \includegraphics[width=0.49\linewidth]{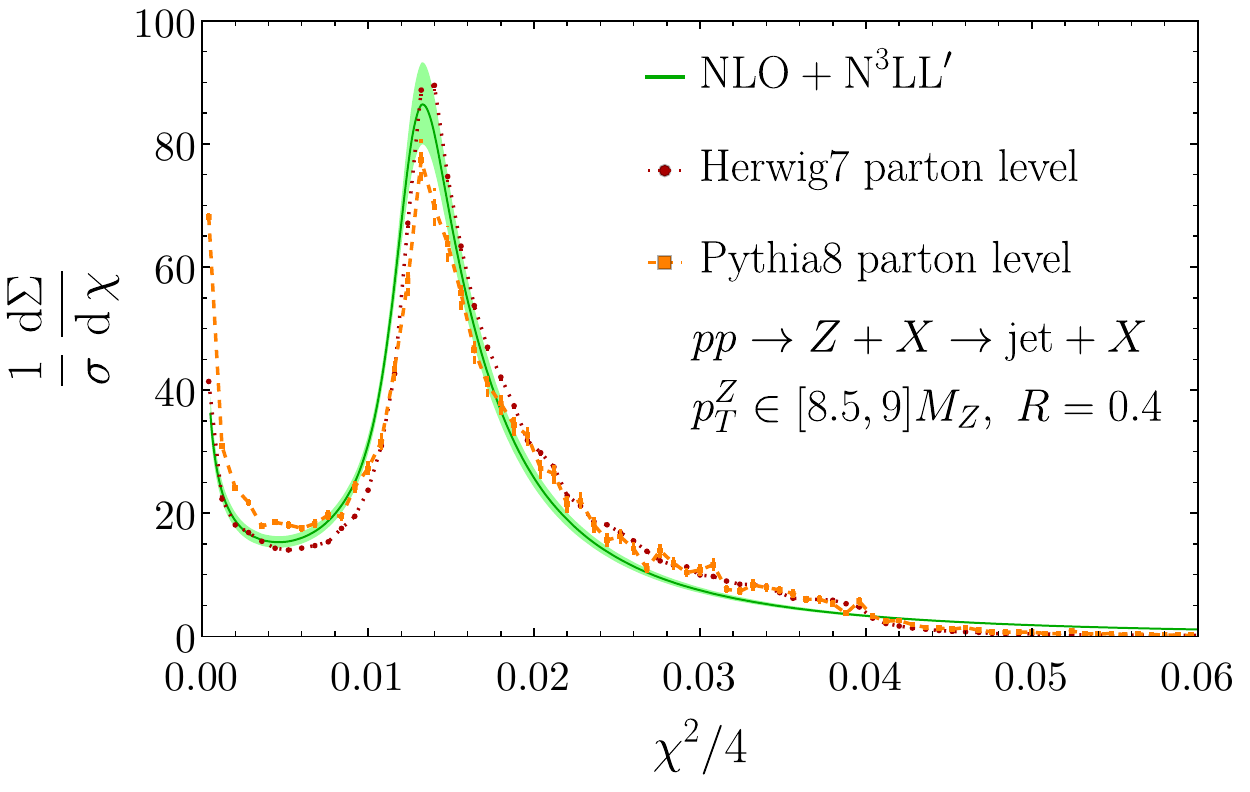}
    \includegraphics[width=0.49\linewidth]{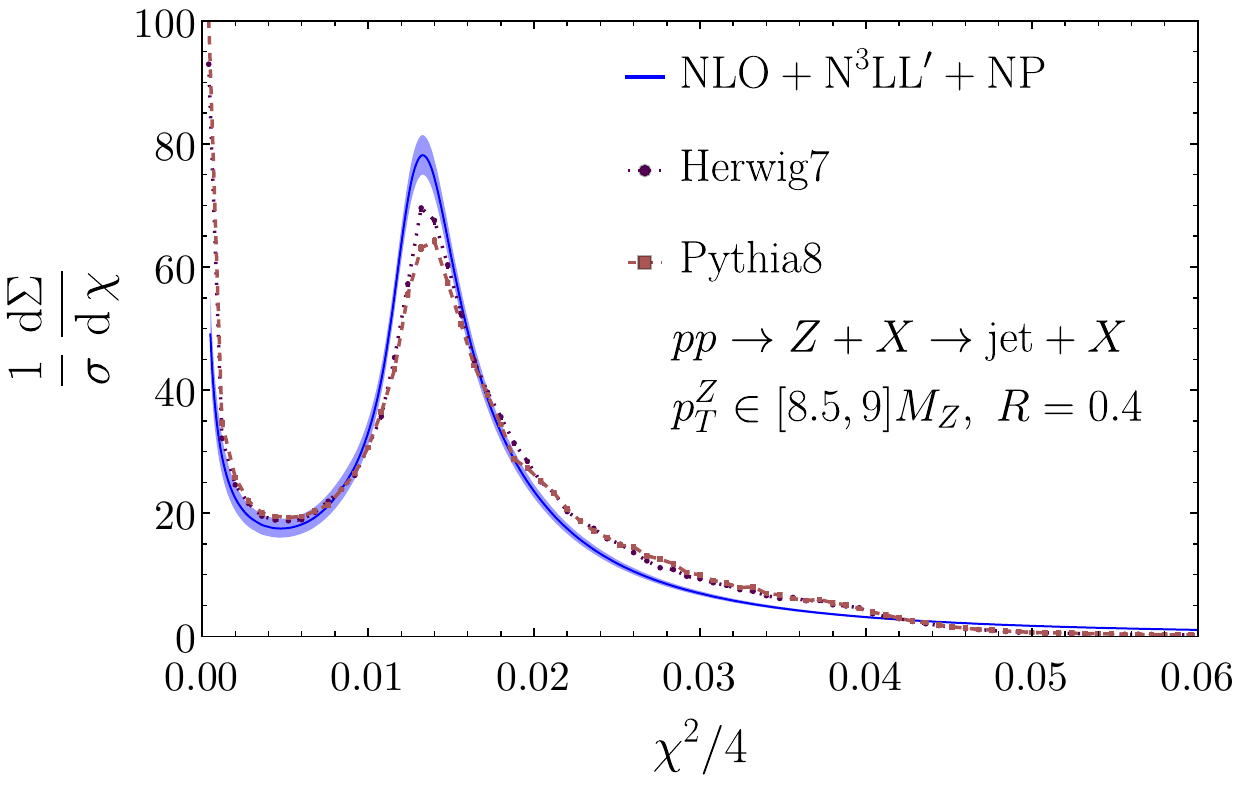}
    \caption{Comparison of the prediction for $\td \Sigma/\td \chi$, computed using \eq{boostonpp}, against the \Herwig and \Pythia event generators for $Z$ jets in $pp$ collisions. The left panel shows parton-level results with hadronisation turned off, while the right panel is the full hadron-level result, including the underlying event. The EEC is evaluated on the constituents of the two leading $R=0.4$ jets in di-$Z$ tagged events, binned in jet $p_T$. Owing to the underlying event, there is a non-negligible mistag rate of $Z$ jets ($\sim 5\%$), which leads to a visible disagreement between the theoretical prediction and the event generators in the right panel. In \fig{ppZjetBackground}, the theoretical curves are improved by including a model of this background, resulting in close agreement.}
    \label{fig:ppZjet}
\end{figure}

We now turn to the evaluation of EEC correlations measured in $Z$-boson–tagged jets in $pp$ collisions, $\Sigma$. \Fig{ppZjet} compares the result of \eq{boostonpp} with predictions from \Pythia and \Herwig, in direct analogy with \fig{eeZZ}. As before, the simulations were performed using \Herwigxx~\cite{Bewick:2023tfi} and \Pythiaxx~\cite{Bierlich:2022pfr}, and the EEC was calculated using Rivet~3~\cite{Bierlich:2019rhm}. The hard subprocesses were generated at leading order with the requirement of two hadronically decaying $Z$ bosons in the final state and each with $p_{T,Z} \geq 300~\text{GeV}$. We use the generators’ default PDFs: \Herwig employs \texttt{CT14lo}~\cite{Dulat:2015mca}, while \Pythia uses \texttt{NNPDF2.3 QCD+QED LO}~\cite{Ball:2013hta}. The full curves include MPI/underlying-event. For each event, the EEC is computed separately on the constituents of each of the two leading anti-$k_T$ jets with radius $R=0.4$, and subsequently binned in jet $p_T$.

In \fig{ppZjet}, and in later plots, we show $\td \Sigma/\td \chi$ as a function of $\chi^2/4$. This choice makes contact with the $e^+e^-$ observable since, in the small-angle limit, $\chi^2/4 \approx z$ and therefore $\chi^2_{\rm peak}/4 \approx (M_Z/P_T^Z)^2$ at large boost. This enables a direct comparison with figures.~\ref{fig:eeZZ} and~\ref{fig:eeZZOPAL}.

To compare \eq{boostonpp} with the Monte Carlo predictions, the equation must be integrated over the $p_T^Z$ and rapidity range of the selected $Z$ jets. The rapidity dependence drops out due to the longitudinal boost invariance of the observable. However, the $p_T^Z$ integration is non-trivial because of the prefactor $\mathcal{N}(p_T^Z)$, which is the normalised transverse momentum spectrum of $Z$-tagged jets. For the results in \fig{ppZjet}, we take $\mathcal{N}(p_T^Z)$ from the \Herwig prediction for the $Z$-jet spectrum within the relevant $p_T^Z$ bin and perform a numerical integration over that bin. Consequently, we would expect better agreement with \Herwig than with \Pythia since the two generators do not predict the spectrum of $p_T^Z$ within each bin.

\begin{figure}
    \centering
    \includegraphics[width=0.6\linewidth]{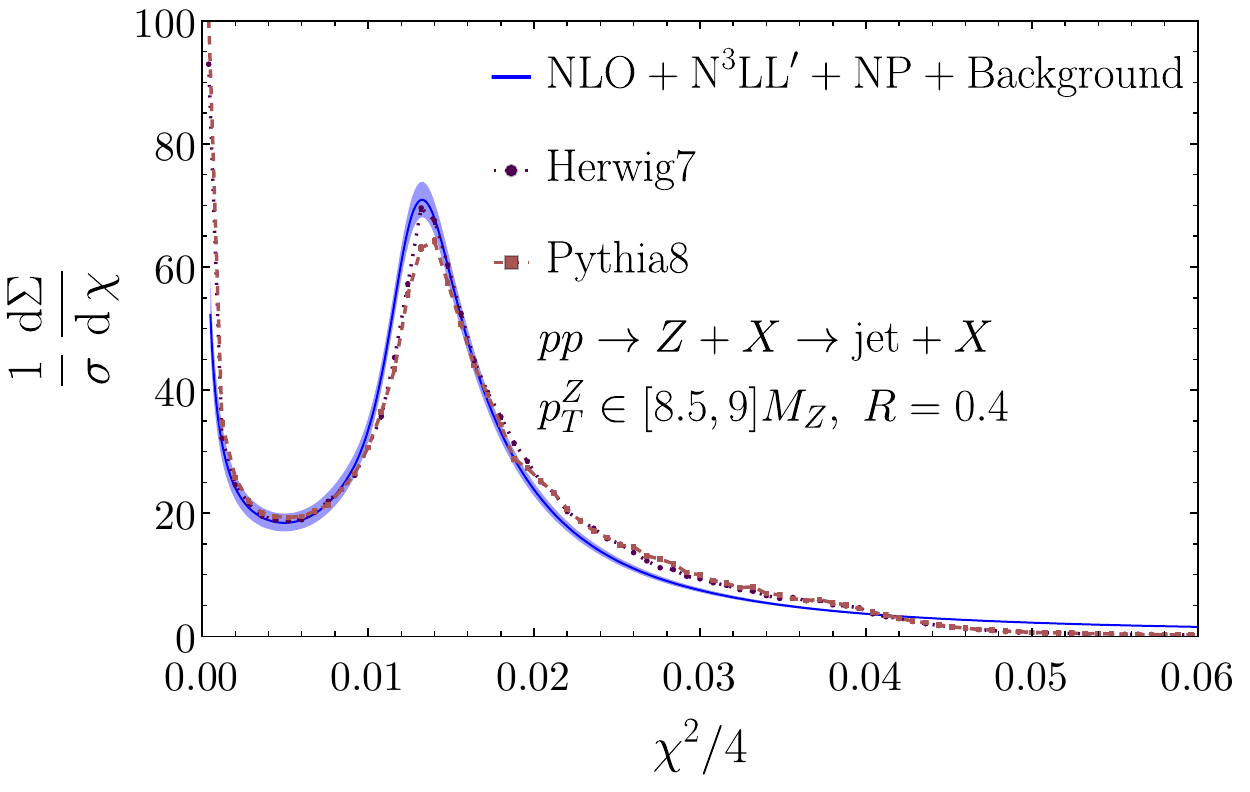}
    \caption{As in \fig{ppZjet}, but including a the fake-$Z$ jet background at leading order within the theoretical prediction. The $p_T^Z$ distribution used to compute the blue curve was extracted from simulations using \Herwig. Consequently, better agreement with the \Herwig generator is expected.}
    \label{fig:ppZjetBackground}
\end{figure}

Our event-generator setup identifies $Z$ jets without an explicit tagger: we simply select the two leading jets in events known to contain two hadronically decaying $Z$ bosons and which pass the fiducial cuts. Consequently, the sample contains a small fraction of ``fake’’ $Z$ jets arising from reconstructed quark or gluon jets that acquire large transverse momentum from initial- and final-state radiation and also the underlying event. Approximately $4\%$ of selected jets fall into this category. At leading order, the EEC contribution from such fake-$Z$ jets follows a simple power law for massless QCD jets,
\begin{align}
    \frac{\td \Sigma_{{\rm Fake}\hbox{-}Z}}{\td \chi \, \td p^{Z}_{T} \, \td \eta_Z} \sim \frac{1}{\chi},
\end{align}
allowing us to incorporate it straightforwardly into the prediction. \Fig{ppZjetBackground} shows the result after including this background component. We find very good agreement between the Monte Carlo simulations and the prediction of \eq{boostonpp} supplemented with the leading-order background.

\begin{figure}
    \centering
    \includegraphics[width=0.6\linewidth]{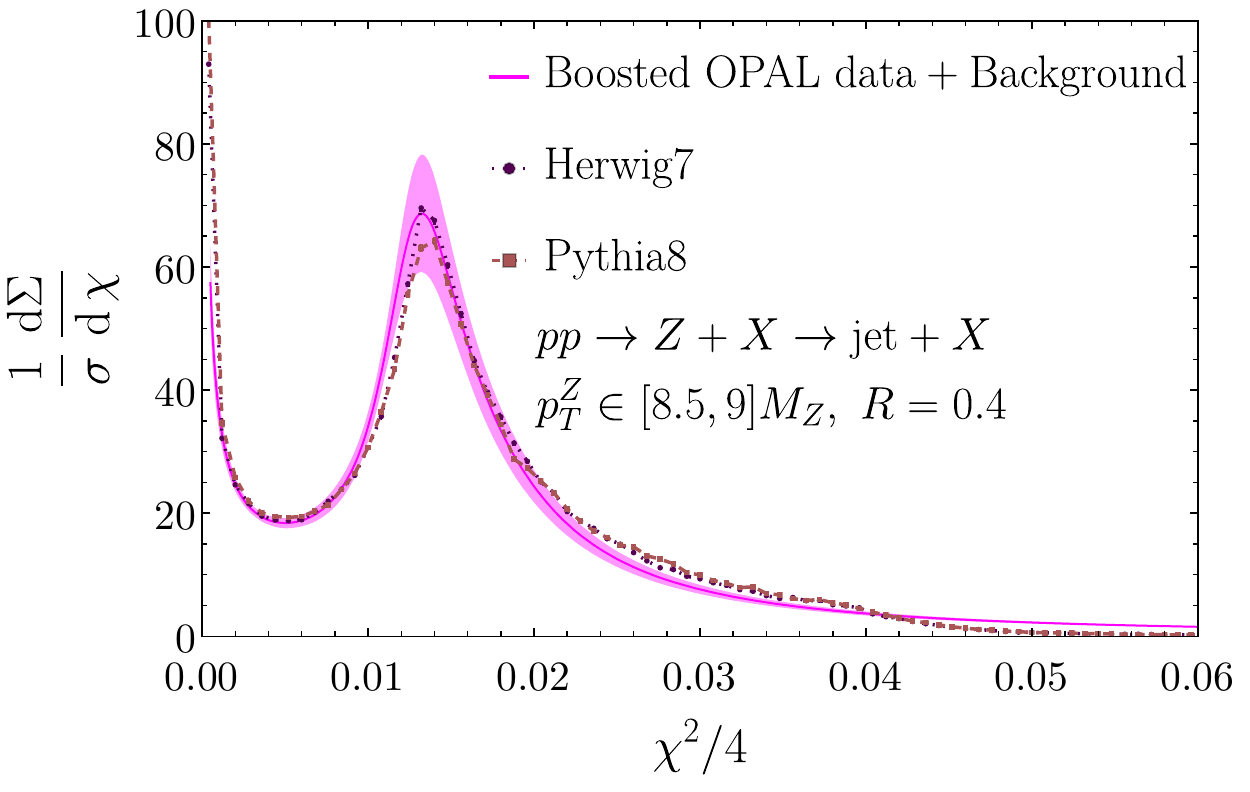}
    \caption{As in the right panel of \fig{ppZjetBackground}, but using the EEC shape function extracted from OPAL data. The shaded region indicates the propagated experimental uncertainties. The $p_T^Z$ distribution used to compute the violet curve was extracted from simulations using \Herwig. Consequently, better agreement with the \Herwig generator is expected. Nevertheless, both \Herwig and \Pythia mostly sit within the propagated uncertainties around the Sudakov peak.}
    \label{fig:ppZjetOpal}
\end{figure}

Finally, \fig{ppZjetOpal} presents the same comparison but using the EEC shape function extracted directly from OPAL data. As in the $e^+e^-$ case (\fig{eeZZOPAL}), the simulated curves lie almost entirely within the propagated experimental uncertainty band around the Sudakov peak.

\subsection{Discussion}

\label{sec:discussion}

The $pp$ and $e^+ e^-$ implementations of this observable share many common features, along with a few notable differences. We first discuss the common ground.

The Sudakov peak originates when the EEC shape function is evaluated in the region $\bar z \rightarrow 1$. Consequently, the position of the peak in the boosted frame can be deduced from the relation
\begin{align}
    \frac{q^2 \, n_1 \cdot n_2}{2 \, q \cdot n_1 \, q \cdot n_2} \approx 1.
\end{align}
The largest contribution arises from a leading-order, on-shell $Z$ decay into two equal-energy partons, which gives the approximate relation
\begin{align}
    \theta_{\rm peak} \approx 2 \arctan\left(\frac{1}{\beta \gamma} \right) ,
\end{align}
where $\beta$ and $\gamma$ denote the velocity and Lorentz $\gamma$-factor of the $Z$ boson. This can be solved to find $z_{\rm peak} \approx 1/\gamma^2$ which holds independently of the magnitude of the boost and so is applicable throughout the $e^+e^-$ implementation of the observable. In the large-boost limit,
\begin{align}
    \theta_{\rm peak} \approx \chi_{\rm peak} = \frac{2}{\beta \gamma} + \mathcal{O}\left( (\beta \gamma)^{-2} \right),
\end{align}
leading to the simple ``rule of thumb'' $\chi_{\rm peak}/4 \approx (M_Z/P_T^Z)^2$.

\begin{figure}
    \centering
    \includegraphics[width=0.485\linewidth]{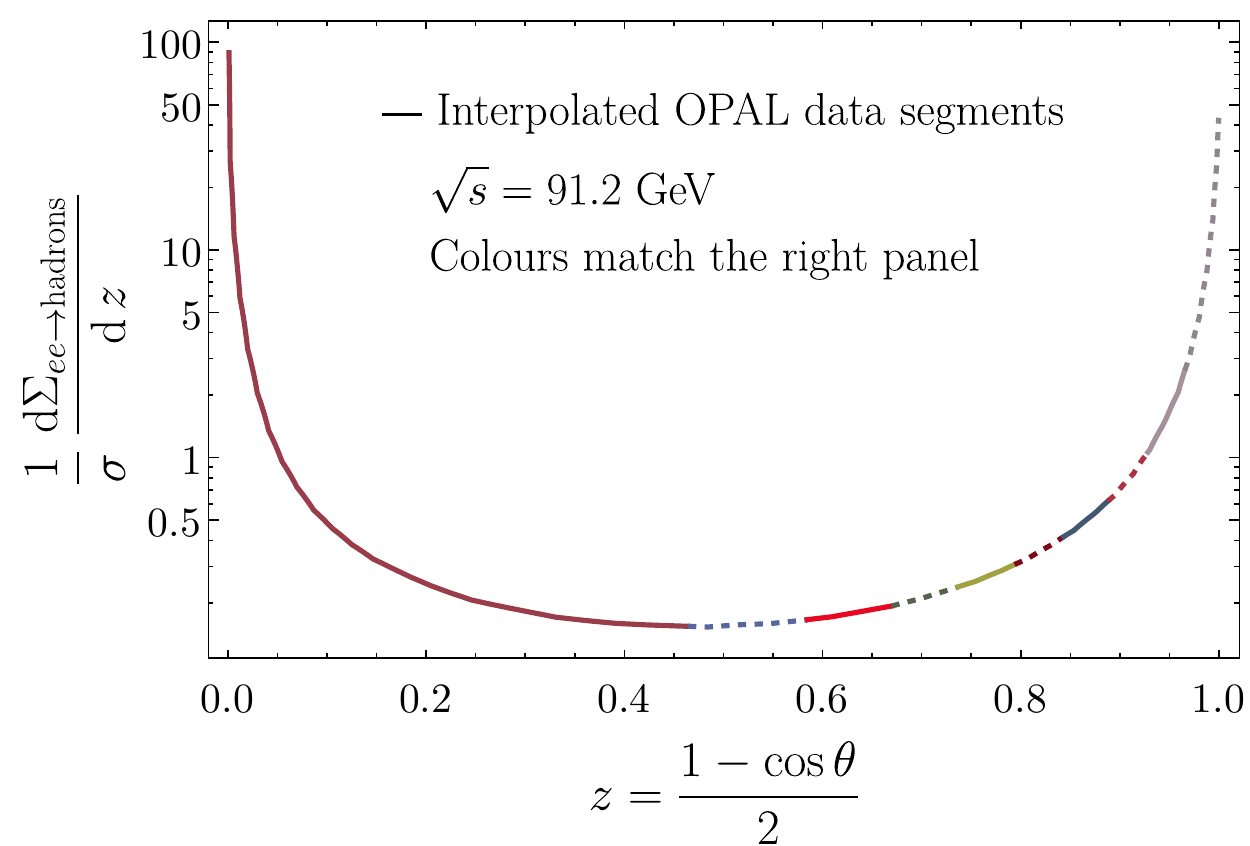} ~~
    \includegraphics[width=0.485\linewidth]{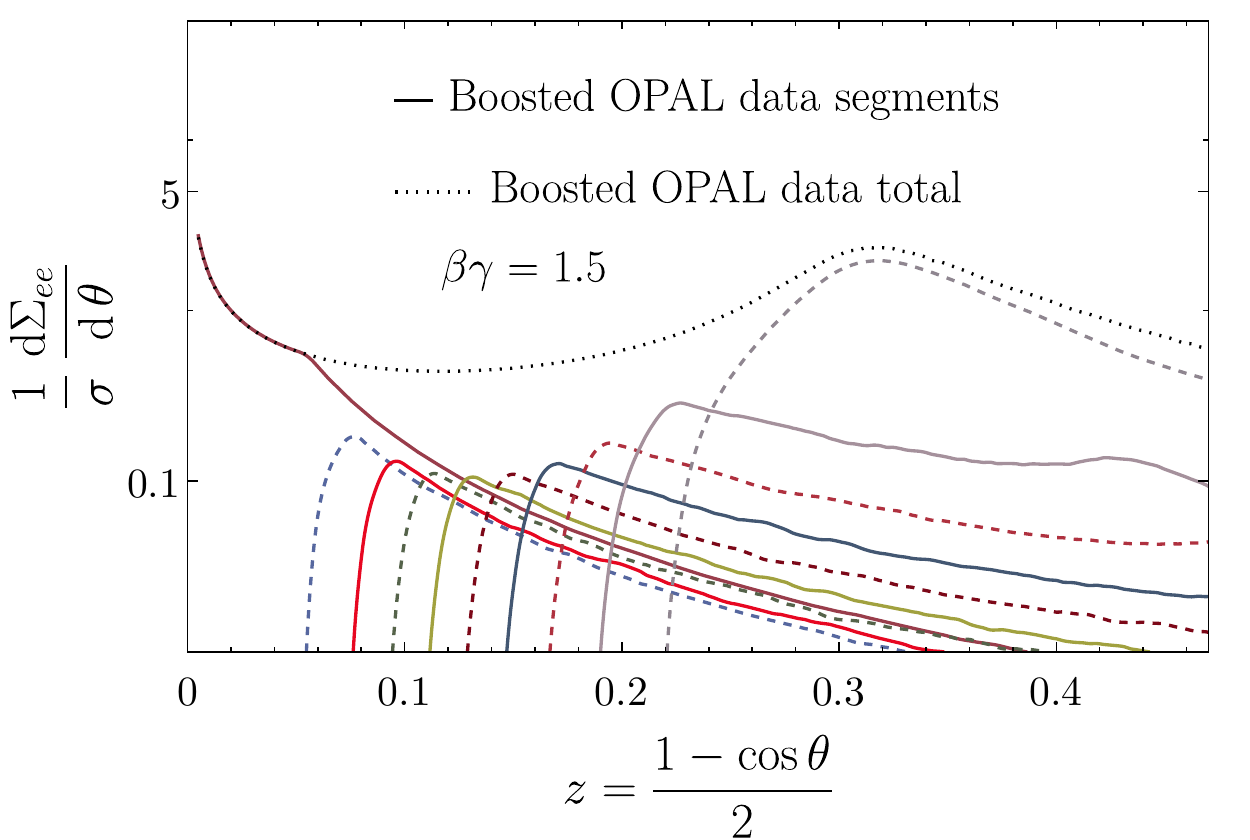}
    \caption{An illustration of the migration of features in the rest-frame towards smaller angles under a boost. Left: the OPAL EEC measurement at the $Z$ pole, divided into ten colour-coded irregular angular segments chosen to achieve a finer resolution in the Sudakov region. For visibility, the segments alternate between solid and dashed curves. Right: the corresponding contributions from each colour coded segment to $\Sigma_{ee}$ for boosted $Z$ bosons with $\beta\gamma = 1.5$. The total boosted result is shown with a black dotted line.}
    \label{fig:OPAL_split_up}
\end{figure}

Boosting the $Z$ boson compresses features from their rest-frame angular positions to smaller angles, following the approximate relation $z_{\rm feature} \approx z^{\rm rest~frame}_{\rm feature}/\gamma^2$. However, integration over the distribution of allowed boosts smears rest-frame features in the boosted frame. \Fig{OPAL_split_up} illustrates this effect. The left panel shows the OPAL EEC measurement at the $Z$ pole, where error bars are omitted and the curve is interpolated. It has been divided into ten segments, each colour coded. The right panel shows the resulting $\Sigma_{ee}$ for $Z$ bosons with a boost of $\beta \gamma = 1.5$ (i.e. $\sqrt{s} \approx 329~$GeV). The black dotted curve is the full result, whilst the solid/dashed coloured curves show the contribution from each individual rest-frame segment. As expected, each segment provides a dominant contribution to $\Sigma_{ee}$ at $z_{\rm segment} \approx z^{\rm rest~frame}_{\rm segment}/\gamma^2$. However, each segment also exhibits a substantial tail to the right. We have observed numerically that, in the large-boost limit, this tail approximately follows a power law:
\begin{align}
    \frac{\Sigma^{\rm segment~tail}_{ee}}{\td z} \sim \frac{1}{z^{5/2}}.
\end{align}
The region to the right of the peak consists solely of these tails and therefore obeys the same approximate power law.

It is evident from \fig{OPAL_split_up} that the peak itself is dominated by the deeply back-to-back region. In the rest frame, this region is governed by the transition from perturbative Sudakov resummation to a non-perturbative scaling law describing free-hadron correlations. Consequently, non-perturbative effects in the distribution are expected to be largest at the peak. This is reflected in the size of the theory uncertainties in figures.~\ref{fig:eeZZboosts},~\ref{fig:ppZjet},~\ref{fig:ppZjetBackground}, which are largest at the peak. By contrast, the slope to the left of the peak below the half width is dominated by Sudakov resummation, where analytical control is strongest. Above the half width and as the slope transitions into the peak, power corrections to the perturbative calculation become larger. In this region, the power corrections which are missing from the analytical calculations are expected to suppress the spectrum facilitating the transition into the non-perturbative free-hadron correlations. Consistent with this expectation, we do observe that the theory curves consistently overestimate this region by a small fraction when compared to the event generator curves whilst the boosted OPAL data finds closer agreement.

A key difference between the $e^{+}e^{-}$ and $pp$ measurements lies in the role of the leptonic and hadronic tensors. In $e^{+}e^{-}$, the leptonic tensor factorises entirely into the overall normalisation. In contrast, in $pp$ the hadronic tensor contains additional information on $Z$ production and therefore its boost. As a result, the $pp$ predictions require a convolution over the $Z$-jet $p_T$ spectrum within each measurement bin. For figures.~\ref{fig:ppZjet},~\ref{fig:ppZjetBackground},~\ref{fig:ppZjetOpal}, this spectrum was taken from the jet $p_T$ spectrum in the relevant bin as simulated by \Herwig.

The relative boost of the $Z$ boson is set by its off-shell momentum at the hard process ($q$ in \eq{start}). In the $pp$ case, we related $q$ to the measured jet transverse momentum, $p_T^Z$, via a leading-power expansion in the jet radius $R$ and $p_T^Z$. However, this expansion has limitations. The neglected power corrections encode pollution within the jet from underlying event, scaling as $R^2$, and momentum lost from the jet by soft physics, scaling as $M_Z / R p_T^Z$. To minimise these corrections, one must take the limit $R \ll 1$ whilst $R \, p_T^Z \gg M_Z$. By comparison to event generators, we found this limit to be physically realised for $R \leq 0.4$ and $p_T^Z > 600~$GeV. Away from this limit, particularly for lower $p_T^Z$ jets where larger $R$ is necessitated to capture the $Z$ decay, the Sudakov peak becomes misaligned between theory predictions and event-generator simulations due to a mischaracterisation of the underlying $Z$ hard $p_T$ spectrum. Addressing this, our factorisation can be systematically improved by introducing a convolution over $q$, without expanding in $M_Z / R p_T^Z$. In particular, \eq{Sigmapp} should be replaced by
\begin{align}	
    &\frac{\td \Sigma(R)}{\td \chi \, \td p^{Z}_{T} \, \td \eta_Z} = \nonumber \\
    & \int \td^4 q ~ \frac{ W_{\rm incl} (p_T^Z, \eta_Z, q^2, P_a \cdot q, P_b \cdot q, R)}{2E_{\rm cm}^2}  \frac{1}{\big[q^2 - M_Z^2\big]^2 +  \big[\Gamma_Z M_Z\big]^2 } \frac{H_{\rm EEC} (q,\chi)}{q_T^2} . \label{eq:qConvo}
\end{align}
Here $W_{\rm incl} (p_T^Z, \eta_Z, q^2, P_a \cdot q, P_b \cdot q)$ denotes the $Z$-jet $p_T$ spectrum as a function of the hard-process and initial-state kinematics ($q, P_a, P_b$). Crucially, since we work at first order in $\alpha_{\rm EW}$, $W_{\rm incl}$ is renormalisation-group invariant and hence can be extracted directly from an independent calculation of the $Z$-jet $p_T$ spectrum or from event-generator simulation. A full implementation of \eq{qConvo} lies beyond the scope of this article, but would extend the range of validity of our approach towards smaller $p_T^Z$.

The issues in inferring the underlying kinematics using only final state measurements mirror the problem encountered in the energy-correlator-based top-mass extraction through the top jet $p_T$ spectrum~\cite{Holguin:2022epo}. This was later solved through the introduction of a \emph{standard candle} strategy~\cite{Holguin:2023bjf}. In that approach, sensitivity to the underlying jet $p_T$ spectrum is replaced by a correlated measurement of the $W$ decay within the top decay chain, thereby removing the need to resum either the sub-leading hadronisation and underlying event effects which contribute to the top-jet $p_T$ spectrum.

\section{Higher-point correlations and re-factorisation  in the Sudakov region}

\label{sec:refact}

In \secn{Obs} we showed that an energy-energy correlator measurement on a boosted $Z$ jet factorises into a hadronic tensor describing the production process and a second hadronic tensor encoding the $Z$ decay together with the EEC measurement, $H^{\rm EEC}$. Subsequently, in \secn{HEEC} we demonstrated that $H^{\rm EEC}$ can be determined using symmetries together with existing results in the literature. This enabled precision predictions to be obtained in a straightforward manner and revealed that the peak structure in the EEC spectrum arises from a Sudakov resummation centred on the Born-level $Z$ decay kinematics. 

In this section, we build on these insights in two directions. First, we generalise the results of \secn{Obs} to higher-point energy-energy correlators. Second, we present a re-factorisation of $H_{\rm EEC}$ for a boosted decay formulated directly in terms of laboratory-frame kinematics. The advantage of this approach, which is complementary to the previous symmetries based analysis, is that it naturally extends to unstable heavy coloured resonances, for which a rest-frame description is less well defined. As such, it provides a useful test-bed for the theoretical description of EEC measurements on top-quark jets.

Thus far, this article has focused on the application of the two-point energy-energy correlator to the substructure of $Z$-boson jets. However, the arguments developed for the two-point correlator generalise straightforwardly to higher-point energy correlators. The corresponding observable for the process $pp \rightarrow Z + X \rightarrow {\rm jet} + X$ is
\begin{align}
    &\frac{\td \Sigma_{N}(R)}{\df^2 \mb n_1 \ldots \df^2 \mb n_N \, \td p^{Z}_{T} \, \td \eta_Z} = \nn \\
    &\quad \sum_{h_1, \ldots ,h_N \in Z\, {\rm jet}}  \int \df \Phi_{h_1 ,\ldots ,h_N } \:  	
    \Bigg(\prod_{j=1}^{N}   \frac{p_{T, {h_j}}}{p^{Z}_{T}} ~ 
    \delta^{(2)} \big(\mb n_{j} - \mb n_{h_j}\big)\Bigg) 
    ~ \frac{\td \sigma_{pp \rightarrow h_1 , \dots , h_N}  \left(R\right)}{\td p^{Z}_{T} \, \td \eta_Z \,  \df \Phi_{h_1,\ldots, h_N} },
\end{align}
where $p_T^Z$ is the measured $Z$-jet transverse momentum, $R$ is the jet radius, and $\eta_Z$ is the pseudo-rapidity of the jet axis. 
The quantity $\td \sigma_{pp \rightarrow h_1 , \dots , h_N}$ is the differential hadronic cross section to produce $N$ hadrons inside the tagged $Z$-jet, inclusive over everything else, with $p_{T,h_j}$ being its transverse  momentum and $\mb n_{h_j}$ the unit vector along its three-momentum. 

Following the discussion in sec.~\ref{sec:ppObs} and generalising eqs.~\eqref{eq:Sigmapp} and \eqref{eq:sigmaEEC}, this observable factorises at leading power as
\begin{align}
    \frac{\td \Sigma_{N}(R)}{\df^2 \mb n_1 \ldots \df^2 \mb n_N \, \td p^{Z}_{T} \, \td \eta_Z}
    = \int \td^4 q ~ \frac{\td \sigma_{\rm ENC}}{\td^4 q \, \df^2 \mb n_1 \ldots \df^2 \mb n_N} ~ \delta(q_T - p_T^Z) ~ \delta(y_q - \eta_Z),
\end{align}
where
\begin{align}	
    &\frac{\td \sigma_{\rm ENC}}{\td^4 q \, \df^2 \mb n_1 \ldots \df^2 \mb n_N}
    = \\
    &\quad \nn \frac{ W_{\rm incl} (q^2, P_a \cdot q, P_b \cdot q)}{2 E_{\rm cm}^2}
    \frac{1}{\big[q^2 - M_Z^2\big]^2 +  \big[\Gamma_Z M_Z\big]^2 }
    \frac{H_{\rm ENC} (q,\mb n_1, \ldots , \mb n_N)}{q_T^N}  \, .
\end{align}
Here $W_{\rm incl}$ is defined as in sec.~\ref{sec:ppObs}, and
\begin{align}\label{eq:HENC_def}
    &H_{\rm ENC} \big(q, \mb n_1 , \ldots \mb n_N\big) \\
    &\quad = \sum_{h_1, \ldots h_N \in \text{$Z$ jet}} 
    \int \df \Phi_{h_1 ,\ldots ,h_N } \: 
    \Bigg(\prod_{j=1}^{N}   p_{T, {h_j}}\, 
    \delta^{(2)} \big(\mb n_{j} - \mb n_{h_j}\big)\Bigg) 
    \frac{\df \sigma_{Z\ra h_1, \ldots ,h_N} (q)}{\df \Phi_{h_1 ,\ldots ,h_N }} \nn \,,
\end{align}
where the $N$-hadron distribution in the $Z$-initiated jet is defined by
\begin{align}\label{eq:N_had_dist}
    &\frac{\df \sigma_{Z\ra h_1,\ldots, h_N}(q)}{\df \Phi_{h_1 ,\ldots ,h_N }} \\
    &\quad \equiv 
    \bigg(\frac{q_\mu q_\nu}{M_Z^2} - g_{\mu\nu}\bigg) 
    \sum_{X_n} \int \df^4x \: e^{+\im x \cdot q} 
    \langle 0 | J_Z^{\dagger \mu} (x) | X_n, h_1, \ldots , h_N\rangle 
    \langle X_n, h_1, \ldots , h_N | J_Z^{\nu} (0) | 0 \rangle \, .\nn 
\end{align}
This factorisation holds at leading power in the limit $n_i\cdot n_j \sim (M_Z/p^Z_T)^2 \ll R^2 \ll 1$.

In the following, we derive a factorisation formula that describes the boosted Sudakov structure starting directly from the multi-hadron distribution in \eq{N_had_dist} formulated in the laboratory frame. We will primarily focus on $N=2$, then in \secn{three_body} we will discuss the generalisation to $N>2$. While the factorisation theorem for the back-to-back EEC in the rest frame of the $Z$ boson is well established~\cite{Moult:2018jzp}, the discussion presented here is novel in that it works explicitly with laboratory-frame kinematics, thereby making the role of the boost of the $Z$ boson manifest. The derivation therefore provides a stepping stone towards factorising EEC measurements on top-jets, wherein working with the boosted kinematics will be essential. The derivation employs soft collinear effective theory (SCET)~\cite{Bauer:2000ew,Bauer:2000yr,Bauer:2001ct,Bauer:2001yt,Bauer:2002nz}; however, the majority of technical details are deferred to the appendices. 
Our aim is for the discussion in the main body of the paper to remain broadly accessible to QCD practitioners familiar with transverse-momentum-dependent distribution functions (TMDs) and soft functions.

\subsection{Born configurations of $Z$ decay}
\label{sec:born}

The Sudakov feature arising from $N$-particle correlations emerges when the Born-level $N$-particle configuration is dressed with soft and collinear radiation. We will first look at $N=2$ case, again corresponding to the decay of a $Z$ boson, then in \secn{three_body} we will extend our discussion to $N=3$ or higher. We begin by describing the kinematics of the leading-order $Z$ decay, which provide the skeleton for the radiation pattern underlying the Sudakov peak in the EEC distribution. We subsequently dress this configuration with soft and collinear emissions, which induce small deviations and give rise to the Sudakov structure in the two-point correlator. For simplicity, we adopt lepton-collider kinematic variables (energies and angles) since in the small-angle limit we can equate $\theta \approx \chi$ and $E \approx p_T$ up to $R^2$ corrections. As per \eq{HENC_def}, we continue to assume that the entirety of the $Z$ decay products, and all subsequent radiation, is caught within the clustered jet.

The two-point correlator within a $Z$ jet, defined using \eqs{n1n2_corr}{HENC_def}, is given by
\begin{align}
    H_{\rm E2C} (q, z_{12}) = \sigma_0 \int \df^2 \mb n_1 \df^2 \mb n_2 \: \delta\bigg(z_{12} - \frac{n_1\cdot n_2}{2}\bigg) \langle \cE(n_1) \cE(n_2)\rangle_q \,,
\end{align}
where $\sigma_0$ denotes the Born cross section, and
\begin{align}\label{eq:Hee_h12_0}
    \langle \cE(n_1) \cE(n_2)\rangle_q  = \sum_{h_1,h_2 \in \text{jet}} \int \df \Phi_{h_1,h_2}\:  \Bigg(\prod_{i=1}^2 E_{h_i} \delta^{(2)}(\mb n_i - \mb n_{h_i})\Bigg)\frac{1}{\sigma_0}
    \frac{\df \sigma(q) }{\df \Phi_{h_1,h_2} }\,.
\end{align}
Here the hadrons $h_{1,2}$, carrying energies $E_{1,2}$ along the directions $\mb n_{1,2}$, are required to lie inside the jet. The invariant phase space and the di-hadron distribution are given, respectively, by \eqs{Phi_N}{N_had_dist} for $N=2$:
\begin{align}\label{eq:Hee_h12}
    \frac{\df \sigma(q) }{\df \Phi_{h_1,h_2} } &= \bigg( \frac{q_\mu q_\nu}{M_Z^2}  - g_{\mu\nu} \bigg) \sumint_{X_n} 
    (2\pi)^d \delta^{(d)} \big(P_{X_n} + p_{h_1} + p_{h_2} - q\big)
    \nn \\
    &\quad \times 
    \langle 0 | J_Z^{\dagger \mu} (0) | X_n , h_1 ,h_2 \rangle \langle h_1 , h_2 ,X_n | J_Z^\nu (0) | 0 \rangle \, .  
\end{align}
Here $X_n$ denotes additional emissions within the jet.

We now consider an on-shell $Z$ boson travelling along the direction $\mb n$, in the laboratory frame, with momentum
\begin{align}
    q^\mu = \big(Q , P_Z \mb n\big) \,,  \qquad P_Z^2 = Q^2 - M_Z^2 \, ,
\end{align}
which decays at leading order into a $q_a(p_a)\,\bar q_b(p_b)$ pair along the directions $\mb n_{a,b}$, such that
\begin{align}
    &p_a = E_{a} n_a^\mu \,,&
    &p_{b} = E_b n_b^\mu \,, &
    &n_{a,b} \equiv (1, \mb n_{a,b}) \, .&
\end{align}
At leading order, these directions coincide with $\mb n_{1,2}$ in \eq{Hee_h12_0}, with $h_1=q_a$ and $h_2=\bar q_b$ (or vice versa), although they may differ at higher orders due to additional radiation.

Significant simplifications arise by working in a coordinate system defined by the light-like vectors
\begin{align}
    \tn_a^\mu = \frac{1}{\sqrt{z_{ab}}}\big(1,\mb n_a\big) \,, \qquad   \tn_b^\mu = \frac{1}{\sqrt{z_{ab}} } \big(1, \mb n_b\big) \, ,
\end{align}
which satisfy $\tn_a \cdot \tn_b = 2$. In this basis, any momentum $p$ can be decomposed as
\begin{align}
    p^\mu = \big(p^+, p^-, \mb p_\perp\big)_{(ab)} = \underbrace{\tn_a\cdot p}_{p^+} \frac{\tn_b^\mu}{2} + \underbrace{\tn_b \cdot p}_{p^-} \frac{\tn_a^\mu}{2} + p_{\perp(ab)}^\mu \, .
\end{align}
This decomposition implicitly defines the transverse component $p_{\perp(ab)}^\mu$, with its boldface version corresponding to a Euclidean metric of positive norm.

This construction is commonly referred to as the ``parton-frame'' decomposition~\cite{Collins:1981uw,Collins:2011zzd}, since in the frame in which the two partons are back-to-back $\mb n_a = -\mb n_b$. In this decomposition, the Born-level partons $a$ and $b$ carry momenta
\begin{align}
    p_a^\mu = \big(0, \omega_a, \mb 0_\perp \big)_{(ab)} \,, \qquad 
    p_b^\mu = \big(\omega_b, 0 , \mb 0_\perp\big)_{(ab)} \,, \qquad
    q^\mu = \big(\omega_a, \omega_b , \mb 0_\perp \big)_{(ab)}\,,
\end{align}
with
\begin{align}
    &\omega_a = q\cdot \tn_b = \frac{Q - P_Z \mb n\cdot \mb n_b}{\sqrt{z_{ab}}} \,,& 
    &\omega_b = q\cdot \tn_a = \frac{Q - P_Z \mb n\cdot \mb n_a}{\sqrt{z_{ab}}} \,.& 
\end{align}
In the limit $P_Z = 0$, one has $E_a = E_b = M_Z/2$ and $z_{ab} = 1$, corresponding to the rest frame of the $Z$ boson. In this special case, the back-to-back EEC factorisation theorem is already well established~\cite{Moult:2018jzp}. Our interest here lies in the more general situation with $P_Z \neq 0$. Importantly, using $Q^2 = M_Z^2 = \omega_a \omega_b$, we find that the Born cross ratio defined in \eq{bar_z_def} is fixed to its maximal value,
\begin{align}
    \bar z_{ab}  \equiv \frac{M_Z^2 n_a \cdot n_b}{2 \, q\cdot n_a\, q\cdot n_b} = \frac{M_Z^2}{\omega_a \omega_b} = 1 \, ,
\end{align}
as required.

\subsection{Formation of the Sudakov feature}
\label{sec:hadron_frame}

Beyond leading order, the hadronic energy flux in the directions $\mb n_{1,2}$ generally differs from the flux along $\mb n_{a,b}$ due to recoil induced by the additional radiation.
As a consequence, the corresponding cross ratio $\bar z_{12}$ also deviates from unity. We are interested in the kinematic regime where $1-\bar z_{12} \ll 1$.

To describe this situation beyond leading order, we introduce the ``hadron-frame'' decomposition, in which the reference axes are aligned with the measured hadron directions $\mb n_{1}=\mb n_{h_1}$ and $\mb n_{2}=\mb n_{h_2}$. This choice differs slightly from the parton frame: the hadron frame corresponds directly to the experimentally measured configuration, while the parton frame more naturally captures the Born-level kinematics.

In the hadron frame decomposition, the reference vectors are
\begin{align}\label{eq:n12_def}
    \tn_{1}^\mu = \frac{1}{\sqrt{z_{12}}} \big(1, \mb n_{1} \big) \,, \qquad \tn_{2}^\mu = \frac{1}{\sqrt{z_{12}}} \big(1, \mb n_{2}\big) \,, \qquad z_{12} =\frac{n_1\cdot n_2}{2} =  \frac{1- \mb n_1 \cdot \mb n_2}{2} \, .
\end{align}
These vectors satisfy $\tn_1 \cdot \tn_2 = 2$ and $\tn_{1,2}^2 = 0$. The hadron momenta are exactly aligned with these reference directions,
\begin{align}\label{eq:p_h12}
    p_{h_1}^\mu = \sqrt{z_{12}} E_{h_1} \tn_1^\mu \,, \qquad 
    p_{h_2}^\mu = \sqrt{z_{12}} E_{h_2} \tn_2^\mu \, ,
\end{align}
where $E_{h_{1,2}}$ denote the hadron energies. The $Z$ boson momentum decomposes as
\begin{align}
    q^\mu =   \underbrace{q\cdot \tn_2}_{\omega_2} \frac{\tn_1^\mu}{2} + \underbrace{q\cdot \tn_1}_{\omega_1} \frac{\tn_2^\mu}{2}+ q_\perp^\mu =  \big( \omega_2, \omega_1, q_{\perp(12)}^\mu \big)_{(12)} \, .
\end{align}

In the hadron frame decomposition, the $Z$ boson acquires a non-zero transverse momentum, the magnitude of which parametrises the deviation from the $\bar z = 1$ configuration. Using
\begin{align}
    \mb q_{\perp(12)}^2  = q\cdot \tn_1 \, q\cdot \tn_2 - M_Z^2 = \omega_1 \omega_2 - M_Z^2 \, ,
\end{align}
the cross ratio for detectors placed along $n_{1}$ and $n_{2}$ is given by
\begin{align}\label{eq:bar_z_12}
    \bar z_{12} =   \frac{M_Z^2 n_1\cdot n_2}{2 \, q\cdot n_1 \, q\cdot n_2} =\frac{M_Z^2}{\omega_1\omega_2} =  1-  \frac{\mb q_{\perp(12)}^2}{\omega_1 \omega_2} \, .
\end{align}
Thus, the transverse momentum $\mb q_{\perp(12)}^2$ is fixed by the measurement of $\bar z_{12}$,
\begin{align}\label{eq:qperp_z12}
    \mb q_{\perp(12)}^2 = M_Z^2 \big(1- \bar z_{12}\big)/\bar z_{12}\,.
\end{align}

We consider the regime in which $q_T \sim \lambda \sqrt{\omega_1 \omega_2} \sim \lambda M_Z$, with a power-counting parameter $\lambda \ll 1$, such that $1-\bar z_{12} \sim \lambda^2$. In this limit,
\[
\mb q_{\perp(12)}^2 = M_Z^2 (1-\bar z_{12}) \big(1 + \cO(\lambda^2)\big) \, .
\]
For small transverse momentum, momentum conservation in the hadron frame implies that the state $X_n$ consists of emissions that are either collinear to the directions $\mb n_{a,b}$ or soft,
\begin{align}
    P_{X_n} = P_{X_{a}} + P_{X_b} + P_{X_s} \, .
\end{align}
The corresponding power counting in the hadron frame is
\begin{align}\label{eq:power_counting}
    &P_{X_a} \sim M_Z \big(\lambda^2, 1 ,\lambda \big)_{(12)} \, ,& 
    &P_{X_b} \sim M_Z \big(1, \lambda^2, \lambda\big)_{(12)} \,,& 
    &P_{X_s} \sim M_Z \big(\lambda, \lambda, \lambda \big)_{(12)} \, .&
\end{align}
Note that, in the laboratory frame, the emissions in $X_s$ are collinear to $n$ but carry lower energies than those in $X_{a,b}$. As a result, at leading power the dominant contribution to the EEC arises from emissions in $X_{a,b}$.

To develop the physical picture, it is convenient to define the partonic momenta as
\begin{align}\label{eq:kabXh}
    p_a = P_{X_a} + p_{h_1} \,, \qquad p_b = P_{X_b} + p_{h_2} \, .
\end{align}
Beyond leading order, the partons $a$ and $b$ initiate jets composed of $X_{a,b}$ together with the hadrons $h_{1,2}$. While the measured hadrons are aligned with the $n^\mu_{1,2}$ measurement axes, the remaining collinear radiation in $X_{a,b}$ recoils against the soft radiation $X_s$, as both carry transverse momentum of $\cO(\lambda)$. Consequently,
\begin{align}\label{eq:perp_12_conserve}
    q_{\perp(12)}^\mu = p_{a\perp(12)}^\mu + p_{b \perp(12)}^\mu + P_{X_s \perp(12) }^\mu \sim \lambda M_Z \, ,
\end{align}
where the subscript $(12)$ indicates that this relation is given in the hadron frame decomposition.

\subsection{Sudakov Factorization and consistency with EEC shape function}

\label{sec:fact}

The Sudakov feature in the EEC cross section can be described by considering the factorisation of the production of two identified hadrons inside the $Z$ jet, as given in \eq{Hee_h12}, valid in the limit $1-\bar{z}_{12} \ll 1$. Following the steps detailed in \app{fact}, the di-hadron factorisation theorem is
\begin{align} \label{eq:dihad_fact}
        &	\frac{1}{\sigma_0}\frac{\df \sigma(q)}{\df \Phi_{h_1,h_2}} = \frac{16(2\pi)^d \zeta_1 \zeta_2}{M_Z^{(d-2)}\Omega_{d-2}r_\eps}\sum_f    \frac{|a_f|^2 + |v_f|^2}{\sum_{f'} (|a_{f'}|^2 + |v_{f'}|^2)} \, \cH_f^{(0)}(M_Z^2)     \\
        & \quad  \times \nn 
        \int \df^{d-2} b_\perp \: e^{+\im b_\perp \cdot q_{\perp(12)}} \Big[\cD_{ h_1/f}\big(\zeta_1, b_\perp\big) \cD_{ h_2/\bar f}\big(\zeta_2, b_\perp \big) S_{n_1n_2} (b_\perp) + \big(n_1, f \lra n_2, \bar f\big)\Big] \nn \, ,
\end{align}
where $\Omega_{d-2}$ and $r_\eps$ are given in \eq{reps}. 
Here $\cH_f^{(0)}$ is the leading-power hard function for the dijet process $\ee \to f\bar f$, evaluated at the hard scale $M_Z$. The second line involves standard transverse-momentum-dependent (TMD) fragmentation functions, whose operator definitions are given by
\begin{align}\label{eq:frag_func}
    \cD_{ h_1/f} (\zeta ,b_\perp) 
    &= \frac{\Theta(\omega_a)}{4\zeta_1 N_c} \sumint_{X_a} \int \dfbar x^+  e^{+\im \frac{1}{2} x^+  \omega_a}\Tr  \Big\langle   0\Big | \frac{\tnslash_2}{2} \chi_{\tn_1}  (x^+, 0, b_\perp )\Big  | X_a , h_1\Big \rangle  \Big \langle X_a, h_1 \Big| \overline \chi_{\tn_1} (0) \Big |0\Big \rangle  \nn \,, \\	
    \cD_{ h_2/\bar f} (\zeta ,b_\perp) 
    &= \frac{\Theta(\omega_b)}{4\zeta_2 N_c}\sumint_{X_b} \int \dfbar x^- e^{+\im \frac{1}{2} x^- \omega_b } \Tr  \Big\langle   0\Big | \overline  \chi_{\tn_2}  (0, x^-, b_\perp )\Big  | X_b , h_2\Big \rangle  \Big \langle X_b, h_2 \Big| \frac{\tnslash_1}{2} \chi_{\tn_2} (0) \Big |0\Big \rangle \, ,
\end{align}
where
\begin{align}\label{eq:zeta12_def}
    &\zeta_1 \equiv \frac{2\sqrt{z_{12}} E_{h_1}}{\omega_a}\,,&
    &\zeta_2 \equiv \frac{2\sqrt{z_{12}} E_{h_2}}{\omega_b}\,,
\end{align}
are the energy fractions of the hadrons, defined using their momenta in \eq{p_h12}. The SCET collinear quark gauge-invariant building block appearing in \eq{frag_func} is defined as
\begin{align}
    &\chi_n(x) \equiv W_n^\dagger(x) \xi_n(x) \, ,&
    &W_n^\dagger(x)   = \mb P \exp \bigg(+\im g \int_0^\infty \df s \: \bn\cdot A_n \big(x^\mu + s \bn^\mu\big)\bigg) \, .
\end{align}
Here Wilson line $W_n$ enforces gauge invariance of the collinear field and is defined via an auxiliary light-like vector $\bn$, satisfying $\bn \cdot n = 2$ . The step functions $\Theta(\omega_a)$ and $\Theta(\omega_b)$ in \eq{frag_func} therefore ensure that the hadrons are produced from quark and anti-quark fields, respectively.

In the factorisation theorem above, the leading power effects of soft partons are captured entirely by the $q_T$-soft function in impact parameter space, defined as
\begin{align}\label{eq:qT_soft}
    S_{n_1n_2}(b_\perp ) &= \frac{1}{N_c} \langle 0 | \big(\bar S_{n_2}^\dagger S_{n_1}\big)^{b\bar a} (b_\perp) \big(S_{n_1}^\dagger  \bar S_{n_2}\big)^{a\bar b}(0) |0\rangle \, , 	
\end{align}
where the soft Wilson lines are defined by
\begin{align}
    &S_{n}^\dagger (x)= \mb P \exp \bigg(+\im g \int_0^\infty \df s \: n\cdot A_s \big(x^\mu + s n^\mu\big)\bigg)\,,\\
    \nn
    &\bar S_n (x)= \mb{ \overline{P} }\exp \bigg(-\im g \int_0^\infty \df s \: n\cdot A_s \big(x^\mu + s n^\mu\big)\bigg)\, .
\end{align}
Here $\mb P$ and $\mb{\overline{P}}$ denote path- and anti-path-ordering of the fields and $x^\mu$ is the position of the current. The differences in the two reflect coupling of soft gluons to quarks  and anti-quarks.

Let us now insert this result into \eq{Hee_h12_0} to obtain the corresponding expression for the EEC. We begin by simplifying the phase space appearing in \eq{Hee_h12_0}:
\begin{align}
    \int \df \Phi_{h_1,h_2} \Bigg(\prod_{i=1}^2 E_{h_i} \delta^{(d-2)} (\mb n_i - \mb n_{h_i})\Bigg)  &=
    \prod_{i=1}^2 	\int \dfbar^d p_{h_i} \delta_+(p_{h_i}^2) \: E_{h_i}\delta^{(d-2)} (\mb n_i - \mb n_{h_i})  \\
    &\nn = \prod_{i = 1}^2   \frac{1}{2(2\pi)^{d-1}}\int E_{h_i}^{\,d-2} \df E_{h_i} \, , 
    \\
    &= \nn 
    \frac{1}{4} \Bigg(\frac{\omega_a \omega_b}{4 (2\pi)^2 z_{12}}\Bigg)^{d-1} \int \df \zeta_1 \df \zeta_2 \: \zeta_1^{d-2} \zeta_2^{d-2}  \, ,
\end{align}
where, in the final line, we used the definitions in \eq{zeta12_def}. It then follows that
\begin{align}\label{eq:HEEC_dihad}
    \langle \cE(n_1) \cE(n_2)\rangle_q =\frac{M_Z^{2(d-1)}}{4^d (2\pi)^{2(d-1)} z_{12}^{d-1}} \sum_{h_1,h_2 \in \text{jet}} 
    \int \df \zeta_1 \df \zeta_2 \: \zeta_1^{d-2} \zeta_2^{d-2}  \, \frac{1}{\sigma_0}\frac{\df \sigma(q)}{\df \Phi_{h_1,h_2}} \, .
\end{align}

In the next step, we perform a small-$b_\perp$ expansion. Since the observable is infrared safe, the transverse-momentum-dependent fragmentation functions (TMDFFs) can be matched onto collinear fragmentation functions and perturbative matching coefficients. From \eqs{dihad_fact}{HEEC_dihad}, we encounter integrals of the form
\begin{align}
    \int_0^1 \df \zeta \: \zeta \times \zeta^{d-2} \cD_{h/f} \big(\zeta, b_\perp \big) \, .
\end{align}	
Using the standard definitions and matching relations between TMD and collinear fragmentation functions, discussed in detail in \app{tmd}, this expression can be written as
\begin{align}
    \int_0^1 \df \zeta \: \zeta^{d-1} \int_\zeta^1 \frac{\df \xi}{\xi^{d-1}} \sum_j \cC_{j/f} \bigg(\frac{\zeta}{\xi}, b_\perp\bigg) d_{j/h} (\xi) \, ,
\end{align}
where $j$ denotes an intermediate parton and $\cC_{j/f}$ are perturbative matching coefficients. Changing variables according to $\zeta \ra \zeta \xi$, and recalling that the coefficients $\cC_{j/f}$ have support only for $\zeta \leq 1$, we find that, after summing over hadrons,
\begin{align}\label{eq:JEECdef}
    \sum_h \int_0^1 \df \zeta \:  \zeta^{d-1} \cD_{h/f} \big(\zeta, b_\perp \big) &=
    \bigg( \sum_j \int_0^1 \df \zeta \: \zeta^{d-1}	\cC_{j/f}(\zeta, b_\perp) \bigg)  
    \int_0^1 \sum_h  \df \xi\: \xi d_{h/j} (\xi)\\
    &=\sum_j \int_0^1 \df \zeta \: \zeta^{d-1}	\cC_{j/f}(\zeta, b_\perp)
    \equiv J_{\rm EEC}^f\big(b_\perp \big)\nn \, ,
\end{align}
where, in the final step, we have used the momentum sum rule for collinear fragmentation functions to define the EEC jet function.

Putting these ingredients together, we obtain the factorisation formula for the EEC,

\begin{align} \label{eq:n1n2_corr_fact}
        \langle \cE(n_1) \cE(n_2)\rangle_q &=\frac{2 M_Z^{2}}{ (8\pi)^{(d-2)} z_{12}^{d-1}\Omega_{d-2} r_\eps}
    \sum_f  \frac{|a_f|^2 + |v_f|^2}{\sum_{f'} (|a_{f'}|^2 + |v_{f'}|^2)} \, \cH_f^{(0)}(M_Z^2) \\
    \nn 
    &\quad \times 
    M_Z^{d-2 }
    \int \df^{d-2} b_\perp \: e^{+\im b_\perp \cdot q_{\perp(12)}} 
    \Big[ J_{\rm EEC}^f\big( b_\perp\big)  J_{\rm EEC}^{\bar f}\big( b_\perp\big)  S_{n_1n_2} (b_\perp) \Big] \, .
\end{align}
Here we have rendered the integrand dimensionless by extracting appropriate powers of $M_Z$, while the overall mass dimension is carried by the explicit factor of $M_Z^2$ in the first line. Furthermore, in $d$ dimensions the celestial dimension of $\cE(n)$ is $-(d-1)$, such that \eq{shape} can be written as
\begin{align}
    \langle \cE(n_1) \cE(n_2)\rangle_q  = \frac{M_Z^2 \cF_{\cE\cE}(\bar z_{12}, \lqcd^2/M_Z^2)}{(2\pi)^{d-2} (n_1\cdot n_2)^{d-1}} \, .
\end{align}
Using \eqs{n12_def}{n1n2_corr_fact}, this leads to the identification
\begin{align}
   \cF_{\cE\cE}\bigg(\bar z_{12} , \frac{\lqcd^2}{M_Z^2}\bigg) &= 	 \frac{2^{4-d}}{\Omega_{d-2}r_\eps}
   \sum_f \frac{|a_f|^2 + |v_f|^2}{\sum_{f'} (|a_{f'}|^2 + |v_{f'}|^2)} \, \cH_f^{(0)}(M_Z^2)\\
   &\quad \times\nn 
   M_Z^{d-2}
   \int \df^{d-2} b_\perp \: e^{+\im b_\perp \cdot q_{\perp(12)}} 
   \Big[ J_{\rm EEC}^f\big( b_\perp\big)  J_{\rm EEC}^{\bar f}\big( b_\perp\big)  S_{n_1n_2} (b_\perp) \Big] \, .
\end{align}
We can make the dependence on $\bar z_{12}$ more explicit by noting that the second line is independent of the direction of $q_{\perp(12)}^{\mu}$. Performing the azimuthal integration therefore yields
\begin{align}
    \cF_{\cE\cE}\bigg(\bar z_{12} , \frac{\lqcd^2}{M_Z^2}\bigg) &= 
    \frac{4^{\eps}}{r_\eps} M_Z^{2 - 2\eps} 
    \sum_f \frac{|a_f|^2 + |v_f|^2}{\sum_{f'} (|a_{f'}|^2 + |v_{f'}|^2)} \, \cH_f^{(0)}(M_Z^2) \label{eq:4p3bottomline}
    \\
    &\quad \times\nn 
    \int_0^\infty \df b_T  \: b_T^{1-2\eps} \,
    J_0\Big(M_Z b_T \sqrt{1- \bar z_{12}}\Big)
    \Big[ J_{\rm EEC}^f\big( b_T\big)  J_{\rm EEC}^{\bar f}\big( b_T\big)  S_{n_1n_2} (b_T) \Big] \, . 
\end{align}
Thus, we have explicitly verified that the Sudakov factorisation of the EEC in the laboratory frame is consistent with defining the shape function through reparametrisation and Lorentz invariance.

\subsection{Three-body decays}
\label{sec:three_body}

We have so far studied the two-point EEC in detail, both from the perspective of the symmetries of the EEC shape function and through a refactorisation of the EEC measurement for a decaying, boosted $Z$ boson. We now turn to how this discussion extends to higher-point correlations, such as the three-point EEC, which is of particular relevance for top-quark decays~\cite{Holguin:2022epo,Holguin:2023bjf}.

As in \secn{HEEC}, $H_{\rm ENC}$ is determined at leading power by the Wightman function of $N$ energy-flow operators,
\begin{align}	
    \langle \cE(n_1) \cdots \cE(n_N) \rangle_q
    \equiv \frac{1}{\sigma_0}\bigg(\frac{q_\mu q_\nu}{M_Z^2} - g_{\mu\nu}\bigg)
    \int \td^4 x \: e^{\im q\cdot x}
    \, \langle 0\big | J^{\dagger\mu}_{Z}(x)
    \cE(n_1) \cdots \cE(n_N)
    J^\nu_{Z} (0) \big | 0 \big \rangle ,
\end{align}
with $n_i = (1, \mb n_i)$. We now generalise the EEC shape function introduced in \secn{HEEC} to an $N$-point correlator.

The set of dynamical invariants upon which $\langle \cE(n_1) \cdots \cE(n_N) \rangle_q$ depends consists of $q^{2}$, $q \cdot n_{j}$, and $n_{j} \cdot n_{k}$, giving a total of $(N^2+N+2)/2$ scalars (although dot products of the $n_i$ are not themselves Lorentz scalars). Among these, there are $N(N-1)/2$ reparameterisation-invariant cross-ratios,
\begin{align}
    \bar{z}_{jk}
    =\frac{q^{2} \, n_{j}\cdot n_{k}}{2 \, q \cdot n_{j} ~ q \cdot n_{k} }.
\end{align}
These cross-ratios are Lorentz scalars. Following the arguments of \secn{HEEC}, reparameterisation invariance of the energy-flow operators then requires
\begin{align}
    \langle \cE(n_{1}) \cdots \cE(n_{N}) \rangle_{q}
    = \frac{(q^{2})^{N/2} ~
        \mathcal{F}_{\cE^{N}}\!\left( \{\bar{z}_{jk}\}, \; M_Z^2/q^2, \; \Lambda_{\rm QCD}^2/q^2 \right)}
    {4 \pi^{2} \:
        \mathcal{D}(\{n_{j} \cdot n_{k}\})},
\end{align}
where
\begin{align}
    n_{j} \rightarrow \rho \, n_{j}
    \quad : \quad
    \mathcal{D}(\{n_{j} \cdot n_{k}\} ) \mapsto \rho^{3} \, \mathcal{D}(\{n_{j} \cdot n_{k}\} ).
\end{align}
Symmetry under permutations of the $n_{j}$ further implies
\begin{align}
    \mathcal{D}(\{n_{j} \cdot n_{k}\} )
    = \prod_{j<k} (n_{j} \cdot n_{k})^{\frac{3}{N-1}}.
\end{align}
Thus, the \textit{ENC shape function} is manifestly Lorentz invariant and can be used to determine the $n$-point correlator in any frame, in direct analogy with the EEC shape function discussed in \secn{HEEC}.

We now focus on the structure of the three-point energy correlator (E3C). As for the EEC, soft gluon radiation perturbs the Born configuration and necessitates Sudakov resummation. In the case of the EEC, the Born configuration consists of two back-to-back partons in the centre-of-mass frame, leading to a Sudakov peak as $z \rightarrow 1$. For the E3C, the Born configuration instead comprises three partons lying in a plane, and resummation of soft radiation emitted out of this plane generates an analogous Sudakov feature. This feature has been proposed as a means of measuring the top-quark mass using E3C measurements on the constituents of top-tagged jets. However, the increased complexity of a three-body decay complicates the identification of the soft configurations that must be resummed in the boosted regime.

In \Refcite{Gao:2024wcg}, out-of-plane configurations of the E3C were studied in $e^{+}e^{-}$ collisions at the $Z$ pole. A single variable $\tau$ was introduced to quantify the degree of aplanarity, with $\tau = 0$ corresponding to purely planar configurations. Sudakov resummation in the $\tau \rightarrow 0$ limit was achieved through the factorisation
\begin{align}
    &\int \td^4 x ~ e^{i \, x \cdot q_{\rm rest}}
    \int \td^2 \v n_{1,2,3} ~\frac{
        \langle\cE(n_1)\cE(n_2)\cE(n_3)\rangle_{q_{\rm rest}}}{M_{Z}^3}
    \delta \!\left(\tau - |(\v n_1 \times \v n_2 )\cdot \v n_3| \right) \nonumber \\
    &= \int_0^1 \td u \, \td v \: H(u,v,M_Z,\mu)
    \int \frac{\td b}{2\pi \xi}
    \, 2 \cos\!\left(\frac{b \tau}{\xi}\right)
    J^{q}(b,\mu,\nu)\,
    J^{\bar{q}}(b,\mu,\nu)\,
    J^{g}(b,\mu,\nu)\,
    S_{3}(b,\mu,\nu), \label{eq:3pointrefactorised}
\end{align}
where $q_{\rm rest} = (M_Z,0,0,0)$ and
$\xi = 4\sqrt{uvw}/[(1-u)(1-v)(1-w)]$ with $u+v+w=1$. The specific details of this factorisation are not relevant for our purposes. Rather, our goal is to construct a generalisation of $\tau$ that depends only on the conformal cross-ratios $\bar{z}_{12}$, $\bar{z}_{23}$, and $\bar{z}_{13}$. Such a definition will allow the Sudakov region of a boosted three-body decay to be immediately identified in any frame, and consequently, the generalisation of \eq{3pointrefactorised} to an arbitrary frame will be transparent. Specifically, by analogy to \eq{4p3bottomline} and the EEC shape function, the factorisation in a general frame is achieved from the rest frame result by replacing $z \rightarrow \bar{z}$.

To this end, consider the delta function
\begin{align}
    \delta\!\left(\tau - |(\v n_1 \times \v n_2 )\cdot \v n_3| \right)
    = 2\tau ~\delta\!\left(\tau^2 - \det(N N^{\rm T}) \right),
\end{align}
where $N$ is the matrix of vectors $N = (\vec{n}_1,\vec{n}_2,\vec{n}_3)$. Since $\det(N N^{\rm T})$ depends only on dot products, it can be expressed entirely in terms of the cross-ratios,
\begin{align}
    \det(N N^{\rm T})
    = \frac{1}{2}
    \left|
    \begin{matrix}
        2 & (1-\bar z_{12}) & (1-\bar z_{31}) \\
        (1-\bar z_{12}) & 2 & (1-\bar z_{23}) \\
        (1-\bar z_{31}) & (1-\bar z_{23}) & 2 \\
    \end{matrix}
    \right|.
\end{align}
The soft region relevant for Sudakov resummation in a boosted top decay is then isolated by the measure
\begin{align}
    \td \bar{z}_{12} \, \td \bar{z}_{23} \, \td \bar{z}_{31}
    ~ \delta\!\left(\tau^2 - \det(N N^{\rm T}) \right)
    \prod_{1 \leq i < j \leq 3}
    \delta\!\left(\bar z_{ij}
    - \frac{q_t^2 \, n_i \cdot n_j}{(n_i \cdot q_t)(n_j \cdot q_t)}\right),
\end{align}
where $q_t$ denotes the off-shell top-quark momentum. Geometrically, this measure corresponds to isolating the squared volume of a warped parallelepiped whose three edges are aligned with the vectors connecting the three correlator axes to the axis defined by $q_t$. Sudakov resummation becomes relevant when this volume approaches zero.

\section{Conclusions and Outlook}
\label{sec:conclusions}

This article presents the first comprehensive theoretical study of energy-correlator measurements on the hadronic decays of boosted heavy neutral particles. In particular, we introduce and study two measurements of two-point energy correlators (EECs) on jets formed from $Z$ decays: one applicable to $pp$ collisions, such as those at the LHC, and another applicable to high-energy $e^+ e^-$ collisions, as will be realised at the FCC-ee.

These measurements both lead to characteristic peaks in the EEC spectrum, whose position is determined by the boost of the $Z$ boson. We demonstrate that both observables are uniquely determined (at leading order in $\alpha_{\rm EW}$) by the \emph{EEC shape function} of an off-shell $Z$ decay. This shape function is explicitly Lorentz invariant. Consequently, a measurement of one of these observables in a given collider system can be directly related to the corresponding measurement in another. Of particular significance, this allows the EEC on boosted $Z$ decays to be computed directly from measurements of the EEC performed in $e^+ e^-$ collisions at the $Z$ pole. Furthermore, this approach is also readily generalisable to EEC measurements on other boosted neutral boson decays.

The EEC measured in $e^+ e^-$ collisions at the $Z$ pole has been extensively studied. By drawing on existing results in the literature, we are therefore able to compute both observables introduced here at N$^3$LL$'$ accuracy, with a straightforward extension to N$^4$LL accuracy~\cite{Electron-PositronAlliance:2025fhk}. We compare these predictions with simulations from \Herwig and \Pythia, finding close agreement. More strikingly, the EEC on boosted $Z$ boson decays can be predicted directly by boosting existing measurements of the EEC in $e^+ e^-$ collisions at the $Z$ pole. We demonstrate this approach by predicting the EEC spectrum for boosted $Z$ jets in both $pp$ and $e^+ e^-$ collisions through a direct boost of the OPAL measurement~\cite{OPAL:1990reb}. We find exceptional agreement between these OPAL-based predictions and the corresponding event-generator simulations.

Furthermore, we have complemented this symmetry-based picture with an explicit SCET factorisation of the EEC in the laboratory frame, starting directly from the di-hadron distribution in boosted $Z$ jets. This analysis demonstrates how the boosted Sudakov structure arises from a standard factorisation involving TMD fragmentation functions and a universal soft function. It thereby clarifies the origin of the peak in the EEC spectrum as resulting from Sudakov resummation about the boosted Born configuration.

Finally, we have extended the discussion beyond two-point correlators by formulating a Lorentz-invariant shape-function description for general $N$-point energy correlators. We showed how the Sudakov-sensitive region of the boosted three-point correlator can be characterised entirely in terms of conformal cross-ratios, enabling a frame-independent identification of the soft region relevant for resummation in boosted three-body decays, such as those of the top quark.

Looking ahead, the framework developed in this work opens several promising directions for future study. The simplicity and predictive power of energy correlators on colour-singlet decays make boosted $Z$ bosons an ideal laboratory for precision jet physics across collider environments, from the LHC and HL-LHC to future lepton colliders such as FCC-ee. Moreover, this work has explicitly identified the peaked feature that emerges in the spectrum of energy-correlator measurements on heavy jets as arising from a specific soft function, ensuring that it is amenable to high-precision calculation. Such peak features have also been proposed as a means to extract a precise measurement of the top-quark mass from boosted top-jet substructure. Extending the present analysis to these systems, as well as to higher-point energy correlators, therefore offers a clear path towards systematically improvable and experimentally robust probes of QCD dynamics in complex jet environments.

\acknowledgments

We thank authors of~\cite{Gao:2026xuq} who made us aware of their work as we completed this manuscript and for arranging for coordinated submissions. We thank Gregory Korchemsky and Kyle Lee for useful discussions. 
We thank Damiano Barcaro for spotting various typos and suggesting improvements.
I.M. is supported by the DOE Early Career Award DE-SC0025581, and the Sloan Foundation. J.H. is supported by the Leverhulme Trust as an Early Career Fellow. S.S. would like to thank the UK Science and Technology Facilities Council (STFC) for the award of a studentship. 
A.P. is funded by the European Union (ERC, TOPMASS, 101165601). Views and opinions expressed are however those of the author(s) only and do not necessarily reflect those of the European Union or the European Research Council. Neither the European Union nor the granting authority can be held responsible for them.

\appendix

\section{Derivation of the dihadron factorization}
\label{app:fact}

In this appendix we describe in detail the derivation of the dihadron factorisation theorem using soft-collinear effective theory (SCET) valid in the limit $1-\bar{z}_{12} \ll 1$. While the back-to-back EEC factorisation theorem in the rest frame of the $Z$ boson is well-known~\cite{Moult:2018jzp}, the derivation we present here is new, and makes the connection between the lab and rest frame of the $Z$ boson immediate. 

\subsection{Leading power operators}
In the effective theory, we only allow for approximately on-shell states satisfying
$p_i^2 \sim \lambda^2 M_Z^2$ where $\lambda^2 \sim 1-\bar{z}_{12} \ll 1$.
The space of such states is described in terms of a \textit{label momentum}
$\tilde p^\mu$, which is exactly on shell, together with small fluctuations
parameterised by residual momenta $k^\mu$, such that
\begin{align}
    &\text{Momentum states in \SCETb:}& 
    &| \tilde p_i , k_i \rangle\,,& 
    &\tilde p_i^2 = 0 \,,&
    & k_i^\mu \lesssim \lambda p_i^\mu \, ,&
    &(\tilde p_i + k_i)^2 \sim M_Z^2 \lambda^2\, .&
\end{align}
Here we are considering \SCETb, in which fluctuations at scales below
$M_Z^2 \lambda^2$ are not retained.
The condition $\tilde p_i^2 = 0$ implies that the label momentum is either light-like or zero.
The label and residual momenta corresponding to the collinear scalings in
\eq{power_counting} are fixed in the hadron frame decomposition to be
\begin{align}\label{eq:p_ab}
    &\tilde p_a^\mu = \omega_a \big(0,1,0\big)_{(12)} \,,& 
    &k_a^\mu \lesssim \omega_a \big(\lambda^2, \lambda, \lambda \big)_{(12)} \,,&
    &\omega_a \sim M_Z \,, &\\
    \nn 
    &\tilde p_b^\mu = \omega_b\big(1,0,0\big)_{(12)} \,,&
    &k_b^\mu \lesssim \omega_b \big(\lambda , \lambda^2, \lambda \big)_{(12)} \, , &
    &\omega_{b}\sim M_Z\,  ,&
\end{align}
where $\tilde p_{a,b}$ are the label components of the partonic momenta $p_{a,b}$ defined in \eq{kabXh}.
The label momenta $\tilde p_i$ for collinear states (with $\tilde p_i \neq 0$)
parameterise equivalence classes $p_i \sim p_i'$, defined by the condition
$p_i \cdot p_i' \sim M_Z^2 \lambda^2$.
At leading power, the large components are equivalent, $\omega_1 \sim \omega_a$,
with differences $\omega_1 - \omega_a \lesssim M_Z \lambda$ absorbed into the residual momenta.
Note that, traditionally, one also includes $\cO(\lambda)$ transverse components in the label
momentum and defines the residual momentum to be strictly $\cO(\lambda^2)$.
Here we avoid this convention, since the $\cO(\lambda^2)$ residual components play no special
role in the derivation. See \Refcite{Stewart:2013SCETnotes} for further details.
We split the $Z$-boson momentum $q$ as
\begin{align}\label{eq:q_split}
    &q^\mu = \tilde q^\mu + q_r^\mu \, ,&
    &\tilde q^\mu = \big(\omega_b, \omega_a , \mb 0_\perp \big)_{(12)} \,,&
    &q_r^\mu = \big(\omega_2 -\omega_b, \omega_1 -\omega_a, \mb q_{\perp(12)} \big)_{(12)} \, .&
\end{align}
For soft momenta, only $\cO(\lambda)$ residual components are present,
\begin{align}
    &\tilde p_s^\mu = 0\,,&
    & k_s^\mu  = \big(k_s^+,k_s^-,\mb k_{s\perp}\big)_{(12)}\,,&
    &k_s^\pm \sim |\mb k_{s\perp(12)}| \sim \lambda M_Z \, .&
\end{align}

As a first step, the electroweak current can be matched onto the following leading-power
hard operators built from SCET fields with definite label momenta:
\begin{align}\label{eq:hard_scatter_ops}
    J_Z^\mu \ra J_Z^{(0)\mu}(x)	=e^{-\im x \cdot \cP } \sum_{\tilde p_1'  ,\tilde p_2'} \sum_f (\Gamma_\perp^\mu)^{\alpha \beta}		 
    C^{(0)}	(2 \tilde p_1' \cdot \tilde p_2') ~ \overline \chi_{\tilde p_1'}^{(f)\alpha \bar a}(x)	\big(S_{n_1'}^\dagger \bar S_{n_2'}\big)^{a\bar b}	\chi_{\tilde p_2'}^{(f)\beta b}	(x)\,	,
\end{align}
where $\Gamma_\perp^\mu = -|e|\gamma_\perp^\mu \big(v_f - a_f \gamma_5\big)$.
Here $a,b$ are colour indices and $\alpha,\beta$ are Dirac indices.
The directions $n_i'$ and energies $\omega_i'$ parameterise the label momenta
$\tilde p_i^{\prime\mu} = \frac{1}{2}\omega_i' n_i^{\prime\mu}$.
Here the operator $\cP^\mu$ in the exponential extracts the label momentum.
The $\perp$ component is defined relative to the directions $n_1^{\prime\mu}$ and
$n_2^{\prime\mu}$ in a light-cone decomposition analogous to \eq{n12_def}, satisfying
$n_1'\cdot n_2' = 2$.
The decomposition of the sum over label momenta in terms of the directions $n_i$ and large components $\omega_i$ and the label-momentum conserving $\delta$-functions are defined as
\begin{align}\label{eq:sum_tpi}
  &  \sum_{\tilde p_i} \equiv \sum_{\mb n_i} \sum_{\omega_i} \,, &
  &  \delta_{\tilde p_i , \tilde p_i'} \equiv \delta_{\mb n_i, \mb n_i'}\delta_{\omega_i,\omega_i'}\,,&
\end{align}
and the field carrying label $\tilde p_i$ is defined as
\begin{align}
    \chi_{\tilde p_i}(x) \equiv \delta_{\tilde p_i^\mu, \cP^\mu }
    \sum_{\mb n_i'} \sum_{\omega_i'} \chi_{n_i',\omega_i'} (x)
    = 
    \chi_{n_i,\omega_i} (x)\, .
\end{align}
The subscript $\omega$ in $\chi_{n,\omega}$ indicates that the field carries a definite
label momentum $\omega$ in the direction $n$, corresponding to the momentum of the
primary quark annihilated by the field $\xi_n$ together with an arbitrary number of
$\bn$-polarised gluons from the Wilson line $W_n$:
\begin{align}\label{eq:chi_n_omega}
    \chi_{n,\omega} (x) = \delta \big(\omega - \bn \cdot \cP\big) W_n^\dagger (x) \xi_n(x) \, .
\end{align}
Our convention is such that $\omega > 0$ corresponds to $\chi_{n,\omega}$ annihilating an
incoming quark, while $\omega < 0$ corresponds to the creation of an outgoing anti-quark.
We note that the $x$ dependence in \eq{hard_scatter_ops} tracks only the residual components.
Label-momentum conservation is always implemented first as a ``superselection rule'' via the $e^{-\im x \cdot \cP}$ prefactor in \eq{hard_scatter_ops},
before considering residual momenta.
Nevertheless, we wish to keep track of label-momentum conservation explicitly, which
motivates the appearance of the momentum-conserving $\delta$-function in \eq{Hee_h12}.

\subsection{Simplifying the collinear matrix elements}

We begin by inserting the leading-power current in \eq{hard_scatter_ops} into the matrix
element in \eq{Hee_h12}.
Working in the hadron frame, this yields
\begin{align}\label{eq:JZ_me}
    &\langle 0 | J_Z^{\dagger\mu} (0) |X_n, h_1, h_2\rangle
    \langle X_n, h_1, h_2 | J_Z^\nu (0) | 0\rangle  \\
    &\nn \quad =
    \big(\Gamma_{\perp}^\nu\big)^{\alpha \beta}
    \big(\Gamma_\perp^{\mu}\big)^{\beta' \alpha'}
    \sum_f \sum_{\tilde p_1' , \tilde p_1''}
    \sum_{\tilde p_2' , \tilde p_2''}
    C^{(0)}\big(\omega_1 ' \omega_2' \big)
    C^{(0)\dagger} \big(\omega_1'' \omega_2'' \big)	\\
    &\quad\quad \times \nn
    \Big \{ 	
    \langle  0 | \chi_{\tilde p_1'' }^{\alpha' a'} (0)| X_a, h_1 \rangle
    \langle X_a, h_1 | \overline \chi^{\alpha \bar a }_{\tilde p_1'} (0) | 0 \rangle \\
    &\quad\quad \quad \times \nn    
    \langle 0 | \overline \chi_{\tilde p_2'' }^{\beta ' \bar b'}(0) |X_b, h_2\rangle
    \langle X_b, h_2 | \chi^{\beta b}_{\tilde p_2'} (0)| 0 \rangle \\
    &\quad\quad \quad \times
    \langle 0 | \big(\bar S_{n_2''}^\dagger S_{n_1''}\big)^{b'\bar a'}
    | X_s \rangle
    \langle X_s | \big(S_{n_1'}^\dagger \bar S_{n_2'}\big)^{a\bar b} | 0 \rangle
    + \big(h_1, X_a \lra h_2, X_b\big)
    \Big\}\, ,\nn
\end{align}
where all quark fields carry the same flavour index, which we have suppressed for notational
simplicity.

In the first step, we simplify the structure of the collinear matrix elements. 
Following \eq{kabXh}, we   introduce the identities 
\begin{align}\label{eq:p_a_Xa_h1}
    1 &= \int \dfbar^d p_a \: (2\pi)^d \delta^{(d)} \big(p_a - P_{X_a} - p_{h_1}\big) 
    =\sum_{\tilde p_a}  \delta_{\tilde p_a , \tilde P_{X_a} + \tilde p_{h_1}} \int \dfbar^{d} k_a \: (2\pi)^d \delta^{(d)} \big(k_a - k_{X_a} \big) 
    \,,\nn 
    \\ 
    1 &= \int \dfbar^d p_b \: (2\pi)^d \delta^{(d)} \big(p_b - P_{X_b} - p_{h_2}\big)  = \sum_{\tilde p_b} 
    \delta_{\tilde p_b , \tilde P_{X_b} + \tilde p_{h_2}} \int \dfbar^{d} k_b \: (2\pi)^d \delta^{(d)} \big(k_b - k_{X_b} \big) 
    \, .
\end{align}
We will first deal with the label components and then the residual momenta. 

Let us  begin by considering the collinear matrix element in \eq{JZ_me} including the momentum conservation in \eq{p_a_Xa_h1}:
\begin{align}\label{eq:coll_me_0}
    & \sum_{   \tilde p_1', \tilde p_1''}  
    \langle  0 | \chi_{\tilde p_1'' }^{\alpha' a'} (0)| X_a, h_1 \rangle \langle X_a, h_1 | \overline \chi^{\alpha \bar a }_{\tilde p_1'} (0)	| 0 \rangle \\
    \nn 
    &\quad =  \sum_{\tilde p_a,  \tilde p_1', \tilde p_1''} 	\delta_{\tilde p_a,\tilde P_{X_a} +\tilde p_{h_1} }  \int \dfbar^d k_a \: (2\pi)^d \delta^{(d)}\big(k_a - k_{X_a}\big)  
    \langle  0 | \chi_{\tilde p_1'' }^{\alpha' a'} (0)| X_a, h_1 \rangle \langle X_a, h_1 | \overline \chi^{\alpha \bar a }_{\tilde p_1'} (0)	| 0 \rangle \, .
\end{align}
Using the fact that the matrix element $ \langle X_a, h_1 | \overline \chi^{\alpha \bar a }_{\tilde p_1'} (0)	| 0 \rangle$ is itself proportional to $\delta_{\tilde p_1' , \tilde P_{X_a} +\tilde p_{h_1}}$, we can set $\tilde p_1' \ra \tilde p_a$ and remove the corresponding sum, leading to the following expression:
$$
\sum_{  \tilde p_a , \tilde p_1''} 	  \int \dfbar^d k_a \: (2\pi)^d \delta^{(d)}\big(k_a - k_{X_a}\big)  
\langle  0 | \chi_{\tilde p_1'' }^{\alpha' a'} (0)| X_a, h_1 \rangle \langle X_a, h_1 | \overline \chi^{\alpha \bar a }_{\tilde p_a} (0)	| 0 \rangle \,  .
$$
Using \eq{sum_tpi}, the conjugate matrix element therefore becomes
\begin{align}
    \sum_{\tilde p_1''} \langle  0 | \chi_{\tilde p_1''}^{\alpha' a'} (0)| X_a, h_1 \rangle
    &= \sum_{\mb n_1''}
    \sum_{\omega_1''}  
    \langle 0|  \chi_{\tilde p_1'' }^{\alpha' a'} (0)| X_a, h_1 \rangle \\
    &\nn =  \langle 0|  \chi^{\alpha' a'}_{\tn_1}(0) | X_a, h_1 \rangle 
    \, ,
\end{align}
where firstly, we have used the superselection rule $\mb n_1'' \ra \mb n_1$ to remove the sum, and secondly we used the fact that the overall label momentum is set to be $\omega_a$, while writing
\begin{align}
    \sum_{\omega}    \chi_{n, \omega} = \chi_n \, .
\end{align}

Finally, we combine everything and bring in the phase space integral associated with $X_a$:
\begin{align}\label{eq:coll_a_prelim}
    &
    \sumint_{X_a}(2\pi)^d \delta^{(d)}\big(k_a - k_{X_a}\big)  
    \langle 0|  \chi^{\alpha' a'}_{\tn_1}(0) | X_a, h_1 \rangle 
    \langle X_a, h_1 | \overline \chi^{\alpha \bar a }_{\tilde p_a} (0)	| 0 \rangle 
    \nn \\
    &\quad  = 
     \delta_{\mb n_a,\mb n_1}  		\hat \cG_{h_1/f}^{\alpha' \alpha a' \bar a }(\omega_a, k_a , p_{h_1}) 
    \, .
\end{align}
Here the matrix element is non-zero only unless $\mb n_1$ and $\mb n_a$ are in the same equivalence class and we have defined the correlator
\begin{align}\label{eq:frag_func_res}
    \hat \cG_{h_1/f}^{\alpha' \alpha  a' \bar a}(\omega_a, k_a, p_{h_1}) &\equiv \Theta(\omega_a)\int \df^d x \: e^{ \im k_a \cdot x} \sumint_{X_a} 
    \langle 0|  \chi^{\alpha' a'}_{\tn_1}(x) | X_a, h_1 \rangle 
    \langle X_a, h_1 | \overline \chi^{\alpha \bar a }_{\tn_1 , -  \omega_a} (0)	| 0 \rangle \,  ,
\end{align}
where we employed the sign convention discussed below \eq{chi_n_omega} such that the $\overline \chi_{n_1}$ field with negative label  leads to production of an hadron with positive  outgoing momentum. 
The $\Theta(\omega_a)$ arises from the momentum conservation in \eq{mom_cons_label}. 
We have indicated  this here explicitly to ensure that we are considering quark fragmentation function. 
Here, as usual, the position $x$ tracks the flow of residual components, on top of the label momentum injected by the field $\overline \chi_{\tn_1,-\omega_a}$.    

Next, by noting that the collinear matrix elements are diagonal in the colour space, we can simplify the correlator by factorising the colour structure:
\begin{align}
    \hat \cG_{h_1/f}^{\alpha'\alpha a' \bar a}\big(\omega_a ,k_a ,p_{h_1}\big) = \frac{\delta^{a'\bar a}}{N_c} 	\hat \cG_{h_1/f}^{\alpha'\alpha  }\big(\omega_a ,k_a, p_{h_1}\big) 	\,.
\end{align}
We therefore arrive at the following expression for the collinear matrix element in \eq{coll_me_0}, including the phase space integral over the unobserved state $X_a$,
\begin{align}\label{eq:coll_me_2}
    &\sumint_{X_a} \sum_{   \tilde p_1', \tilde p_1''}  
    \langle  0 | \chi_{\tilde p_1'' }^{\alpha' a'} (0)| X_a, h_1 \rangle \langle X_a, h_1 | \overline \chi^{\alpha \bar a }_{\tilde p_1'} (0)	| 0 \rangle \\
    &\quad = \nn 
    \frac{\delta^{a' \bar a}}{N_c} \sum_{\omega_a} \sum_{\mb n_a}\: \delta_{\mb n_a, \mb n_1}  \int \dfbar^d k_a \:
    \hat \cG_{h_1/f}^{\alpha'\alpha  }\big(\omega_a ,k_a, p_{h_1}\big)  \, .
\end{align}

\subsection{Multipole expansion}
We now turn to the momentum conserving $\delta$-function in \eq{Hee_h12}. Since the labels $\mb n_{a,b}$ are already fixed to be $\mb n_{1,2}$, we can uniquely decompose the momentum conservation in the hadron frame into separate conservation of label and residual components:
\begin{align}\label{eq:mom_cons_label}
    \delta_{\mb n_a,\mb n_1}  	 \delta_{\mb n_b,\mb n_2}  	 \delta^{(d)} (p_a + p_b + P_{X_s} - q)  &= 2 \: 
    \delta_{\mb n_a,\mb n_1} \delta_{\omega_a, \tn_2 \cdot \tilde q} \: \delta \big(\tn_2 \cdot (k_a + k_b + P_{X_s} - q_r) \big)  \nn \\
    &\nn \quad \times	\delta_{\mb n_b,\mb n_2}   \delta_{\omega_b, \tn_1 \cdot \tilde q}  \: 
    \delta \big(\tn_1 \cdot (k_a + k_b + P_{X_s} - q_r) \big)
    \\
    &\quad \times \delta^{(d-2)} \big(k_{a\perp}  + k_{b\perp} + P_{X_s\perp} - q_\perp\big)  \, .
\end{align}
Thus, we see that in the sum in \eq{coll_me_2} the label momentum is completely fixed. 

In the next step, we deal with the collinear residual components $k_a$ and $k_b$. Firstly, inside the residual momentum conservation in \eq{mom_cons_label}, the  components $\tn_1 \cdot k_a, \tn_2 \cdot k_b \sim \lambda^2 M_Z$  can be dropped relative to $\tn_1\cdot P_{X_s}, \tn_2\cdot P_{X_s} \sim \lambda M_Z$, such that integration over these components can be moved entirely into the collinear matrix elements considered above:
\begin{align}\label{eq:mom_cons_label_1}
    \delta_{\mb n_a,\mb n_1}  	 \delta_{\mb n_b,\mb n_2}  	 \delta^{(d)} (p_a + p_b + P_{X_s} - q)  &= 2 \: 
    \delta_{\mb n_a,\mb n_1} \delta_{\omega_a, \tn_2 \cdot \tilde q} \: \delta \big(\tn_2 \cdot k_a + \tn_2 \cdot P_{X_s} -\tn_2 \cdot q_r \big)  \nn \\
    &\nn \quad \times	\delta_{\mb n_b,\mb n_2}   \delta_{\omega_b, \tn_1 \cdot \tilde q}  \: 
    \delta \big(\tn_1 \cdot k_b + \tn_1 \cdot P_{X_s} - \tn_1 \cdot q_r \big)
    \\
    &\quad \times \delta^{(d-2)} \big(k_{a\perp}  + k_{b\perp} + P_{X_s\perp} - q_\perp\big)  \, .
\end{align}
Next, the components $\tn_2 \cdot k_a , \tn_1\cdot k_b \lesssim \lambda M_Z$ can be dropped relative to the $\omega_a,\omega_b \sim M_Z$ label momenta in the collinear matrix elements in \eq{coll_a_prelim}. As a result, one can freely integrate over the $\cO(\lambda)$ longitudinal components of the residual collinear momentum:
\begin{align}
    &\int \dfbar \tn_2 \cdot k_a \int \dfbar \tn_1 \cdot k_b \: (2\pi)^d \delta^{(d)} \big(k_a + k_b + P_{X_s} - q_r\big) \big|_{\tn_1 \cdot k_a \ra 0 , \tn_2 \cdot k_b \ra 0}\\
    &\quad \ra  \nn 
    \int \dfbar \tn_2 \cdot k_a \int \dfbar \tn_1 \cdot k_b \:  \delta \big(\tn_1 \cdot k_b + \tn_1 \cdot P_{X_s} - \tn_1 \cdot q_r \big) \delta \big(\tn_2 \cdot k_a + \tn_2 \cdot P_{X_s} -\tn_2 \cdot q_r \big)\\
    &\qquad \quad \quad \times 2(2\pi)^d
    \delta^{(d-2)} \big(k_{a\perp}  + k_{b\perp} + P_{X_s\perp} - q_\perp\big) \, , \nn
    \\
    &\quad = 2 (2\pi)^{(d-2)} \delta^{(d-2)} \big(k_{a\perp} + k_{b\perp} + P_{X_s\perp} - q_\perp\big)  \nn \\
    &\quad=2 \int \df^{d-2} b_\perp \: e^{-\im b_\perp \cdot \big(k_{a\perp} + k_{b\perp} + P_{X_s\perp} - q_\perp \big)}	 \nn 
    \, .
\end{align} 
Now, in \eq{coll_a_prelim} we bring in the integral over the remaining residual components to define
\begin{align}\label{eq:frag_func_prelim}
    \hat \cG_{h_1/f}^{\alpha'\alpha }\big(\zeta_1,   b_\perp \big) &= \frac{\Theta(\omega_a)}{2N_c \zeta_1}\frac{1}{2\pi} \int \frac{\dfbar \tn_1\cdot k_a}{2} \int \dfbar^{d-2} k_{a\perp} \: e^{-\im b_\perp \cdot k_{a\perp}} \,
    \hat \cG_{h_1/  f}^{\alpha'\alpha }\big(\omega_a, k_a, p_{h_1}\big) \big|_{\tn_2 \cdot k_a \ra 0}\, , \nn \\
    \hat \cG_{h_2/ \bar   f}^{\beta \beta' }\big(\zeta_2,   b_\perp \big) &= \frac{\Theta(\omega_b)}{2N_c \zeta_2} \frac{1}{2\pi}\int \frac{\dfbar \tn_2\cdot k_b}{2} \int \dfbar^{d-2} k_{b\perp} \: e^{-\im b_\perp \cdot k_{b\perp}} \,
    \hat \cG_{h_2/\bar f}^{\beta \beta'}\big(\omega_b, k_b , p_{h_2}\big) 	\big|_{\tn_1 \cdot k_b \ra 0}	 \, . \end{align}
The prefactors in the definitions of the collinear objects are chosen so as to give the standard formulae for transverse-momentum-dependent  fragmentation functions (TMDFFs). Note that here TMDFFs are defined in the hadron frame decomposition. Their connection to the standard formulation in the parton frame and matching to collinear fragmentation functions is discussed below in \app{tmd}. 
Finally, these manipulations also lead to the soft function in \eq{qT_soft}  by contracting the colour indices of the soft matrix element with those in \eq{coll_me_2}:
\begin{align}
    \sum_{a,a',b,b'}\delta^{ \bar b' b} \delta^{a ' \bar  a}	S_{n_1,n_2}^{b' \bar a' a \bar b}\big(b_\perp \big)
    = 	N_c 	S_{n_1,n_2}(b_\perp )	\,  .
\end{align}

For unpolarised hadrons, the spin structure can be simplified by decomposing the fragmentation functions into scalar densities:
\begin{align}
    \hat \cG_{h/f} (\zeta, b_\perp) = \Bigg(\cD_{ h/f} - H^{\perp(1)}_{ h/f} M_h \bslash_\perp \Bigg) \frac{\tnslash_1}{4} \,.
\end{align}
While the second term is by itself vanishing when averaged over the azimuthal angle of $b_\perp$, it nevertheless contributes when combined with the corresponding term in $\hat \cG_{h_2/\bar f}$ for the opposite jet. However, it leads to a power suppressed contribution in the limit $b_\perp \ra 0$. Thus, we will ignore this term from the outset and focus on the first scalar density, which is given by
\begin{align}
    \cD_{ h_1/f} (\zeta ,b_\perp) &= \Tr \Big[\hat \cG_{h/f}(\zeta,b_\perp) \frac{\tnslash_2}{2}\Big] \, .
\end{align}
Using \eqs{frag_func_res}{frag_func_prelim}, this fragmentation function admits the following operator definition
\begin{align}\label{eq:frag_func_label}
    &\cD_{ h_1/f} (\zeta ,b_\perp) = \frac{\Theta(\omega_a)}{2\zeta_1 N_c} \int \frac{\dfbar x^+}{2} \: \sumint_{X_a} \Tr  \Big\langle   0\Big | \frac{\tnslash_2}{2} \chi_{\tn_1}  (x^+, 0, b_\perp )\Big  | X_a , h_1\Big \rangle  \Big \langle X_a, h_1 \Big| \overline \chi_{\tn_1, -\omega_a} (0) \Big |0\Big \rangle \,, 
\end{align}
and likewise for the anti-quark fragmentation function:
\begin{align}
    &	\cD_{ h_2/\bar f} (\zeta ,b_\perp)  = \frac{\Theta(\omega_b)}{2\zeta_2 N_c}  \int \frac{\dfbar x^-}{2}\:  \sumint_{X_b} \Tr  \Big\langle   0\Big | \overline  \chi_{\tn_2}  (0, x^-, b_\perp )\Big  | X_b , h_2\Big \rangle  \Big \langle X_b, h_2 \Big| \frac{\tnslash_1}{2} \chi_{\tn_2, -\omega_b} (0) \Big |0\Big \rangle \, .
\end{align}
Upon performing the residual momentum integration, the transverse position of the $\chi_n(0)$ operator is set to $b_\perp$. In \eq{frag_func} we wrote this in the standard form by also bringing in the label momentum into the Fourier transform and neglecting the $\cO(\lambda)$ residual components $\tn_{2,1}\cdot k$ relative to $\omega_{a,b}$. 
To see this, first note that~\cite{Bauer:2002nz}
\begin{align}
	\int \frac{\df x^+}{2} \: e^{\im \left(\omega - \omega ' + r^- \right) \frac{x^+}{2}}  = \delta_{\omega, \omega'} \int  \frac{\df x^+}{2} \:e^{\im \frac{r^- x^+}{2}} \, ,
\end{align}
where $r^- \sim \lambda Q$ is a residual component. In \eq{frag_func_label} we have no residual momentum $\tn_2\cdot k_a$ left after multipole expansion but we leave it here for clarity.  
This is a statement that the momentum conservation first must be implemented at the level of discrete label momenta $\omega$ and $\omega'$. 
Now, for a generic collinear field $\Phi_{n_i}$, writing $\Phi_{n_i,\omega_i} = \delta (\omega_i - \bn_i\cdot \cP) \Phi_{n_i}$, we have 
\begin{align} 
	\int \frac{\df x^+}{2} \: e^{\im x\cdot r_i}\Phi_{n_i,\omega_i} (x)&= 	\int \frac{\df x^+}{2} \: e^{\im \left(\omega_i - \bn_i\cdot \cP  \right) \frac{x^+}{2} + \im x \cdot r_i}  \sum_{\omega_i'} \Phi_{n_i,\omega_i'} (x)\nn
	\\
	&\nn = 	\int \frac{\df x^+}{2} \: e^{\im  \frac{x^+\omega_i}{2} + \im x\cdot r_i}  \sum_{\omega_i'} e^{-\im \omega_i' x^+/2}\Phi_{n_i,\omega_i'} (x)\\
	& =
	\int \frac{\df x^+}{2} \: e^{\im  \frac{x^+\omega_i}{2} + \im x\cdot r_i}\, \hat \Phi_{n_i} (x)	\,. 
\end{align}
In the final line, $\hat \Phi_{n_i}(x)$ denotes the standard field one begins with in the derivation of the collinear Lagrangian that also includes the label momentum in the phase~\cite{Stewart:2013SCETnotes}. At this stage one can combine the label $\omega$ and residual momentum $r^-$ back into one continuous variable. Applying these steps to \eq{frag_func_label} leads to \eq{frag_func} where, for simplicity, we dropped the hat on $\chi_{n}$ fields. 

Finally, we now compute  the leading-order Born cross section:
\begin{align}
    \sigma_0^{(f)} =  \int \df\Phi_{h_1,h_2} \: (2\pi)^d \delta^{(d)}(p_{h_1} + p_{h_2} - q) \bigg[ 2 (d-2)N_c\Big(|a_f|^2 + |v_f|^2\Big) e^2 M_Z^2 \bigg]\,, 
\end{align}
where the quantity in the brackets is the Born matrix element. The phase space integral can be performed by decomposing the hadron momenta in $n$-$\bn$ coordinate system defined relative to the direction of the $Z$ boson. Defining 
\begin{align}
    z_1 = \frac{\bn \cdot p_{h_1}}{Q + P_Z} \,,\qquad 
    z_2 = \frac{\bn \cdot p_{h_2}}{Q + P_Z}  \,,
\end{align}
the phase space integral becomes
\begin{align}
     \int \df\Phi_{h_1,h_2} \: (2\pi)^d \delta^{(d)}(p_{h_1} + p_{h_2} - q) &= \frac{1}{2} \int_0^1 \df z_1 \int \dfbar^{d-2} p_{h_1\perp} \delta \Big(\mb p_{h_1\perp}^2 - M_Z^2 z_1(1-z_1)\Big) \nn \\
     &= \frac{1}{4} \int_0^1 \frac{\df z_1}{\big[M_Z^2 z_1(1-z_1)\big]^\eps} \int \frac{\df \Omega^{(1)}_{d-2}}{(2\pi)^{d-2}} \,,
\end{align}
where $p^\mu_{h_1\perp}$ and the associated $d-2$-dimensional solid angle $\Omega^{(1)}_{d-2}$ are defined relative to $n$-$\bn$ axes in the transverse plane. Thus, we have
\begin{align}
    \sigma_0^{(f)} = \frac{1}{2} (d-2)N_c\Big(|a_f|^2 + |v_f|^2\Big) e^2 M_Z^{2-2\eps} \frac{\Omega_{d-2}}{(2\pi)^{d-2}} r_\eps  \, ,
\end{align}
where
\begin{align}\label{eq:reps}
    &\Omega_{d-2} = \frac{2\pi^{1-\eps}}{\Gamma(1-\eps)} \,,& 
    &r_\eps \equiv \frac{\Gamma^2(1-\eps)}{\Gamma(2-2\eps)} \,.&
\end{align}

Combining all the pieces, we find
\begin{align}
    \frac{1}{\sigma_0}\frac{\df \sigma(q)}{\df \Phi_{h_1,h_2}} &= 8(2\pi)^2 \zeta_1 \zeta_2 \sum_f |M_0^{(f)}|^2 \cH_f^{(0)}(M_Z^2)   \\
    &\quad \times \nn 
    \int \df^{d-2} b_\perp \: e^{+\im b_\perp \cdot q_\perp} \Big[\cD_{ h_1/f}\big(\zeta_1, b_\perp\big) \cD_{ h_2/\bar f}\big(\zeta_2, b_\perp \big) S_{n_1n_2} (b_\perp) + \big(1\ra 2\big)\Big] \nn \, .
\end{align}
where $|M_0^{(f)}|^2$  is given by
\begin{align}\label{eq:M0f_def}
    |M^{(f)}_0|^2 &= \frac{1}{\sum_f \sigma_0^{(f)}}N_c\bigg( \frac{q_\mu q_\nu}{M_Z^2}  - g_{\mu\nu} \bigg) \Tr \Bigg[
    \frac{\tnslash_1}{4} \Gamma_\perp^{\nu} \frac{\tnslash_2}{4} \Gamma_\perp^\mu 
    \Bigg]\nn \\
    &=  \frac{1}{\sum_f \sigma_0^{(f)}} (d-2) N_c e^2 \big(|a_f|^2 + |v_f|^2\big)\nn
    \\
    &= \frac{2 (2\pi)^{d-2}}{M_Z^{2-2\eps} \Omega_{d-2} r_\eps}  \frac{|a_f|^2 + |v_f|^2  }{\sum_{f'} (|a_{f'}|^2 + |v_{f'}|^2)}   \, . 
\end{align}

The above result, after matching to collinear fragmentation functions and employing the sum rules, reproduces the well-known factorisation for the back-to-back EEC derived in \Refcite{Moult:2018jzp}.  However, it is worth remarking on the differences with the derivation presented here. In the original derivation of \Refcite{Moult:2018jzp}, the reference vectors were aligned along the jet axes, and the $\perp$-component of the soft momenta was defined with respect to the thrust axis. The $z_{12}$ measurement is then given by a combination of transverse hadron momenta with respect to the jet axes and the soft momenta with respect to the thrust axis (as per eq.~(3.7) of \Refcite{Moult:2018jzp}). On the other hand, in the derivation above we chose the reference vectors to lie along the outgoing hadrons, in order to make the boosting properties immediate. This choice directly gives access to the relative angle $z_{12}$, without referencing to jets or the thrust axis. This reflects both the highly inclusive nature of the EEC and the underlying boosted Born kinematics,  better facilitating its generalisation to higher-point correlations and top jets.

\section{Fragmentation functions}
\label{app:tmd}

Here we compile relevant formulae and identities related to fragmentation functions. These were first presented by Collins and Soper in \Refcite{Collins:1981uw} (see also \Refcite{Collins:2011zzd}).

\subsection{Definitions in parton and hadron frames}

When describing the transverse-momentum-dependent (TMD) fragmentation functions, we must be specific as to which frame we are considering. 
Fragmentation functions are naturally defined in the parton frame decomposition discussed in \secn{born}, whereas the hadron frame discussed in \secn{hadron_frame} is the most convenient for deriving factorisation. Consider a parton with momentum $k$ and an identified hadron with momentum $P$, with components given by 
\begin{align}\label{eq:parton_hadron_frame}
    &\text{Parton frame:}&	&k = \big(k_p^+, \omega, \mb 0_\perp\big) \,,& &P = \big(P_p^+, z\omega , \mb P_\perp \big)\,,&
    \\
    &\text{Hadron frame:}&
    &k = \big(k_h^+, \omega, \mb k_\perp\big)\,,&
    &P= \big(P_h^+, z\omega, \mb 0_\perp \big)\,.&
    \nn
\end{align}
Here the reference vector $n$ is chosen to lie along either the parton or hadron in the two cases. 

For a given vector in the hadron frame decomposition,
\begin{align}\label{eq:frameconvert}
    V_p^+ =V_h^+ +  \frac{\mb k_\perp^2}{\omega^2} V_h^- - \frac{2\mb k_\perp \cdot \mb V_{h\perp}}{\omega}
    \, , \qquad 
    \mb V_{p\perp} = \mb V_{h\perp} - \mb k_{\perp} \frac{V_h^-}{\omega}
    \, .
\end{align}
Hence, from \eq{parton_hadron_frame} we have
\begin{align}
    \mb P_\perp = -z \mb k_\perp \, ,
\end{align}
where it is understood that these are components of two different vectors in two different frames. 

The fragmentation function in the parton frame decomposition is defined as distribution of hadrons carrying momentum fraction $z$ and transverse momentum $P_\perp$ relative to the progenitor parton. The operator definition is given by
\begin{align}
    \cF_{h/q } (z, P_\perp) &= \frac{1}{2zN_c}\int \frac{\dfbar x^+}{2} \dfbar^{d-2} x_\perp\: e^{\im \frac{1}{2} \omega  x^+}  \sumint_X \Tr \langle 0 | \frac{\bnslash}{2} \chi_n (x) | h ,X\rangle \langle h, X | \overline \chi_n (0) | 0 \rangle_{\text{part. fr.}}	\,	.
\end{align}
where we have adopted the notation of \Refcite{Luo:2019hmp}. 
The corresponding TMDFF in the hadron frame is given by
\begin{align}\label{eq:Dhqk}
    \cD_{h/q} (z, k_\perp) &=	
    \frac{1}{2zN_c}\int \frac{\dfbar x_h^+}{2} \dfbar^{d-2} x_{h\perp}\: e^{\im \frac{1}{2} \omega  x_h^+	-	\im \mb k_\perp \cdot \mb x_{h\perp}}  \sumint_X \Tr \langle 0 | \frac{\bnslash}{2} \chi_n (x_h) | h ,X\rangle \langle h, X | \overline \chi_n (0) | 0 \rangle \, .\end{align}
The $x_h$ are hadron frame coordinates. 
We can relate the two definitions by shifting the coordinate $x_{h\perp}$ inside $\chi_n$  to $0_\perp$: 
\begin{align}
    \langle 0 | \chi_n (x^+ , 0, x_{h\perp}) | h ,X \rangle &= e^{-\im x_{h\perp}\cdot P^\perp_{X}} 	\langle 0 | \chi_n (x^+ , 0, 0_\perp ) | h ,X \rangle \, , \\
    &= 
    \nn 
    e^{+\im \mb x_{h\perp} \cdot  \mb k_\perp }	\langle 0 | \chi_n (x^+ , 0, 0_\perp ) | h ,X \rangle	\,	,
\end{align}
where we used the fact that $P_{X}^\perp$ is the same as transverse momentum of the original parton and $h$ has zero transverse momentum in this hadron frame decomposition. This implies numerical equality of the two probabilities:
\begin{align}
    \cD_{h/q}  (z, k_\perp)  = \cF_{h/q} (z, P_\perp	=	-zk_\perp) 	\,	.
\end{align} 

The Fourier transform of the hadron frame TMDFF is given by
\begin{align}\label{eq:Dhqb}
    \cD_{h/q} (z, b_\perp) &= \int \df^{d-2} k_\perp \: e^{+\im \mb b_\perp \cdot \mb k_\perp} \cD_{h/q} (z, k_\perp)\\
    &=
    \nn 
    \frac{1}{2zN_c}\int \frac{\dfbar x_h^+}{2} \: e^{\im \frac{1}{2} \omega  x_h^+}  \sumint_X \Tr \langle 0 | \frac{\bnslash}{2} \chi_n (x_h^+, 0, b_\perp) | h ,X\rangle \langle h, X | \overline \chi_n (0) | 0 \rangle   \, .
\end{align}
Whereas for the parton frame TMDFF we have
\begin{align}
    \cF_{h/q} (z, b_\perp) &= \int \df^{d-2} P_\perp \: e^{+ \im \mb b_\perp \cdot \mb P_\perp } \cF_{h/q} (z, P_\perp) 
    \\
    &= z^{d-2}\int \df^{d-2} k_\perp \: e^{-\im z \mb b_\perp \cdot \mb k_\perp } \cD_{h/q} (z, k_\perp)\nn \\
    &= \nn 
    z^{d-2} \cD_{h/q} (z, -zb_\perp) \, .
\end{align}

\subsection{Matching to collinear fragmentation functions}

The integrated (or collinear) fragmentation functions (FFs) are defined as integral of the TMDFFs in the parton frame decomposition: 
\begin{align}
    d_{h/q} (z) \equiv \int \df^{d-2} P_\perp \: \cF_{h/q} (z, P_\perp)	\,	.
\end{align}
Its operator definition is however written in the hadron frame where the Fourier transform is explicit. By the change of variables $P_\perp \ra -zk_\perp$ in the definition above, one obtains
\begin{align}\label{eq:dhqOp}
    d_{h/q} (z) &= z^{d-2} \int \df^{d-2} k_\perp \cD_{h/q} (z, k_\perp) \\
    &\nn 
    =
    \frac{z^{d-3}}{2  N_c}  \int \frac{\dfbar x^+}{2} \: e^{\im \frac{1}{2}\omega x_h^+} \sumint_{X}\Tr \langle 0 | \frac{\bnslash}{2}\chi_n (x_h^+, 0, 0_\perp) | h, X \rangle \langle h, X | \overline \chi_n (0) | 0 \rangle \, .
\end{align}

The matching relation between TMDFFs and FFs for $b_\perp \ra 0$ is given by
\begin{align}\label{eq:FFmatch}
    \cD_{h/q} (z, b_\perp) = \int_z^1 \frac{\df \xi}{\xi } \: \frac{1}{\xi^{d-2}}\:  \cC_{jq} \bigg(\frac{z}{\xi}, b_\perp \bigg) d_{h/j} (\xi) 
    \, .
\end{align}
Note, that $1/\xi^{d-2}$ is required to compensate the relevant factor in \eq{dhqOp}. In terms of parton frame FF, we have
\begin{align}
    \cF_{h/q} \bigg(z, \frac{b_\perp}{z}\bigg) = \int_z^1 \frac{\df \xi}{\xi} \Big(\frac{z}{\xi}\Big)^{d-2} \: \cC_{jq}\bigg(\frac{z}{\xi}, b_\perp\bigg) d_{h/j} (\xi)	\,	.
\end{align}

\bibliography{qcd}

\end{document}